\documentclass[%
 amsmath,amssymb,
 aps,
]{revtex4-2}

\usepackage{graphicx}% Include figure files
\graphicspath{{./figure/}}
\usepackage{dcolumn}% Align table columns on decimal point
\usepackage{bm}% bold math
\usepackage{here}
\usepackage{color}
\usepackage{ulem}

\begin{document}

\preprint{APS/123-QED}

\title{Molecular dynamics simulations of head-on low-velocity collisions between particles}

\author{Yuki Yoshida$^{1}$} \email{yoshida.yuki.astpl@gmail.com}
\author{Eiichiro Kokubo$^{2}$}
\author{Hidekazu Tanaka$^{3}$}

\affiliation{
$^{1}$Center for Planetary Science, Kobe University, 7-1-48 Minatojimaminami, Chuo-ku, Kobe, Hyogo 650-0047, Japan, \\
$^{2}$Center for Computational Astrophysics, National Astronomical Observatory of Japan, 2-21-1 Osawa, Mitaka, Tokyo 181-8588, Japan, \\
$^{3}$Astronomical Institute, Graduate School of Science, Tohoku University, 6-3, Aramaki, Aoba-ku, Sendai 980-8578, Japan
}

\date{\today}% It is always \today, today,

\begin{abstract}

%E1114: CORの半径依存性の結果がない?
%Y: 入れました。
%Y0621: 現在形で統一

The particle contact model is important for powder simulations.
Although several contact models have been proposed, their validity has not yet been well established.
%E1104: powder simulation とは DEM のこと?
%Y DEMのことを表しています。
%However, their validity is not well established.
%E: JKR theory、but を使わないで文を書きたい。JKR を使うなら簡単な説明が必要。
Therefore, we perform molecular dynamics (MD) simulations to clarify the particle interaction.  
We simulate head-on collisions of two particles with impact velocities less than a few \% of the sound velocity to investigate the dependence of the interparticle force and the coefficient of restitution on the impact velocity and particle radius.
In this study, we treat particles with a radius of 10-100 nm and perform simulations with up to 0.2 billion atoms.
%\textcolor{red}{We focus mainly on the collisions with impact velocities as much as that where the pressure about yield stress occurs and investigate the process of the energy dissipation.}
We find that the interparticle force exhibits hysteresis between the loading and unloading phases. %ここはだめdue to plastic deformation near the contact surface.
Larger impact velocities result in strong hysteresis and plastic deformation.
%%%%%%%%%%%%%%%%%%%%%%%%%%%%%%%%%%%速い衝突で塑性変形がみられた。
%E1104: 変形は接触面近くだけではないのでは?
%Y ヒステリシスが起こる原因の変形は接触面付近が主であると考えられます。確かに変形はこれまでの発表のように高速度衝突時に球表面にも表れますが、ヒステリシスは低速度衝突時でも起こります。
%The results showed hysteresis between loading and unloading.
%The hysteresis is caused by the decrease in elastic displacement due to plastic deformation near the contact surface.
%E: decrease とは?
For all impact velocities and particle radii, the coefficient of restitution is smaller than that given by the Johnson-Kendall-Robert theory, which is a contact model that gives the force between elastic spherical particles.
%E1104: JKR の簡単な説明が必要。
%Y 修正しました
A contact model of inelastic particles cannot reproduce our MD simulations.
In particular, the coefficient of restitution is significantly reduced when the impact velocity exceeds a certain value.
This significant energy dissipation cannot be explained even by the contact models including plastic deformation.
%%%%%%%%%%%%%%%%%%%%%%%%%%%%%%%%%%%%%%%%%%%%%%%%%%%従来の研究からよすされる反発係数よりも
We also find that the coefficient of restitution increases with increasing particle radius.
%E1104: なるべく数字を使いたくない。使う場合は無次元量の意味を説明する?
%Y 数字をつかわないようにしました
We also find that the previous contact models including plastic deformation cannot explain the strong energy dissipation obtained in our MD simulations, although they agree with the MD results for very low impact velocities.
Accordingly, we have constructed a new dissipative contact model in which the dissipative force increases with the stress generated by collisions.
%In particular, the impact velocity dependence shows a significant decrease in the repulsion coefficient with impact velocity for $v_{\rm imp}>0.10$, a decrease that is difficult to reproduce in any extended model.
%E: extended model の説明必要。
The new stress-dependent model successfully reproduces our MD results over a wider range of impact velocities than the conventional models do.
%We also construct a simple model with a dissipative force proportional to the cube of the impact velocity and the contact radius to the 1.5 power to reproduce the MD results well at low impact velocity.
In addition, we proposed another, simpler, dissipative contact model that can also reproduce the MD results.
%A model that considers energy dissipation due to deformation over the entire particle at high-velocity collisions is needed.

\end{abstract}

%Use show keys class option if keyword display desired

\maketitle

%\tableofcontents

\section{Introduction \label{sec:intro}}

Particle collisions are universal processes that occur in many fields of science and technology \cite[e.g.,][]{REYNOLDS20053969,JUTZI2008242,MISHRA2001225}.
In mineral-processing engineering, for example, fine ores form agglomerates such as pellets; particle collision processes are essential for their formation, growth, and fragmentation.
In the fertilizer industry, agglomeration technology is used to form granules.
In astrophysics, dust aggregates---the starting materials of planets---are formed by many silica and water ice particles.
%ここ修正
In this process, submicron-sized dust particles stick together at collision velocities of $\lesssim 100$ m/s to form agglomerates.
%ダストの衝突合体を考える上で、エネルギー散逸を考えることは重要である。
% 惑星形成を考える上で、granular力学の研究は重要である。(引用)
Furthermore, the structure of the small bodies of the solar system is described by the granular mechanic's theories, which have been studied theoretically, experimentally, and numerically \citep[e.g.,][]{2019A&ARv..27....6H}.

Powder simulations have been used to study the physics of agglomerates, including the structures and tensile strengths of agglomerates, the threshold collision velocity for fragmentation, and fragment distributions \cite[e.g.,][]{2008ApJ...684.1310S,2013A&A...559A..62W,2023ApJ...944...38H,2021ApJ...915...22H,2019ApJ...874..159T,2008ApJ...677.1296W}.
%Thus, powder simulation is one of the powerful methods for understanding various physical processes.
In powder simulations, the interaction forces and torques between particles are usually calculated assuming spherical particles.
Several contact models have been proposed for spherical particles.
The Johnson--Kendall--Roberts (JKR) theory \cite{1971RSPSA.324..301J,1995PMagA..72..783D,1996PMagA..73.1279D,1997ApJ...480..647D,1987come.book.....J} is the model often used in the powder simulations \citep[e.g.,][]{2008ApJ...677.1296W,2013A&A...559A..62W,2008ApJ...684.1310S,2021ApJ...915...22H,2023ApJ...944...38H,2019ApJ...874..159T}.
The JKR theory assumes a pressure distribution in which the pressure diverges at the rim of the contact region.
%E1104: the edge of the contact region とはどこ?
%Y: 接触面縁のことですが、rimにしておきました。
This model is consistent with the experiments that use large particles.
Besides the JKR theory, other contact models have been proposed.
The Derjaguin--Muller--Toporov (DMT) theory is another contact model \cite{1975JCIS...53..314D}; it is suitable for small particles and assumes that adhesive forces act around the rim of the contact area between the particles.
The Maugis--Dugdale solution was proposed for medium-sized particles \cite{1960JMPSo...8..100D,1992JCIS..150..243M}.
Several previous studies indicated that although these models describe the basic contact process, their treatments of energy dissipation are insufficient.
In particular, a study using molecular dynamics simulations showed that the kinetic energy of the macroscopic particles is converted into molecular motions, which results in more energy dissipation than in the JKR theory \cite{2012PThPS.195..101T}.
Other molecular dynamics simulations of head-on collisions between non-adhesive particles demonstrated that plastic deformation of the particles results in the dissipation of kinetic energy, and they also obtained the yield velocity at which the plastic deformation begins \cite{2014PhRvE..89c3308T,PhysRevE.92.0324032015}.
Molecular motion and particle deformation result in energy dissipation and affect the physical processes of the powder system, but these effects are not included in the contact models listed above.
Krijt et al. \citep{2013JPhD...46Q5303K} constructed a new dissipation model that adds the viscoelasticity of the particles and the effect of the plastic deformation to the JKR theory.
Krijt et al. showed the validity of their model for micron-sized particles by comparing it to some experiments of collisions with impact velocities lower than a few \% of sound velocities.
It is necessary to check the validity of their model through microscopic molecular dynamics simulations.
%Krijtは衝突速度の音速度のa few%以下の衝突実験と比較している。
%Krijt マクロ サブミクロン粒子はミクロ それをMDを使ってテストする
%そのためには非常に膨大な分子数が必要である。
%(先行MD研究は20nm程度、数ミリオン、我々はサブビリオン個まで)←後に書く

Some contact models do include particle deformation and molecular motions.
For example, crack propagation in the contact area due to the molecular motion has been studied \cite{2004JPhD...37.2557G}.
Dissipation forces due to the viscoelastic deformation were also considered \cite{2013JPhD...46Q5303K,2007PhRvE..76e1302B}.
%E1105: 上の2文は能動態過去形にする。
%Y 修正しました。
The contact models that include these effects are consistent with experiments on collisions between particles larger than a few $\mu$m in size \cite[e.g.,][]{2013JPhD...46Q5303K,2008JAerS..39..373K,1990AerST..12..926W}.
However, their validity for smaller particles has not yet been confirmed.
%, and the validity of those contact models needs to be investigated.
%The molecular dynamics simulations can treat the contact for particles smaller than $\mu$m.
Molecular dynamics simulations are particularly useful to clarify the actual particle interactions for such small particle sizes.

Molecular dynamics (MD) simulations are used to study the physical processes involved by analyzing molecular motion.
%In the MD system, the simulation is performed as a molecular $N$-body system.
In other words, MD simulations are molecular $N$-body simulations.
Such MD simulations have been used primarily in engineering, biology, and even in the study of particle contact dynamics.
In particularly, collisions between nanoparticles have been well investigated \cite[e.g.,][]{2021A&A...647L..13A,2020A&A...641A.159N,2020NRL....15...67N,2018RSPSA.47470723T,2016PhRvE..93f3004M,2017GeoRL..4410822N,2020Icar..35213996N,2017PCCP...1916555N,2019PhRvE..99c2904N, 2021NatSR..1114591U}.
%, as have forces and torques involving rolling and torsional motions \cite{2021NatSR..1114591U}.
%E1105: as 以下不要? 書くなら rolling and torsional motions の説明必要。
%Y: as以下を消去しました。
Collisions between a nanoparticle and an atomic cluster were also simulated, and their rebound or adhesion was studied \cite{2007PhRvB..76k5437A}.
%Y0123: 以下3行新規
Nietiadi et al. \citep{2017PCCP...1916555N,2020NRL....15...67N} simulated the collisions between silica nanoparticles with an impact velocity of less than 1000 m/s and investigated the COR and bouncing threshold.
They used Si and O atoms to model the silica nanoparticles, and covalent bonds bind these atoms.
Most of these studies focused on high-velocity collisions of nanoparticles with impact velocities exceeding 1000 m/s, about 20 \% of the sound velocity, to investigate particle melting and fragmentation, which is beyond the scope of contact models.
Although there have been few MD simulation studies with impact velocities of the order of tens of m/s, a few \% of the sound velocity, the validity of contact models under these conditions has not yet been confirmed.
Tanaka et al. \citep{2012PThPS.195..101T} used the Lennard-Jones potential as the interatomic potential and investigated the collisions between submicron-sized particles with an impact velocity of less than 50 m/s, assuming argon atoms.
They found that the COR obtained from the MD simulations is smaller than that predicted by the JKR theory and that collisions with large velocities cause strong deformation and energy dissipation.
Nietiadi et al. \citep{2019PhRvE..99c2904N, 2020NRL....15...67N} investigated the collisions between nanoparticles of radii less than 20 nm.
In this way, the previous studies mainly simulated collisions between nanoparticles with a radius on the order of 10 nm, consisting of several million molecules, and since low collision velocities result only in coalescence, they investigated collision velocities with about 10\% of the sound velocity, although such collision processes cannot be described by the contact model presented above.
As the particles become larger, they start to bounce back even at a few \% of the sound velocity, and such a collision process can be compared to the contact model.
To validate the model, we should use particles of about 100 nm or less, where bouncing occurs.
Here, we use MD simulations to explore the realistic particle interactions with submicron particles at impact velocities of around 100 m/s.
%We would like to investigate the validity of the contact model with submicron particles at impact velocities of around 100 m/s.
%E1105: JKR の妥当性を確かめたいというよりも、MDシミュレーションで真の粒子間相互作用を明らかにしたい、では? 
%Y 上記のように修正しました
We have also developed a new contact model with dissipation that can reproduce MD simulations.

In this paper, we study particle collisions using MD simulations.
The JKR theory and a dissipative model of Krijt et al. \cite{2013JPhD...46Q5303K} are first introduced in Sec.~\ref{sec:model} because the JKR theory is often used in the powder simulations and the dissipative model is an extended model of the JKR theory.
We next explain our simulation method, the particle models, and the initial condition of the collisions in Sec.~\ref{sec:method}.
%E1105: 方法について足す。
In Sec.~\ref{sec:results}, we present the results of the MD simulations in which particle interactions and the coefficient of restitution are investigated.
%E1105: COR をスペルアウト?
%Y 修正しました
Based on these results, we construct new contact models in Sec.~\ref{sec:expand}.
Finally, we summarize our results and discuss future work in Sec.~\ref{sec:Summary}.

\section{contact model} \label{sec:model}

\subsection{Hertz theory} \label{sub:contact}

In this section, we first explain the models of the interaction between two elastic spheres in contact. 
We start with the Hertz theory.
We consider two particles with radii $R_1$ and $R_2$.
They have Young's moduli $E_1$ and $E_2$, Poisson's ratios $\nu_1$ and $\nu_2$, and masses $m_1$ and $m_2$.
We introduce the reduced particle radius $1/R^* = 1/R_1 + 1/R_2$, the reduced Young's modulus $1/E^* = (1-\nu_1^2)/E_1 + (1-\nu_2^2)/E_2$, and the reduced mass $1/m^*=1/m_1+1/m_2$.
When the spheres are in contact, the contact surface can be treated as a disk with radius $a$.
The mutual approach $\delta$ is the compression length given by
\begin{equation}
    \delta = R_1 + R_2 - X,
\end{equation}
where $X$ is the distance between the centers of the two particles.
In the Hertz theory, connection and disconnection occur at $\delta=0$; thus, $\delta$ is positive for particles in contact. 
%Two particles are compressed in the contact state, and the mutual approach is denoted by $\delta = R_1 + R_2 - X$, where $X$ is the distance between the centers of two particles.
%Using these parameters, we can write the pressure distribution in the contact area as a function of the distance from the center of contact $r$;
%\begin{equation} \label{eq:pressure}
%    p(r) = \frac{E^*}{\pi R^*} \frac{a^2-2r^2+\delta R^*}{\sqrt{a^2-r^2}}\ (0\leq r\leq a).
%\end{equation}
%At the center of the contact area ($r=0$), the pressure is maximum as
%\begin{equation} \label{eq:center_P}
%    p_{\rm c} = p(0) = \frac{E^*}{\pi R^*}\frac{a^2+\delta R^*}{a}.
%\end{equation}
%We can obtain the elastic force $F_{\rm E}$ by integral of $p(r)$ over the contact area as
%\begin{equation} \label{eq:elastic}
%    F_{\rm E} = \frac{2E^*}{3R^*} (3a\delta R^* - a^3).
%\end{equation}
%Based on eq.~(\ref{eq:elastic}), we can derive the Hertz theory and the JKR theory.

The Hertzian force can be written as follows:
\begin{equation} \label{eq:hertz}
    F_{\rm H} = \frac{4E^*a^3}{3R^*},
\end{equation}
where the radius of the contact surface between the two particles, $a$, is given by $\sqrt{R^* \delta}$ in the Hertz theory. 
The Hertzian force is always repulsive and non-adhesive, as a positive force denotes a repulsive one.
%A positive force means a repulsive one. 
%The Hertzian force is always repulsive and non-adhesive.
%In the Hertz theory, $\delta$ and $a$ have the relation as $\delta R^* = a^2$.
%The mutual approach is always $\delta \geq 0$ and the Hertzian force is always positive and non-adhesive.
%In this theory, the moments of connection and disconnection are the same at $\delta = 0$.
The forces are identical in both the loading and unloading phases, and there is no hysteresis.
As a result, the coefficient of restitution is always $e=1$ in the Hertz theory.
We can also obtain the Hertzian potential energy by integrating eq.~(\ref{eq:hertz}):
\begin{equation} \label{eq:pot_H}
    U_{\rm H} = \frac{8E^*}{15} {R^*}^{1/2} \delta^{5/2}.
\end{equation}
The pressure at the center of the contact area increases with $\delta$ and is given by
\begin{eqnarray} \label{eq:hertz-p}
    p_{\rm c, H} = \frac{2E^*}{\pi} {R^*}^{-1/2} \delta^{1/2}.
\end{eqnarray}

\subsection{Johnson--Kendall--Roberts theory} \label{sub:Hertz-JKR}

In the JKR theory \citep{1971RSPSA.324..301J,1987come.book.....J}, $\delta$ and $a$ are related as
\begin{equation} \label{eq:delta-a}
	\delta R^* = a^2 - \sqrt{\frac{4\pi \gamma a {R^*}^2}{E^*}},
\end{equation}
where $\gamma$ is the surface energy per unit area\footnote{Note that $\gamma$ in this paper is equal to $\gamma_L$ and twice $\gamma$ of Krijt et al. \cite{2013JPhD...46Q5303K}.}.
%E1111: これはここに必要?
%Y: footnoteにしました。(PhyRevEはfootnoteは参考文献に組み込まれます。)
The contact radius is obtained from this equation as a function of $\delta$.
%In the powder simulations, the mutual approach $\delta$ is easily obtained and the contact radius is determined by $\delta$ using eq.~(\ref{eq:delta-a}).
The force between the particles is given by 
\begin{equation} \label{eq:JKR}
    F_{\rm J} = \frac{4E^*a^3}{3R^*} - \sqrt{16\pi \gamma E^* a^3}.
\end{equation}
%The maximum force to separate the two contact particles is $F_c = 3\pi \gamma R^*$.
%This formula suggests that the positive force is the repulsive force and the negative one is the attractive force.
A positive value of $F_{\rm J}$ again indicates a repulsive force, whereas a negative one indicates an attractive force.
The potential energy in the JKR theory can be written as follows
\begin{equation} \label{eq:pot_J}
    U_{\rm J} = a^3 E^* \left[ \frac{1}{5}\left(\frac{a}{R^*}\right)^2 - \frac{2}{3}\frac{\delta}{R^*} + \left(\frac{\delta}{a}\right)^2 \right] -2\pi \gamma a^2.
\end{equation}
%The first term in this equation is the elastic potential energy, and the second term is the surface energy.
This equation consists of two terms: the first represents the elastic potential energy, and the second represents the surface energy. 
The maximum absolute value of the attractive force between the two contacting particles, $F_c$, is given by $3\pi \gamma R^*$, and it occurs at $\delta=-0.397\delta_0$, where $\delta_0$ is the stationary mutual approach.
%E1111: F_c -> F_J,c ?
%Y: 他の論文ではF_cが一般的なので、こちらにしたいと思います。
%E1111: delta_0 の説明必要。
%Y: ,where~で説明を入れました。
The stationary point exists where $F_{\rm J}=0$ and is denoted by the subscription 0.
The stationary contact radius $a_0$ and the stationary mutual approach $\delta_0$ are given by
\begin{equation} \label{eq:a0delta0}
a_0 = \left( \frac{9\pi \gamma {R^*}^2}{E^*} \right)^{1/3},\ \delta_0=\frac{a_0^2}{3R^*}.
\end{equation}
%E1111: 式は2つに分ける?
%Y: 分けても分けなくても良いのかなと思いましたが、見やすさですか？
The pressure at the center of the contact area is given
\begin{eqnarray} \label{eq:JKR-p}
    p_{\rm c, J} = \frac{2E^*a}{\pi R^*} - \sqrt{\frac{4\gamma E^*}{\pi a}}.
\end{eqnarray}

%In the JKR theory, the moment of contact is $\delta=0$ and the disconnection occurs when $\delta=-(3/4)^{2/3}\delta_0$.
%The moments of contact and disconnection are not the same.
%This difference leads to the constant energy dissipation $\Delta K_{\rm p}$ as $\Delta K_{\rm p} \simeq 0.773 F_c\delta_0$, and the COR is given as $e = \sqrt{1 - \Delta K_{\rm p}/K_{\rm p,ini}}$ \cite{2007ApJ...661..320W}, where $K_{\rm p}$ is the particle kinetic energy.
%The collisional process in the JKR theory is explained in Sec.~\ref{sub:interaction} detail.
In the JKR theory, contact between the two particles starts at $\delta=0$, and the disconnection occurs at $\delta \simeq -0.825\delta_0$.
The difference in $\delta$ between the connection and disconnection represents hysteresis, which indicates energy dissipation in the amount of $\simeq 0.773 F_c \delta_0$ \cite{2007ApJ...661..320W}, which gives the coefficient of restitution in the JKR theory.
%E1111: K_p の説明必要。ここで説明しないなら \Delta K_p は使わない。
%Y: 使わないことにしました。

\subsection{Krijt model} \label{sub:krijt}

The contact model of Krijt et al. \cite{2013JPhD...46Q5303K} (hereafter referred to as the Krijt model) includes two energy dissipation mechanisms. 
One is the bulk dissipative force due to the viscoelasticity \cite{2007PhRvE..76e1302B,1996PhysRevE.53.5382B}. 
The other is a delayed evolution of the contact radius based on the viscoelastic crack theory \cite{2004JPhD...37.2557G}.
%We explain the dissipation model \cite[e.g.,][]{2013JPhD...46Q5303K,PhysRevE.53.5382B,2007PhRvE..76e1302B}, which is called the Krijt model in this paper.
%The Krijt model includes plastic deformation and viscoelastic dissipation.
%The plastic deformation is represented by adding a dissipative force.
%The viscoelastic dissipation, which is caused at the rim of the contact area, makes the contact radius evolution delay.
%Then, the Krijt model solves the evolution of $a$ independently.
%These two effects cause the hysteresis of interacting force between particles.
%Then, we show the two hysteresis details.

In the Krijt model, the interaction force between two particles in contact consists of two components, and the equation for relative motion is given by
%The Krijt model adds the dissipative force to the equation of motion as
\begin{equation} \label{eq:EoS}
	m^* \frac{{\rm d}^2\delta}{{\rm d}t^2} = -(F_{\rm E} + F_{\rm dis, K}),
%E1111: 添字の Kriit は K にする? dis は d にする?
%Y: Krijt → K にしました。dis はこのままにしようと思います。
\end{equation}
where $F_{\rm E}$ is the elastic force and $F_{\rm dis, K}$ is the dissipative force:
\begin{eqnarray} 
    F_{\rm E} & = & \int_0^a 2\pi r p(r) {\rm d}r, \label{eq:elastic-force} \\
    F_{\rm dis, K} & \simeq & \frac{T_{\rm vis}}{\nu^2} \int_0^a 2\pi r\dot{\delta} \frac{\partial p(r)}{\partial \delta} {\rm d} r = \frac{2T_{\rm vis} E^*}{\nu^2}a\dot{\delta}, \label{eq:krijt-force}
\end{eqnarray}
where $p(r)$ is the pressure distribution across the contact area \cite{1980JCIS...77...91M,2013JPhD...46Q5303K} and $T_{\rm vis}$ is the relaxation time \cite{1996PhysRevE.53.5382B, 2007PhRvE..76e1302B}.
Since $p(r)$ depends on both $\delta$ and $a$, the elastic force $F_{\rm E}$ has hysteresis due to the delayed evolution of $a$.
%E1111: ?
%Y: slackに回答
Some studied suggested that $T_{\rm vis}$ is approximately proportional to the particle radius \cite{2013JPhD...46Q5303K, 2021ApJ...910..130A}.
%The time differentiation $\dot{\delta}$ is equal to the relative velocity of two particles.
%We should note that the dissipation force has the same sign as the relative velocity $\dot{\delta}$ because the dissipation force $F_{\rm dis, K}$ is opposite to the motion.
%This formula shows that the dissipation force is one of the viscous resistance.
We should note that the dissipation force $F_{\rm dis, K}$ has the opposite sign to the relative velocity ${\rm d}X/{\rm d}t$ because $\delta=R_1+R_2-X$.
Thus, this dissipation force acts in the same way as a viscous resistance. 

To describe the evolution of the contact radius, the Krijt model uses the Griffith theory \cite{1921RSPTA.221..163G} instead of eq.~(\ref{eq:delta-a}). 
The Griffith theory describes the crack propagation and gives the rate of evolution of the contact radius, $\dot{a}$, which is proportional to $1/T_{\rm vis}$. 
The detailed formula for $\dot{a}$ is presented in Appendix~\ref{App:a}.
The viscoelastic effect in crack propagation results in a delay in the evolution of $a$ compared with that given by eq.~(\ref{eq:delta-a}) and causes hysteresis in the evolution of the contact radius and the elastic force $F_{\rm E}$ between loading and unloading phases \cite[e.g.,][]{2013JPhD...46Q5303K, WAHL2006178}. 
In our MD simulations for nano- and submicron-sized particles, however, we find that the hysteresis of $F_{\rm E}$ due to the delayed $a$ is much smaller than that of the dissipative force $F_{\rm dis, K}$.
%E1111: ?
%Y: ここでは言葉で触れ、接触半径aはJKR的であるということに触れています。
%The evolution rate of the contact radius $\dot{a}$ is described by the Griffith theory for the crack propagation \cite{1921RSPTA.221..163G}.
%The viscoelastic effect in the crack propagation causes a difference in the contact radius and the elastic force between the closing (loading) and opening (unloading) phases \cite{2004JPhD...37.2557G, 2006JCIS..296..284G}
%Second, the hysteresis of the contact radius evolution comes from the difference between opening and closing crack \cite{2004JPhD...37.2557G, 2006JCIS..296..284G}.
%The viscoelastic crack is described by the Griffith theory \cite{1921RSPTA.221..163G}.
%To consider propagating the crack in the viscoelastic particles, the apparent surface energy $G_{\rm eff}$ is introduced, which differs from $\gamma$.
%The detailed formula of the crack propagation is explained in Appendix~\ref{App:a}.
%The above evolution of contact radius can explain the adhesion hysteresis in normal contact \cite[e.g.,][]{2013JPhD...46Q5303K, WAHL2006178}.
%???However, the evolution of $a$ is almost the same as the JKR theory in this paper, which is explained in Appendix~\ref{App:a}.???
%The viscoelastic effect causes delayed $a$, which increases with increasing particle radius \cite{2013JPhD...46Q5303K,2021ApJ...910..130A}.
%For sub-micron size, this effect does not cause hysteresis well.

Given the initial conditions of the collision, the outcome of the collision can be obtained by numerically solving eq.~(\ref{eq:EoS}).
Krijt et al. incorporated the plastic deformation model of Thornton and Ning \citep{THORNTON1998154} into the outcome to obtain a coefficient of restitution $e$ that includes the effects of plastic deformation.
The plastic deformation of the Thornton and Ning model with $v_{\rm imp}>v_Y$ results in a decrease in $e$, as expressed by the following equation:
\begin{equation} \label{eq:TN}
    e_{\rm TN}= \left(\frac{6\sqrt{3}}{5}\right)^{1/2} \sqrt{1-\frac{v_Y^2}{6v_{\rm imp}^2}} \left[ 1 + 2\sqrt{\frac{6v_{\rm imp}^2}{5v_Y^2}-\frac{1}{5}} \right]^{-1/4},
\end{equation}
where $v_Y$ is the yield velocity, which is the impact velocity at which the plastic deformation begins, and $\hat{v}=v_Y/v_{\rm imp}$.
Takato et al. \citep{2014PhRvE..89c3308T} performed MD simulations of collisions between non-adhesive particles and estimated $v_Y \simeq 26.1$ m/s for $R>15$ nm for Argon particles.
Krijt et al. combined $e$ obtained by numerically solving eq.~(\ref{eq:EoS}) with the energy dissipation due to plastic deformation in the eq.~(\ref{eq:TN}) and obtained the coefficient of restitution $e_{\rm pl}$ including the effect of plastic deformation as:
\begin{equation}
    e_{\rm pl} = \sqrt{e^2 - (1 - e_{\rm TN}^2)}.
\end{equation}
%where $e$ is the coefficient of restitution obtained by numerically solving the equation of motion (eq.~(\ref{eq:EoS})).
%It should be noted that this is a prescription derived from the numerical solution of the equations of motion, and thus does not take into account the sequence of processes from contact to disconnection.

\section{Methods} \label{sec:method}

We have simulated the head-on collisions between two equal-mass particles using the MD simulation code LAMMPS \citep{LAMMPS}.
%0430: LAMMPS 済
The particles are spheres consisting of 12--6 Lennard--Jones (LJ) atoms arranged in a face-centered cubic (fcc) structure.
%E1114: 12-6?
%Y: 他のPhysical Reviewの論文を見てみると指数が-12, -6のLJポテンシャルを 12-6 LJ atomsと表記していたので、これを採用しました。
The LJ atoms are subject to the potential $u(r_{ij})=4\epsilon\{(r_{ij}/\sigma)^{-12} - (r_{ij}/\sigma)^{-6}\}$, where $r_{ij}$ is the distance between the $i$- and $j$-th atoms, $\epsilon$ is the potential depth, and $\sigma$ is the distance at which $u(r_{ij})=0$.
We normalize the parameters using $\sigma, \epsilon$ and the molecular weight, thus expressing them in so-called LJ units.
For example, the unit of temperature is $\epsilon k_{\rm B}^{-1}$, where $k_{\rm B}$ is the Boltzmann constant and $k_{\rm B}=1$ is set in the LJ units.
%In this paper, the parameters are presented in the LJ units.
We prepare a population of molecules arranged in an fcc structure and cut the molecules outside a certain distance from the origin to form a sphere.
The number of atoms $N$ in a particle is from $N\simeq 1.1\times10^5$ to $N \simeq 1.2\times 10^8$.
The particle density is $\rho \simeq 1.09 m \sigma^{-3}$.
The particle temperature $T$ is determined by giving random motion to the atoms: $T=2E_{\rm kin}/3N$ where $E_{\rm kin} = \sum_i^N m_i v_i^2/2$, $m_i$ and $v_i$ are the mass and velocity of the $i$-th atom.
The initial particle temperature is set to $T=10^{-6} \epsilon k_{\rm B}^{-1}$ in the canonical ensemble (NVT), corresponding to, for example, $8\times10^{-5}$ K for argon.
In this paper, we treat the collisions with extremely low temperatures as a first step.
We place two particles at a distance greater than $r_{\rm cut}=5\sigma$ from each other and give an initial velocity for each particle so that they collide head-on.
The orientations of each particle are randomly rotated.
We also vary the impact velocity from $v_{\rm imp}=0.04$ to 0.50.
%Y0430: 分子にランダムな運動を与えて粒子の温度が10-6ε/Kになるように与えた。済
% 初期条件をもっと丁寧に書く
%Y0430: first step として極限で温度が低い場合を調べた。を上に追加する。済
%Y0430: NVEアンサンブルでエネルギー保存の中でシミュレーションを行った。済
We adopt a cutoff distance for the interaction for all atoms at $r_{\rm cut}=5.0\sigma$ and perform the simulations in the microcanonical ensemble ({\it NVE}) for a time of at least $3\times10^4$ with a timestep of $2^{-7}$ in the LJ units.
The positions and velocities of molecules are updated by the Velocity Verlet algorithm.
%We used the MD code LAMMPS \cite{LAMMPS} to simulate the collisions and OVITO \cite{ovito} to visualize the MD simulations.
%0430: 上文を消して、OVITOは図1の所で述べる。済
%The MD simulation is one of $N$-body simulations, so the kinetic and potential energy sum is conserved.
%E1114: N体だからエネルギーが保存するわけではない。
%Y: 論理関係はないですね。今回はエネルギーが保存する系というだけでした。
%0430: ここにLJ unitsの話を持ってくる。そしてTable IIを先に持ってくる。済

We introduce examples of LJ units for three atoms, shown in Table~\ref{tab:LJunits}.
%例としてというニュアンスをいれたい
The LJ lengths are almost the same, and the particle radius corresponds to 10--100 nm.
Water molecules have a complex potential.
%and TIP4P \citep{2005JCP..123..23A} is typically used as a water potential.
Note that the normalizing parameters of water molecules are quantitatively different by a factor of several.
\begin{table}[h]
\caption{A table of the LJ units for Argon \cite{MICHELS1949627}, water \cite{1967Mtog} and Ag\cite{1975PhysSS..30..619H}. Time and velocity of the LJ units are $t=\sigma\sqrt{m/\epsilon}$ and $v=\sqrt{\epsilon/m}$.} \label{tab:LJunits}
\begin{ruledtabular}
\begin{tabular}{cccccc}
    & $m$ (g/mol) & $\sigma$ (nm) & $\epsilon$ (meV) & $t$ (ps) & $v$ (m/s)\\
    \colrule
    Ar  & 39.95 & 0.341 & 10.32 & 2.16 & 158 \\
    H$_2$O & 18.02 & 0.319 & 53 & 0.599 & 533 \\
    Ag & 107.87 & 0.2644 & 345 & 0.476 & 555 \\
\end{tabular}
\end{ruledtabular}
\end{table}

%We give the particle model in the paragraph above.
By simulating a Hertzian contact, we next determine the particle radius $R(=R_{1,2})$ and the reduced Young's modulus $E^*$. 
In the Hertz theory, there are no adhesive forces between particles.
To simulate a Hertzian contact, we therefore remove the attractive term $-(r/\sigma)^{-6}$ between atoms belonging to different particles.
By fitting to the Hertzian force, we can obtain $R$ and $E^*$, which we represent in Table~\ref{tab:model}.
The fitting procedures are shown in Appendix~\ref{App:Hertz}.
A previous simulation investigated the elastic constant of FCC LJ solid and showed $E^*\simeq34.7\epsilon/\sigma^3$ \citep{PhysRevB.48.6795}.
%, although they employed the cutoff $r_{\rm c}=3\sigma$.
Their obtained $E^*$ is smaller than that of our MD simulations due to their shorter cut-off length than ours.
%We expect that the cutoff length affects the Young's modulus and consider that the values of previous works and ours are consistent.
We adopt the surface energy as $\gamma=3.17$ and Poisson's ratio $\nu=0.25$.
The surface energy of the particle is derived as discussed in Appendix~\ref{App:surface_energy}.
The sound velocity of the particles is calculated as $V\simeq\sqrt{E/\rho}\sim 11$.
In this paper, we mainly treat the impact velocities less than a few \% of the sound velocity ($v_{\rm imp}\lesssim 0.3$).
%0430: LJ系での音速をここで与える。そして衝突速度がa few %であることを書く。済

\begin{table}[h]
\caption{The particle radius $R$ and the reduced Young's modulus $E^*$ in the LJ units.} \label{tab:model}
\begin{ruledtabular}
\begin{tabular}{ccc}
    $N$ & $R$ & $E^*$\\
    \colrule
    113637 & 29.6 & 53.2 \\
    914443 & 58.9 & 58.4 \\
    3120599 & 88.4 & 63.7 \\
    14434147 & 147 & 65.5 \\
    115506757 & 294 & 68.3 \\
\end{tabular}
\end{ruledtabular}
\end{table}

\section{Results} \label{sec:results}

\subsection{Interparticle force} \label{sub:interaction}

\subsubsection{Hysteresis} \label{subsub:hysteresis}

\begin{figure*}[h]
\centering
\includegraphics[width=0.3\textwidth]{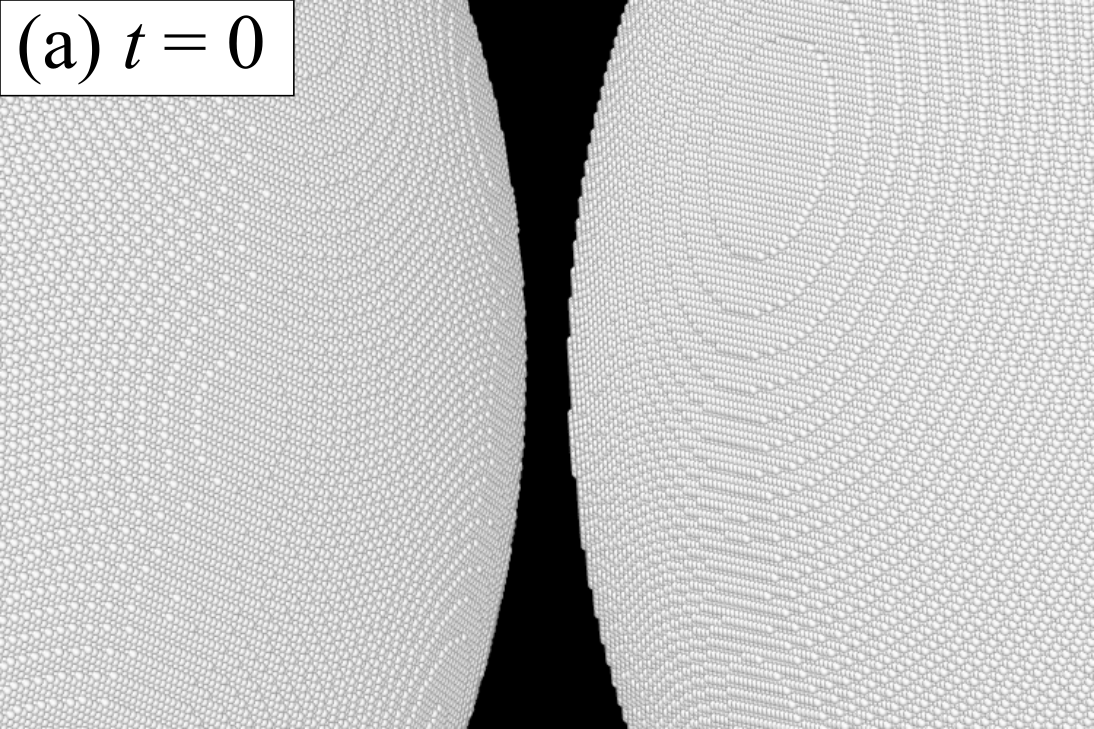}
\includegraphics[width=0.3\textwidth]{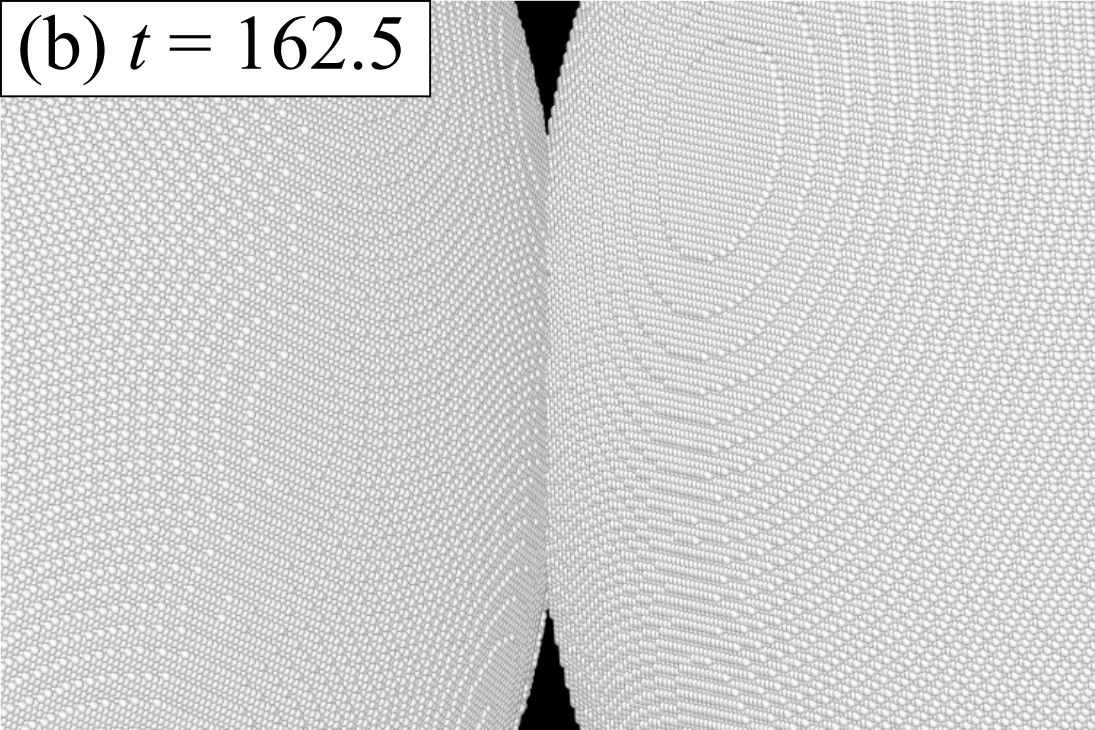}
\includegraphics[width=0.3\textwidth]{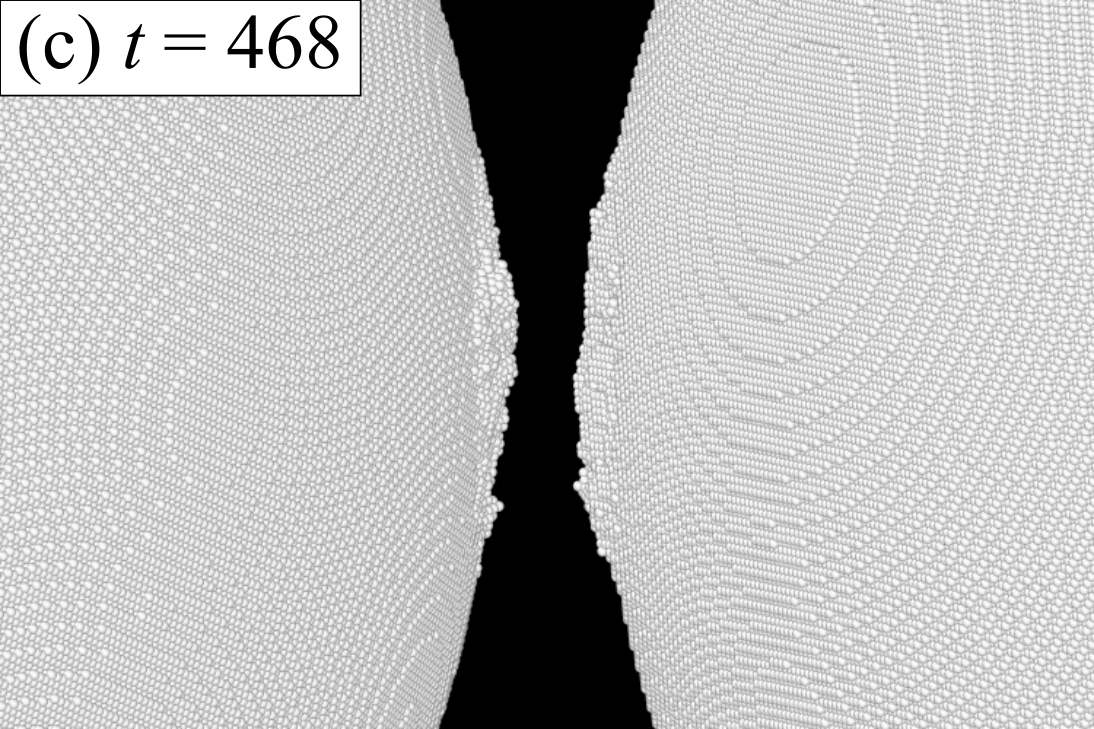}
\caption{\label{fig:snapshot}
    Snapshots of particles with $R=147$ during a collision with $v_{\rm imp}=0.10$ (a) the initial stage, (b) at the highest compression, and (c) after disconnecting.
    These figures are visualized by OVITO \cite{ovito}.
}
\end{figure*}

%We perform MD simulation of particle collisions described in Section \ref{sec:method} and 
We show an example of collision simulation with $R=147\sigma$ and $v_{\rm imp}=0.10 (\epsilon/m)^{1/2}$ in Fig.~\ref{fig:snapshot}.
The particle radius and impact velocity correspond to 50 nm and 15--50 m/s, respectively, in the physical units.
We always use the LJ units, although we omit the symbols such as $\sigma$ and $\epsilon$ from now on.
% 単位を省略するが、LJ単位系を使うので注意
From left to right, the panels show the particles in the initial stage, during the contact, and after the disconnection.
%After the disconnection, some atoms are displaced, and the surface is roughened.
After disconnection, the surfaces become rough because of the displacement of atoms.
%As shown in the panels, we obtain the coordinates of molecules.
We calculate the center of mass of each particle and derive its velocity and acceleration by differentiating it with respect to time.
The interparticle force is then obtained by multiplying the acceleration by the particle mass.
%from the derivation of the center of mass every moment and then multiply the acceleration by the mass to derive the force acting on the particle.

\begin{figure}
    \centering
    \includegraphics[width=0.5\textwidth]{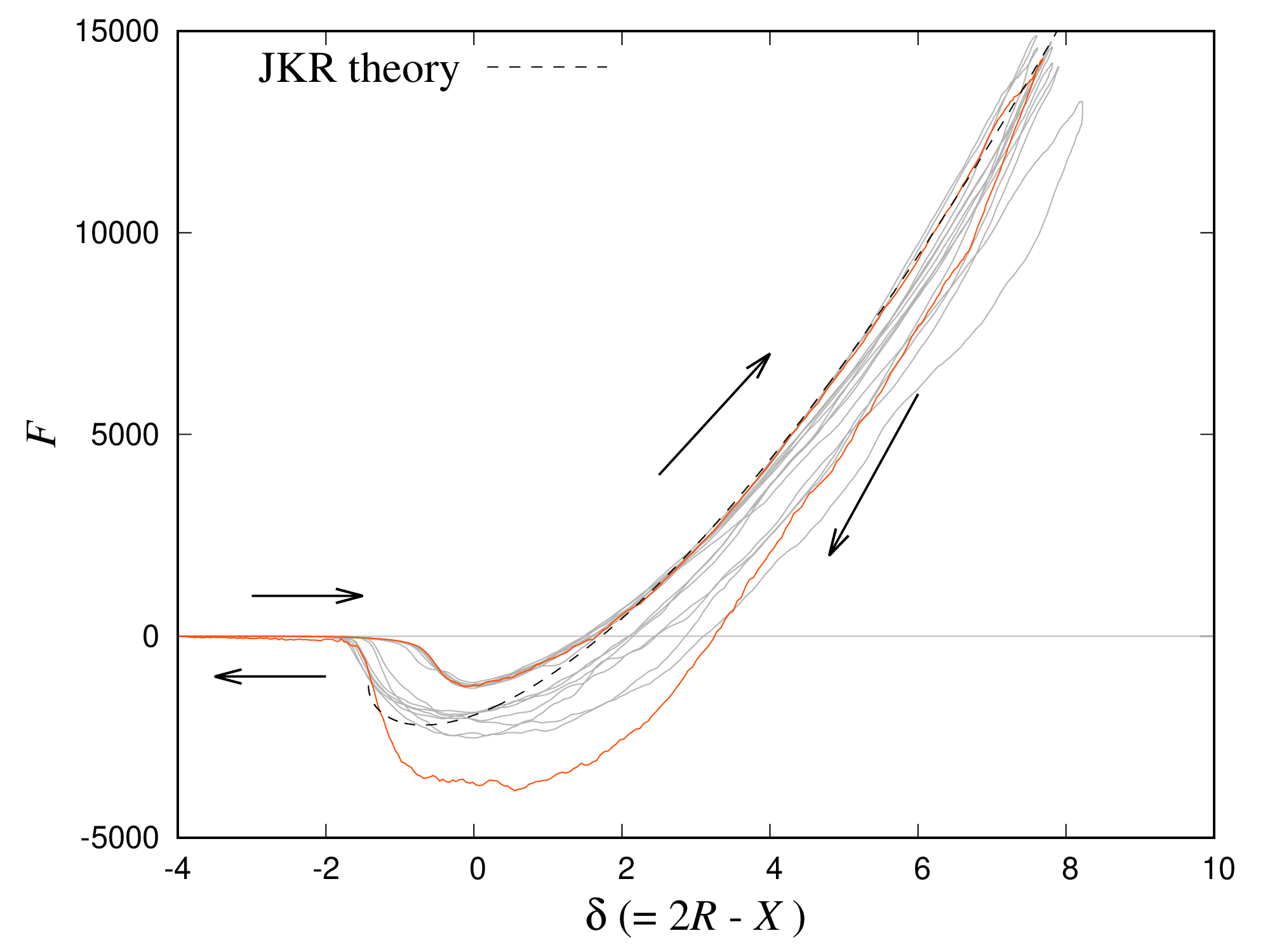}
    \caption{\label{fig:interaction147}
    Interparticle force with $R=147$ as a function of $\delta$ for $v_{\rm imp}=0.10$.
    The black dotted curve is the JKR theory, the red curve shows the example run, and the gray curves show the other runs with different crystal orientations.
    The arrows show the time evolution of the force.
    The force and the length of the mutual approach are plotted in the LJ units.
    }
\end{figure}

Figure~\ref{fig:interaction147} shows the interparticle force as a function of $\delta$.
The red curve shows the case illustrated in Fig.~\ref{fig:snapshot}, the gray curves show the results obtained with different crystal orientations, and the dotted line shows the JKR theory.
The unstable solution of the JKR theory is omitted in the figure since it is not stably realized in the actual contact process. 
%Then, we do not describe the unstable solution in the figure.
In the JKR theory, the two particles are in contact at $(\delta, F) = (0, -8F_c/9) \simeq (0, -1950)$, disconnected at $(\delta, F) = (-(3/4)^{2/3}\delta_0, -5F_c/9) \simeq(-1.33, -1220)$, and are in the stationary state at $(\delta, F)=(\delta_0,0) \simeq (1.61, 0)$ for $R=147$.
%E1119: 何の equilibrium?
%Y: 平衡状態(F=0)とmodelの所で書いていますが、接触、平衡、切断を並べるとややこしそうだったので、上記のように2文に分けました。
%E1205: 整理しました。これいいかな。
%Y: ありがとうございます。大丈夫です。
The maximum attractive force is $F_{c} = 3\pi \gamma R^* \simeq 2200$ for $R=147$.
%, and $-F_{c}$ is the minimum of the JKR theory in Fig.~\ref{fig:interaction147}.
%E1119: -F_c を引力の最小値という言い方は変?
%Y: 引力の最大値が正しい言い方になると思います。
%E1206: 後半必要ですか? the minimum of the JKR theory が何を意味するかわかりにくいです。
%Y わかりにくいなら、消します。maximum attractive forceで理解してくれると思います。
The JKR force in both the loading and unloading phases follows the same curve, although the moments of connection and disconnection are different; this represents the hysteresis in the JKR theory.
%The force of the MD simulations follows the arrows.
%E1119: このような説明は図の説明へ。 arrow を follow するのでなく、arrow は進化経路を示す。
%Y: 力の時間進化という説明を図のcaptionに入れました。
The two particles first approach each other, represented by increasing $\delta$.
A negative force acts on them when they come into contact because of the attractive molecular force.
As they are compressed, the repulsive force increases with increasing $\delta$, which is the loading phase.
After the greatest compression, the two particles rebound, and $\delta$ decreases; this corresponds to the unloading phase.
For $\delta$ becomes $\delta<-(3/4)^{2/3}\delta_0$, the disconnection of the contact occurs; otherwise, the two particles coalesce, oscillating between loading and unloading.
%E1119: この段落でJKRの進化のみを説明する。

%E1119: 次の段落でMD結果の説明をする。
In all the MD simulations, we find a difference between the loading and unloading phases, which is the hysteresis in the MD simulations.
The example case discussed above has the largest hysteresis among eight runs.
The collisional process depends on the orientations of each particle, and particles easily stick to each other when their orientations are similar \citep{2019PhRvE..99c2904N}.
Although the force in the loading phase agrees with the JKR theory, the force in the unloading phase differs from that in the loading phase.
%The hysteresis is mainly caused by the bulk dissipation force but not by the delayed $\dot{a}$.
%E1119: bulk dissipation force とは? delaying a'で通じる? delaying ではなく delayed?
%Y: 調べるとdelayed の方が意味として適切なので、他の部分も全てdelayedにします。
%Y: Bulk dissipation forceはKrijtモデルにおける散逸力のことで、MD結果に書くのは不適なので消去します。
Figure~\ref{fig:interaction147} shows that hysteresis shifts the mutual approach $\delta$ toward larger $\delta$ in the unloading phase.
%\sout{This shift occurs because the pressure crushes the contact surface; that is, plastic deformation occurs.}
%Y0202: 説明追加
%Y0517: ここで塑性変形は言わない
%\sout{Plastic deformation is considered to occur during the loading phase, especially at maximum compression.}
%\sout{During the loading phase, part of the translational kinetic energy is used for the plastic deformation, which may decrease the resilience of the particle.}
%\sout{The decrease in the resilience is expected to weaken the repulsive force in the JKR force (eq.~\ref{eq:JKR}) for the interparticle force and make the force follow the path obtained by the MD simulations for the unloading phase (see Figs.~\ref{fig:interaction147} and \ref{fig:interaction}).}
%\sout{The hysteresis indicates the dissipation of the kinetic energy.}
Particle kinetic energy is dissipated by the area enclosed by the loading and unloading curves.
We note that the total energy of the molecular system is conserved.
%Y0122: 校正の note that because ~ はさすがに間違い?
The kinetic energy of particle translational motion is converted to the internal energy of particles.
% lost, and molecular random motions are thermalized.
%E1119: ここでのエネルギーの話はこれくらいで大丈夫?
%Y: のちに詳しく述べるとも記述しているので大丈夫です。
The energy conversion is discussed in detail in Section~\ref{sub:energy}.

\subsubsection{Dependence on particle radius and impact velocity} \label{sub:force-Rv}

\begin{figure*}
    \centering
    \includegraphics[width=0.8\textwidth]{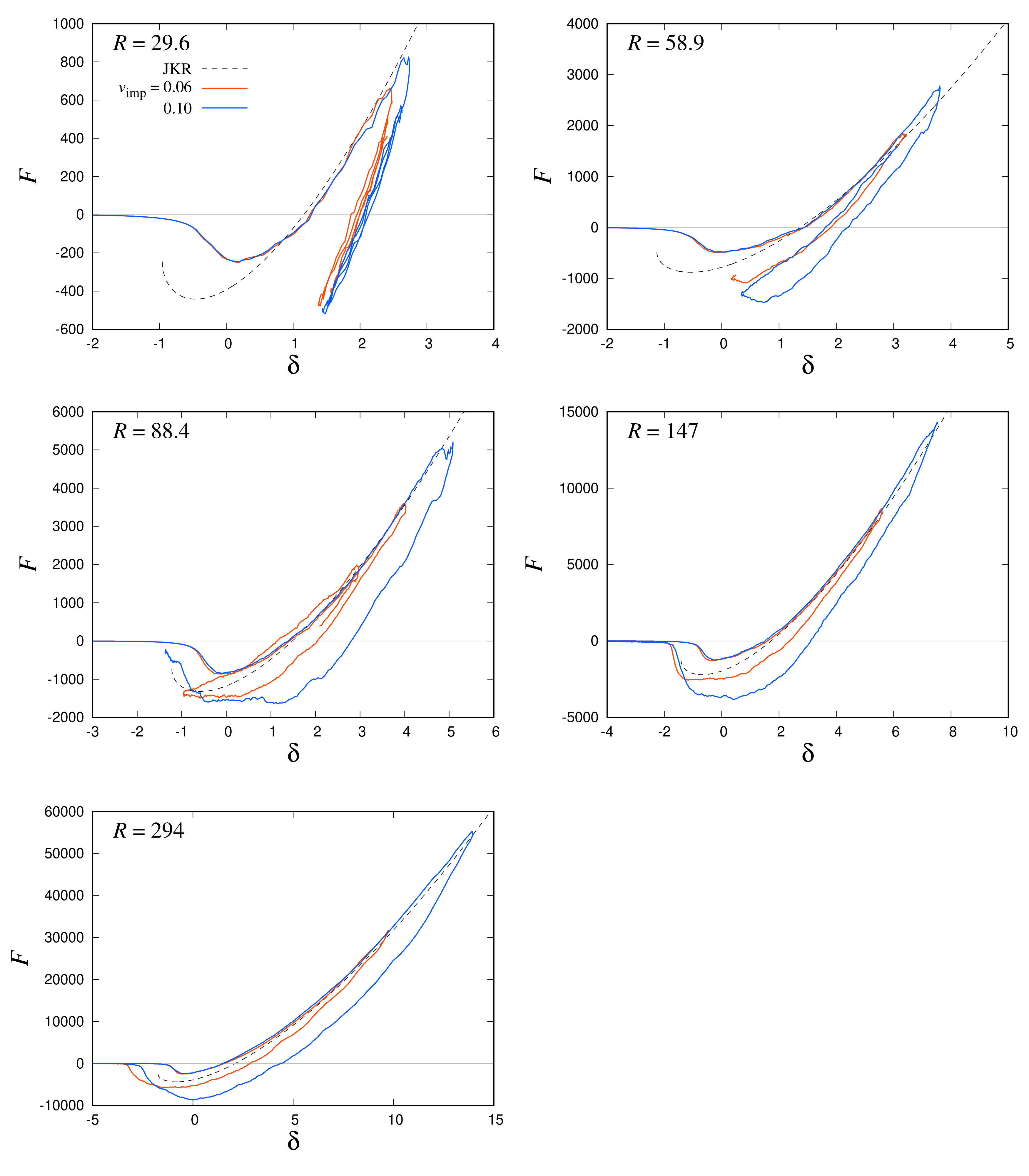}
    \caption{\label{fig:interaction}
    Interparticle force as a function of $\delta$ with $v_{\rm imp}=0.06$ and 0.10.
    The black curves represent the JKR theory.
    The force and the length of the mutual approach are plotted in the LJ units.
%    The force agrees with the JKR theory at the loading phase and the curves are shifted to large displacements at the unloading phase.
%E1119: 結果の説明は本文で。
%Y: hysteresis節に記述しました。
    }
\end{figure*}

%E1119: ここでは前節との重複は避け、半径依存性についてまとめる。
Figure~\ref{fig:interaction} shows the interparticle force as a function of $\delta$ for all radius cases with $v_{\rm imp}=0.06$ and 0.10.
%Although the force in the loading phase agrees with the JKR theory, the force in the unloading phase differs from that in the loading phase.
%E1119: この結果は前の説で言うべきでは。ここでは半径依存性について述べる。
%Y: 前節に入れました。。
%These differences suggest the hysteresis in the collision.
%E1119: 上のコメントに同じ。
%Y: 前節で述べているので、ここは削除します。
%The mutual approach $\delta$ shifts towards a larger $\delta$ in the unloading phase.
%This shift in $\delta$ suggests that plastic deformation occurs.
%E1119: 上の2文の論理がわかりません。
%Y: 「接触面が潰れて、δが大きい状態にシフトするので、これは塑性変形を表している」という説明をします。ここでは半径・速度依存性なので、前節のヒステリシスに説明を入れました。
The particles coalesce for $R\leq88.2$, and the force oscillates at a $\delta$ larger than in the JKR theory.
For $R=29.6$, the force paths of the loading and unloading phases are distinctly different and the force finally oscillates at $\delta\simeq 2$, which indicates the occurrence of the plastic deformation.
%E1119: JKRの何より大きい?
%Y: JKRのdeltaより大きいということです。
%The difference in the force relative to the JKR force decreases with increasing radius.
%MDとJKRの差はRが小さいほど小さく、同時に粒子間力自体も小さくなる。
%Rが小さくなれば差の絶対値は小さくなり、相対的な値 (差/粒子間力) は大きくなる。
%\textcolor{red}{The force difference between the loading and unloading phases normalized by the force at the maximum compression increases as the particle radius decreases}
%This indicates that the effect of hysteresis is larger for a smaller radius.
Focusing on the force difference between the loading and unloading phases normalized by $F_{\rm c}$, we see that it is about $F_{\rm c}$ for $R=29.6$ and about 3-4 times as large as $F_{\rm c}$ for $R=294$.
This trend is roughly consistent with the relationship $F_{\rm dis, K}/F_{\rm c}\propto R^{1/2}$ as $F_{\rm dis, K}\propto a T_{\rm vis} \propto R^{3/2}$ and $F_{\rm c}\propto R$, where we approximate $a\propto R^{1/2}$, if the hysteresis comes from the dissipative force of the Krijt model.
As the radius increases, the difference in force normalized by $F_{\rm c}$ also increases.
However, the effect of hysteresis becomes smaller as the radius increases since the kinetic energy, which is inertia, is proportional to $R^3$.

\begin{figure}[h]
    \centering
    \includegraphics[width=0.5\columnwidth]{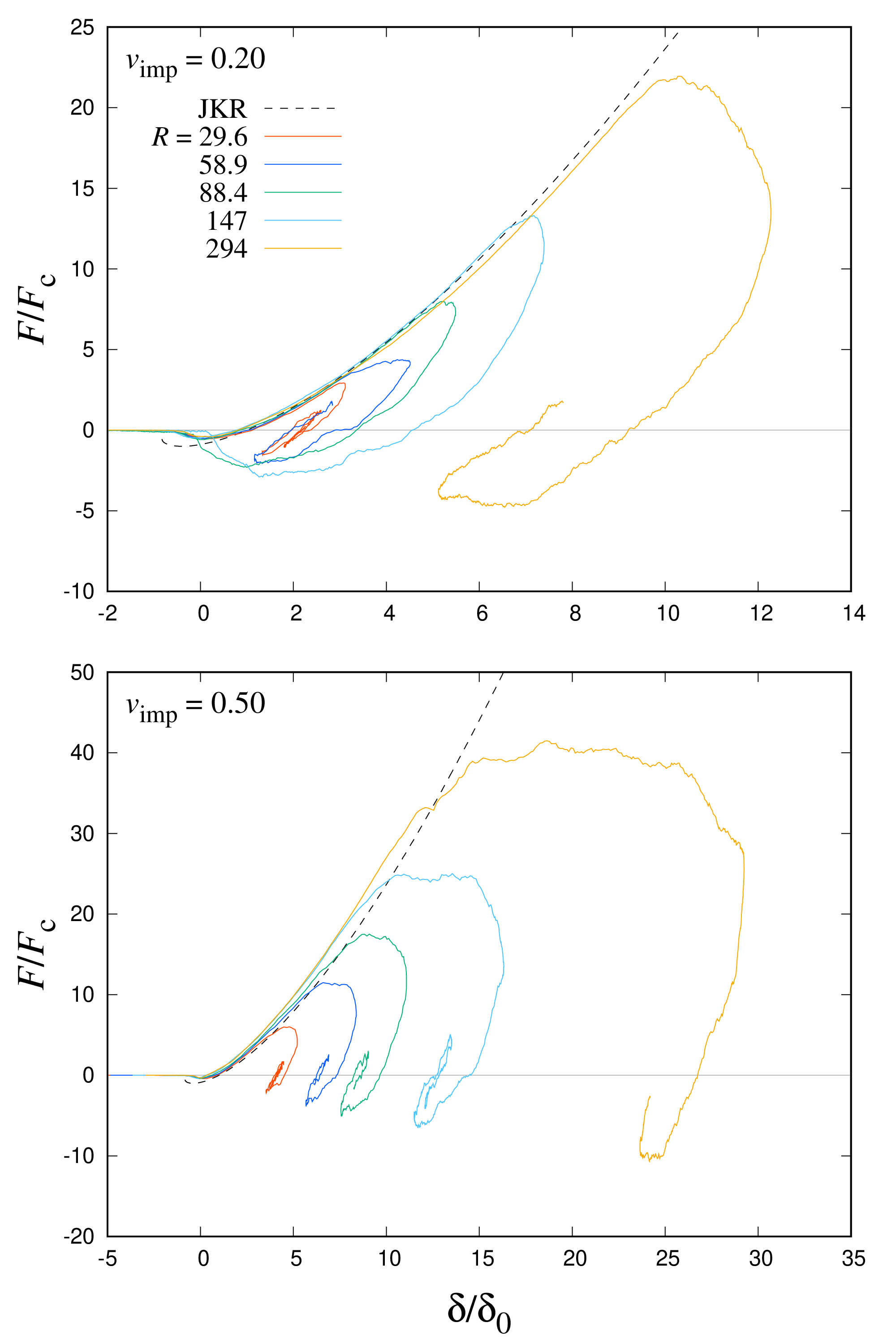}
    \caption{\label{fig:interaction_high}
    Normalized interparticle force as a function of $\delta/\delta_0$ with $v_{\rm imp}=0.20$ and 0.50.
%    The shift from the loading becomes significant.
%E1119: 説明は本文で。
%Y: 承知しました。
    }
\end{figure}

Figure~\ref{fig:interaction_high} shows the interparticle force as a function of $\delta$ normalized by $F_c$ and $\delta_0$ for all radius cases with $v_{\rm imp}= 0.20$ and 0.50.
The hysteresis is clearly apparent in Fig.~\ref{fig:interaction_high}.
Higher impact velocities cause significant shifts in $\delta$ in the unloading phase, resulting in stronger hysteresis.
The significant shifts in $\delta$ for $v_{\rm imp}\geq0.2$ indicate plastic deformation.
%E1119: 何の(いつの) delta? delta は距離の変数として使われている。
%Y: unloading時のdeltaなので、上記のように修正しました。
In all radius cases, the particles coalesce for $v_{\rm imp}=0.50$ with the remaining at large $\delta$, which suggests strong plastic deformation.
%The degree of hysteresis increases with the impact velocity.
%The significant shifts in $\delta$ indicate plastic deformation.
Though this figure shows the collision results with a particular orientation, such large shifts in $\delta$ occur and particles stick even for other orientations.
%0430: expectじゃなくて何通りかやればよい。我々はobserveした。済

\begin{figure}[t]
    \centering
    \includegraphics[width=0.4\textwidth]{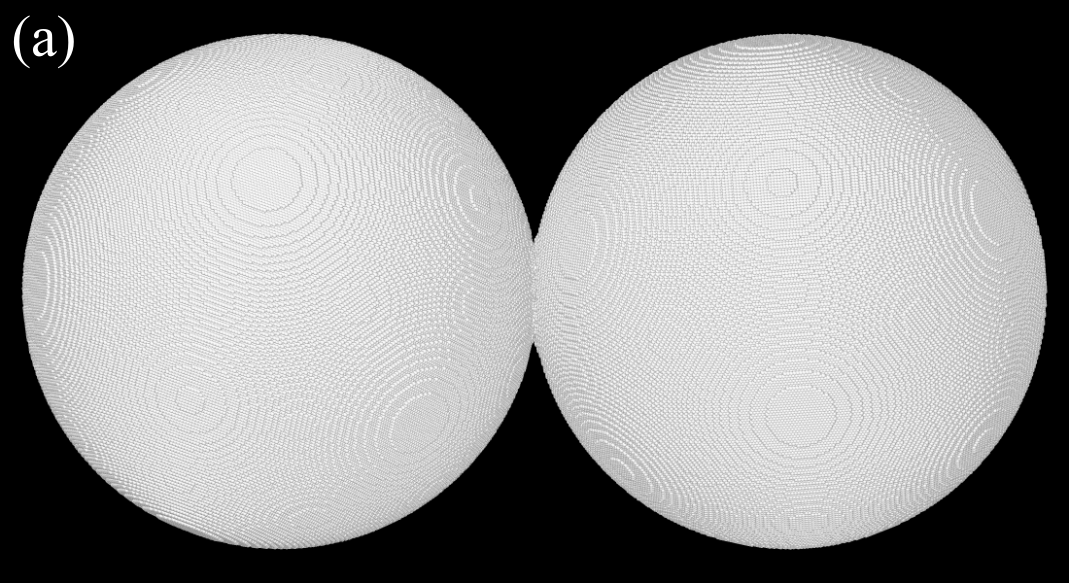} 
    \includegraphics[width=0.4\textwidth]{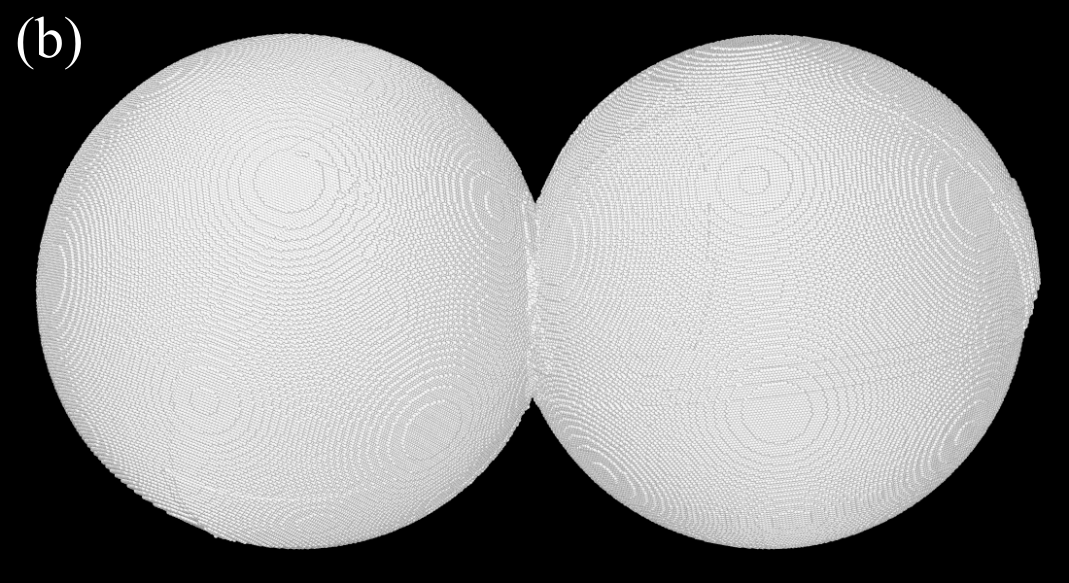}
    \includegraphics[width=0.4\textwidth]{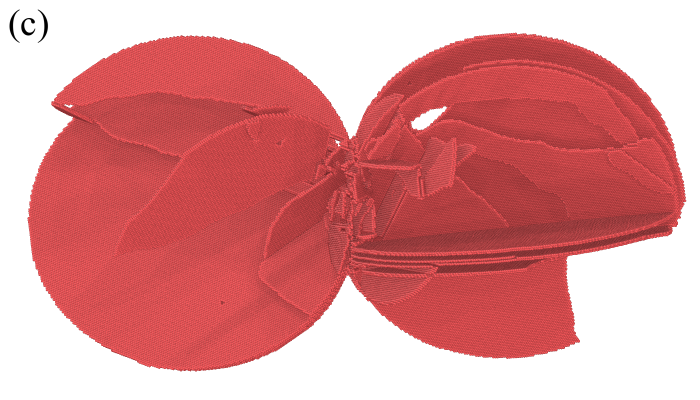}
    \caption{\label{fig:deform}
    Images of particles of $R=88.4$ with (a) $v_{\rm imp}=0.10$ and (b) $v_{\rm imp}=0.50$ at the end of simulations.
    Panel (c) shows the results of dislocation analysis of panel (b) and red atoms show those that do not have an FCC lattice structure due to molecular movement by collision.}
\end{figure}

Figures~\ref{fig:deform}a and b show images of particles with $v_{\rm imp}=0.10$ and 0.50 at the ends of the simulations for $R=88.4$.
The particles remain spherical for $v_{\rm imp}=0.10$.
In contrast, a bump is formed on the particle surface, and the contact area is increased for $v_{\rm imp}=0.50$.
This deep sticking causes a significant shift in $\delta$ in the unloading phase.
%E1119: 前節の delta へのコメント参照。
%Y: at the unloading phaseを追記
The bump on the right-most particle in Fig.~\ref{fig:deform}b is called an antipodal deformation.
The atoms are pushed from the contact area to the opposite surface, forming the bump.
%The left-most particle also exhibits deformation and cracks on the surface.
Figure~\ref{fig:deform}c shows the atoms that do not arrange in an FCC lattice structure, indicating cracks in the particles.
The cracks are seats and extend from the contact surface in a planar state, which can be observed as the crack lines on the surface in Figure~\ref{fig:deform}b.
In particular, the crystal structure around the contact surface is extremely broken.

\subsubsection{Plastic deformation} \label{sub:pladef}

Plastic deformation occurs due to stress acting on the contact area.
The interparticle force in Figs.~\ref{fig:interaction} and \ref{fig:interaction_high} shows possible plastic deformation at all impact velocities for $R = 29.6$ and at all radii for $v_{\rm imp} = 0.2$.
In this section, we estimate the pressure and verify whether it is sufficient to cause plastic deformation.

The onset of plastic deformation is determined by the yield strength $Y$.
Takato et al. \citep{2014PhRvE..89c3308T} performed MD simulations of collisions between non-adhesive particles and found a yield velocity of $v_Y\simeq 0.17$ for $R>44$, which corresponds to $Y=0.15E^*$. %E: この修正でOK?
On the other hand, the yield strength $Y$ is theoretically given as $Y\lesssim 0.38E^*$ for defect-free material \cite{1953Iron.174..25P, 2013JPhD...46Q5303K}.
Spheres begin to deform plastically when the pressure at the center of the contact area reaches $p_{\rm c} = 1.6Y$ \cite{1987come.book.....J}.
Then, the plastic deformation of particles is expected to occur for $p_{\rm c}\geq0.24E^*$ numerically and $p_{\rm c}\geq0.6E^*$ theoretically.
We compare these values with maximum pressures between particles in our MD simulations. % although these values differ by a factor of 2.5. %E: このestimated pressureは何? -> Y: 我々のMD計算で推定される最大圧力と比較する。という風に説明を追加。2.5倍云々は逆に不要と考え、削除

The pressure at the center of the contact area can be estimated by eq.~(\ref{eq:hertz-p}).
Since from the energy conservation, the initial kinetic energy equals the Hertzian potential energy of eq.~(\ref{eq:pot_H}) at the maximum compression, the maximum $\delta$ can be estimated as
%Y0122: 校正ではfrom energy dissipationが2回出てきているので、多分ミス
\begin{equation} \label{eq:delta_max}
    \delta_{\rm max} = \left( \frac{15}{32} \frac{mv_{\rm imp}^2}{E^* {R^*}^{1/2}} \right)^{2/5}.
\end{equation}
Combining eqs.~(\ref{eq:hertz-p}) and (\ref{eq:delta_max}), we can estimate the maximum pressure as $p_{\rm c, H}\simeq 0.26E^*$ for $v_{\rm imp}=0.20$.
We note that the estimated maximum pressure using the Hertz theory does not depend on the particle radius.
%For $v_{\rm imp} = 0.10$, we can estimate $\delta_{\rm max}\simeq 1.47, 6.80$ and 13.3 and $p_{\rm c, H}/E^* \simeq 0.200$, 0.194 and 0.191 for $R=29.6, 147$ and 294, respectively.
For small particles, the Hertz theory underestimates $\delta_{\rm max}$ and $p_{\rm c}$, and the JKR theory estimates them more accurately; for $v_{\rm imp}=0.06$, $p_{\rm c, J}\simeq0.29E^*$ with $R=29.6$.
%The maximum pressure is the same as that predicted by the JKR theory, although we determine the parameters numerically; when $v_{\rm imp}=0.1$, $p_{\rm c, J}\simeq0.28E^*$ for $R=29.6$.
%Collisions of smaller particles result in a larger pressure at the same impact velocity.
%E1119: 適宜式を引用する。依存性の理由を述べる。
%Y: (4), (13)式より依存性を説明しました。
%For $R=147$, $v_{\rm imp} = 0.20$ yields $p_{\rm c, H}/E^* \simeq 0.25$, which is larger than that for $v_{\rm imp}=0.10$.
%Eq.~(\ref{eq:hertz-p}) and (\ref{eq:delta_max}) lead to $p_{\rm c, H}\propto v_{\rm imp}^{1/5}$, which shows that collisions with higher impact velocities result in larger pressures.
%E1119: 適宜式を引用する。依存性の理由を述べる。
%Y: (4), (13)式より依存性を説明しました。

The pressure estimated above is less than the theoretical onset pressure of plastic deformation ($0.6E^*$) but larger than that obtained by the previous MD simulation.
The displacement of $\delta$ shown in Figs.~\ref{fig:interaction} and \ref{fig:interaction_high} indicates the occurrence of the plastic deformation.
%Theoretically, however, plastic deformation occurs at $v_{\rm imp}\gtrsim1.8$, which is outside the impact velocity range of this study.
%In contrast, plastic deformation can occur using the onset pressure suggested by Takato et al, which is consistent with our results.
%On the other hand, Takato et al. showed that plastic deformation occurs around $v_{\rm imp}\simeq0.165$, which is consistent with our results.
Therefore, in this study, we use their onset pressure of plastic deformation.
%Plastic deformation does not occur in comparing with theoretical values, $0.6E^*$, but occurs in comparison with previous MD simulations, $0.24E^*$.
%Theoretical values indicate that plastic deformation occurs at $v_{\rm imp}\gtrsim1.8$, which is outside the impact velocity range of this study.
%In this study, the value obtained by MD simulations of Takato et al. is considered to be a suitable reference pressure for plastic deformation.

%Thus, a larger pressure results in both larger hysteresis and larger plastic deformation.
%E1119: 論理関係が逆。変形が大きいからヒステリシスが大きい。
%Y: 変形とヒステリシスを並列にしたつもりでしたが、わかりにくいのでandにしました。

%図\ref{fig:interaction}と\ref{fig:interaction_high}は$X$と$\delta$の図。
%loading時はJKRと一致するが、unloading時はずれる。
%変位が大きい状態へシフトしている。
%これは圧縮による変形を示唆している。
%高速度衝突時ではそれが顕著であるが、低速度衝突時でも変化量は少ないが同様のシフトが見られた。
%合体する時は、loadingとunloadingを繰り返しつつ減衰する。

%大きな変形は図\ref{fig:deform}で確認できる。
%この図は$v_{\rm imp}=0.10$と0.50の結果を比較している。
%$v_{\rm imp}=0.10$ではほぼ球であるが、$v_{\rm imp}=0.50$では、粒子の右側に大きな凸が見られている。
%凸が接触面と反対側に生じており、antipodalと呼ばれる。
%もちろん図で左側の粒子にも変形はみられる。

%Y0206: 以下新規
\subsubsection{Contact radius}

\begin{figure}[h]
	\includegraphics[width=0.7\columnwidth]{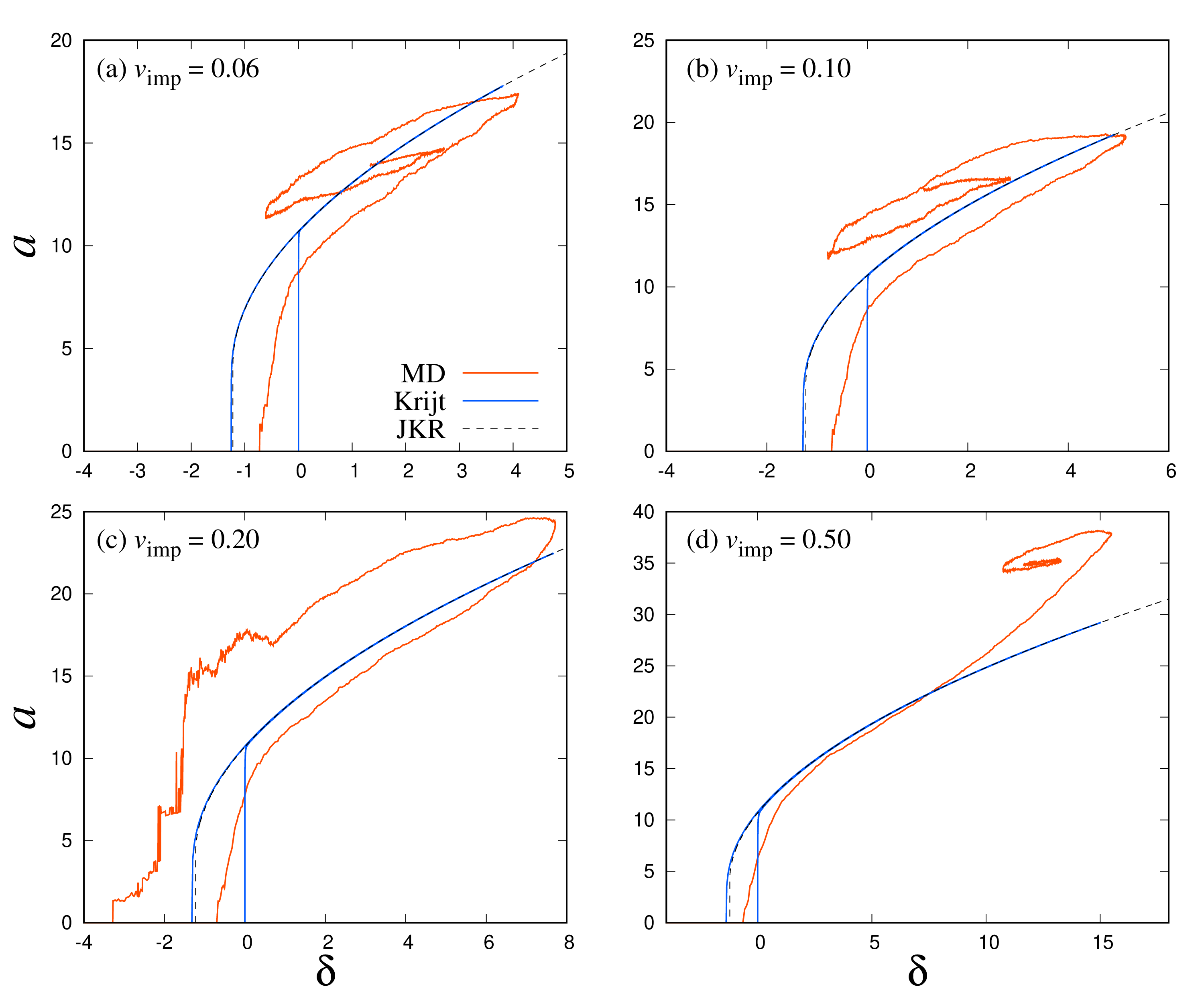}
	\caption{ \label{fig:contR}
    Contact radius as a function of $\delta$ for $v_{\rm imp}=0.06, 0.10, 0.20$ and 0.50 with $R=88.4$.
    The black dotted curves show the JKR theory, the red curves show the MD simulations, and the blue curves show the Krijt model with $T_{\rm vis}=0.075$, respectively. 
	}
\end{figure}

%我々は、接触面付近(幅Δx=1)の分子に対して、面内一様分布を仮定して、分子分布から接触半径aを求めた。
%接触半径はギレーション半径のsqrt2倍で求められる。
%Y0219: 下記2文を追加
%We calculate the contact radius $a$ from the molecular distribution near the contact surface (width $\Delta x=1$), assuming a uniform molecular distribution for the $yz$-plane; $a=\sqrt{2 \sum_i |{\bm r}_i - {\bm r}_{\rm M}|^2 /n}$, where $n$ is the number of molecules in width $\Delta x$, ${\bm r}_i$ and ${\bm r}_{\rm M}$ are the position vectors of the $i$-th atom and the center of mass of the region in the $yz$-plane.
%We obtain the contact radius using the gyration radius derived from the distribution of molecules in a thin region of the thickness 1$\sigma$ containing the contact surface.
%Let ${\bm r}_i$ be the position vectors projected onto the contact surface for each molecule in this region.
%Furthermore, assuming that the position vectors are uniformly distributed within the contact radius on this surface, we can evaluate the contact radius $a$ as $a=\sqrt{2 \sum_i |{\bm r}_i - {\bm r}_{\rm M}|^2 /n}$, where ${\bm r}_{\rm M}$ is the position of the center of mass of the molecules in the region and $n$ is the number of these molecules.
We calculate the contact radius using the gyration radius derived from the molecule distribution in a thin layer of the thickness 1$\sigma$ containing the contact surface.
Let ${\bm r}_i$ be the position vectors projected onto the contact surface for molecules in this layer.
The gyration radius is given by $r_\mathrm{g} = \sqrt{\sum_i |{\bm r}_i-{\bm r}_{\rm M}|^2 /n}$,  where ${\bm r}_{\rm M}$ and $n$ are the position of the center of mass and number of the molecules in the layer, respectively.
Then, we obtain $a = \sqrt{2} r_\mathrm{g}$. 
Figure~\ref{fig:contR} shows the contact radius as a function of $\delta$ for $R=88.4$.
The contact radius of the MD simulations is smaller than that of the JKR theory in the loading phase and larger in the unloading phase, indicating the delayed $a$ and the hysteresis.
The hysteresis of $a$ in the MD simulations is larger than that of the Krijt model, although the behavior of the contact radius evolution is similar to that of the Krijt model.
%This indicates that some extra force other than the JKR force is acting between the particles, possibly corresponding to the dissipation force predicted by the Krijt model.
%E0206: ここの論理がよくわかりません。
%Y0208: 上記のように修正

%The large contact radius is formed around the greatest compression for the high-velocity collisions of $v_{\rm imp}\geq0.2$.
%As the impact velocity increases, the mutual approach $\delta$ increases, which geometrically means that the area where the two particles overlap increases.
%However, since the volume does not change much, the atoms in the overlapping region move to form a larger contact surface, which is not considered in the model.

For the high-velocity collisions of $v_{\rm imp} \geq 0.2$, the contact radius increases around the most compressed point. 
For $v_{\rm imp} = 0.5$, the contact radius increases significantly around the most compressed point due to plastic deformation (see Fig.~\ref{fig:contR}).
Such an effect of plastic deformation is not considered in the JKR theory.

In the crack propagation model \cite[e.g.,][]{2004JPhD...37.2557G,2013JPhD...46Q5303K}, $a$ is delayed due to the viscoelasticity, but the degree of delay predicted by the model is insufficient compared to that in the MD simulations.
The model has the uncertainty of the relaxation time $T_{\rm vis}$, which may cause the deference between the MD simulations and the model.

\subsection{Coefficient of restitution} \label{sub:restitution}

\subsubsection{Dependence on the impact velocity}

\begin{figure}[h]
	\includegraphics[width=0.6\columnwidth]{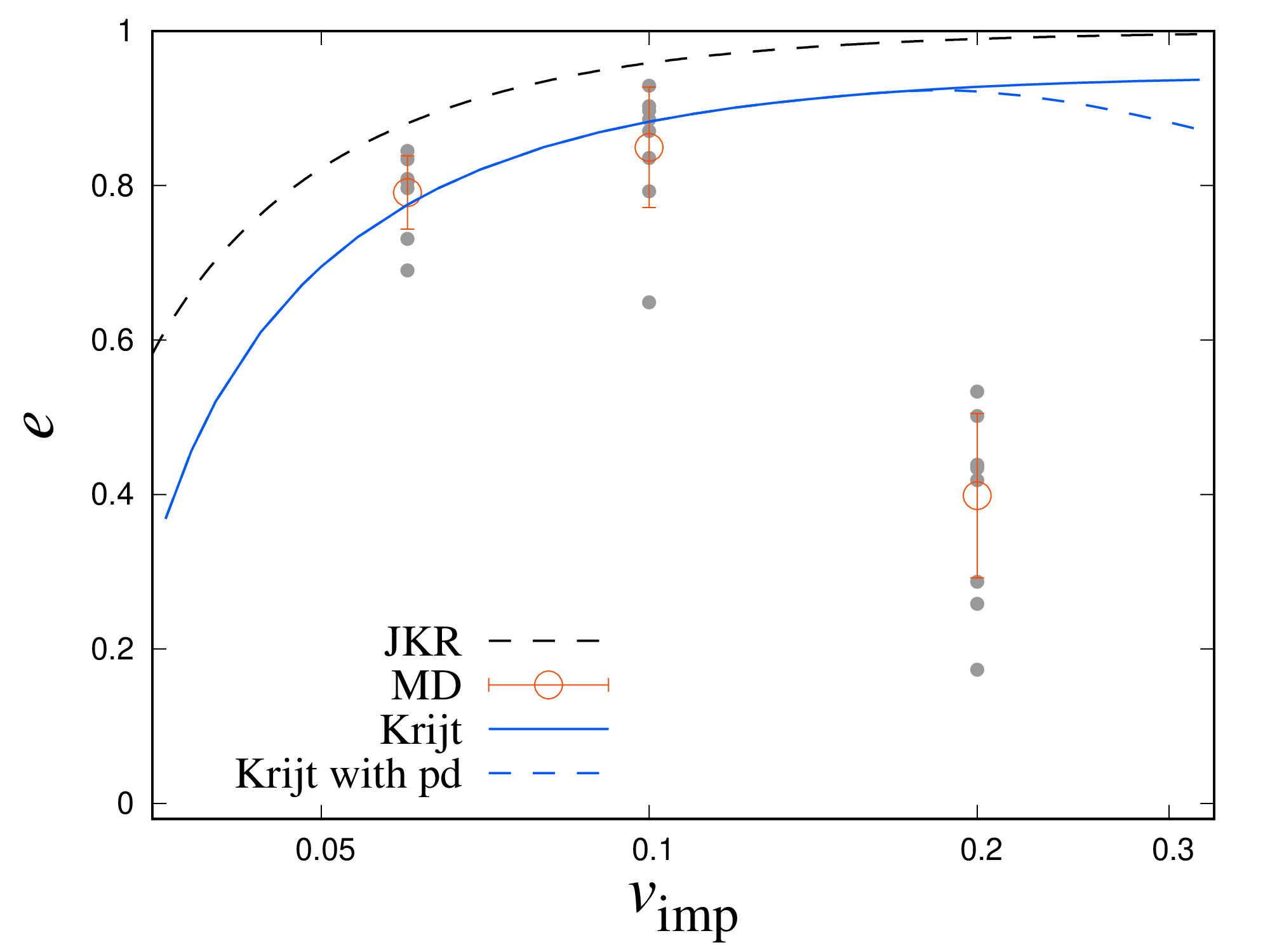}
	\caption{ \label{fig:v-e147}
        Coefficient of restitution  for $v_{\rm imp}=0.06, 0.10$ and 0.20 with $R=88.4$.
        The gray points are the results of 8 runs with different crystal orientations, and the red points with error bars are the root mean square (RMS) of the 8 results with their deviations.
        The dashed black and solid blue curves show the JKR theory and the Krijt model with $T_{\rm vis}=0.075$, respectively.
        The dashed blue curve shows the Krijt model with plastic deformation (pd) for $Y = 0.12 E^*$ based on the model of Thornton and Ning \citep{THORNTON1998154}.}
\end{figure}

The coefficient of restitution (COR) $e$ provides an appropriate method for evaluating the effects of the kinetic energy dissipation of the translational motion.
Figure~\ref{fig:v-e147} shows the COR for $v_{\rm imp}=0.06, 0.10$ and 0.20 for particles with $R=147$.
%The gray points show the results of 8 runs with different crystal directions.
%E1120: 図の説明へ移動する。
%Y: 承知しました。
A previous study suggested that the crystal orientation within the particles may significantly affect the collision results \cite{2019PhRvE..99c2904N}, consistent with Fig.~\ref{fig:v-e147}.
%The red points with error bars are the root mean square (RMS) of 8 results and their deviations.
%E1120: 図の説明へ移動する。
%Y: 承知しました。
Figure~\ref{fig:v-e147} also shows that the COR of the MD simulations is smaller than that obtained from the JKR theory.
As shown in Section \ref{sub:interaction}, the hysteresis is larger for collisions with higher impact velocities, which leads to larger energy dissipation and smaller values of $e$.
Since the energy dissipation is independent of the impact velocity in the JKR theory, $e$ approaches 1.0 as the impact velocity increases.
The Krijt model predicts larger energy dissipation than does the JKR theory because it includes the energy dissipation due to the hysteresis described in Section~\ref{sub:krijt}.
%E1205: ヒステリシスもエネルギー散逸の結果なので論理がおかしい。
%Y: ヒステリシスによるエネルギー散逸という風にしました。
In Fig.~\ref{fig:v-e147}, we set $T_{\rm vis}=0.075$ to fit the MD results with $v_{\rm imp}=0.06$.
This model can also reproduce the MD results for $v_{\rm imp}=0.10$ within the error bars, although it does not match for $v_{\rm imp}=0.20$.
This occurs because high-velocity collisions have large hysteresis, as shown in Fig.~\ref{fig:interaction_high}, causing $e$ to decrease as the impact velocity increases.
A stronger energy dissipation occurs in the MD simulations than that caused by the model of plastic deformation.

Here, we compare our MD simulations with previous studies for velocity dependence.
Nietiadi et al.\citep{2020NRL....15...67N} treated silica particles of $10-20$ nm radius.
%0430: 比較は音速の何%か、半径はnmで表す(我々はargon)。LJ unitsはだめ。
In their results, the COR $e$ has a peak similar to our MD simulations, but the peak is located where $v_{\rm imp}$ is 5-8 \% of sound velocity, which is several times as large as that of our results, about 1 \% of sound velocity.
%Silica has an anisotropic potential and is a covalent material, which may have caused the difference in the peak, indicating that molecular potential has a significant influence on collision results.
This difference is thought to be due to the difference in potentials: silica has covalent bonds and spherically asymmetric potential, whereas LJ atoms have non-covalent bonds and spherically symmetric potential.
The LJ atoms tend to move tangentially to the contact surface and deform easily even at low impact velocities, and translational energy is easily converted to kinetic energy of atomic random motion.
Therefore, the impact velocity where $e$ peaks in our simulations is smaller than that of previous studies.
% 0430: LJ分子は球対称ポテンシャルなので、転移が起きやすい(tangentialな動きがしやすい)、したがって、塑性変形が起こりやすく、低い速度でも散逸が起きやすい。
%0430:「だから我々がこのように違う、そのようになっている」という言い方 済
Nietiadi et al.\citep{2019PhRvE..99c2904N} used LJ atoms as we do and examined $e$ for $R<90$ and showed the peak of $e$ at the same impact velocity.
%0430: 非常に少ない差である 済
Tanaka et al.\citep{2012PThPS.195..101T} examined the coefficient of restitution for $R\simeq88$ only, and their results are consistent with that of this study.
Few previous studies examined the particle collisions with $R>100$ and thus cannot be compared to our results.
Our results indicate a strong dependence of the translational kinetic energy dissipation on the radii for $29<R<150$.

\subsubsection{Dependence on radius}

Figure~\ref{fig:v-e} compares the COR obtained for $R=58.9, 88.4, 147$ and 294 as a function of $v_{\rm imp}$ with the results from the JKR theory and the Krijt model.
For $R=29.6$, all particle collisions result in coalescence.
In the JKR theory, the collisions result in energy dissipation of $\simeq 0.773F_c \delta_0 \propto R^{4/3}$, which means that collisions with smaller particles result in smaller $e$ because the translational kinetic energy is proportional to $R^3$.
%The red points are the RMS of 20 results for $R=58.9$ and 88.4 and 8 results for $R=147$ and 294 with different crystal directions, and error bars are their deviation.
%E1120: 図の説明へ。前の図の説明参考に。
%Y: 図の説明に入れました。
In the MD simulations, $e$ decreases, and its deviation increases as the radius decreases.
As shown in Fig.~\ref{fig:interaction}, the hysteresis is larger for smaller radii and higher impact velocities, which explains the trend of the COR in Fig.~\ref{fig:v-e}.

%Y0123 ここから新規
The coefficient of restitution $e$ obtained by the MD simulations for $R=88.4$ is the same as Tanaka et al. \citep{2012PThPS.195..101T}.
Nietiadi et al. \citep{2020NRL....15...67N,2017PCCP...1916555N} also found that $e$ has the peak and weakly depends on the particle radius for radii between 10-20 nm, which corresponds to $R\simeq$ 30 and 60 in the LJ units.
%E0123: comaring ... の意味がよくわかりません。 10 nm と 20 nm のCORがあまり変わらないということを言いたい?
%Y0123: その通りです。上記のように修正
However, our MD simulations show that the results of collisions with $R=29.6$ and 58.9 are quite different; all collisions with $R=29.6$ result in coalescence while $e$ reaches $\sim0.3$ for $R=58.9$.
The difference between the previous studies and our study comes from the interaction of molecules consisting of particles.
%E0123: structure と言うより相互作用?
%Y0123: 相互作用に変更
The previous studies used silica nanoparticles in which covalent bonds bond Si and O atoms.
Such a particle is expected to be more rigid and to have a larger Young's modulus than that in particles composed of LJ atoms.
Comparison to the previous studies suggests that the molecular bonds affect particle collisions well.

\begin{figure}[h]
    \centering
    \includegraphics[width=0.6\columnwidth]{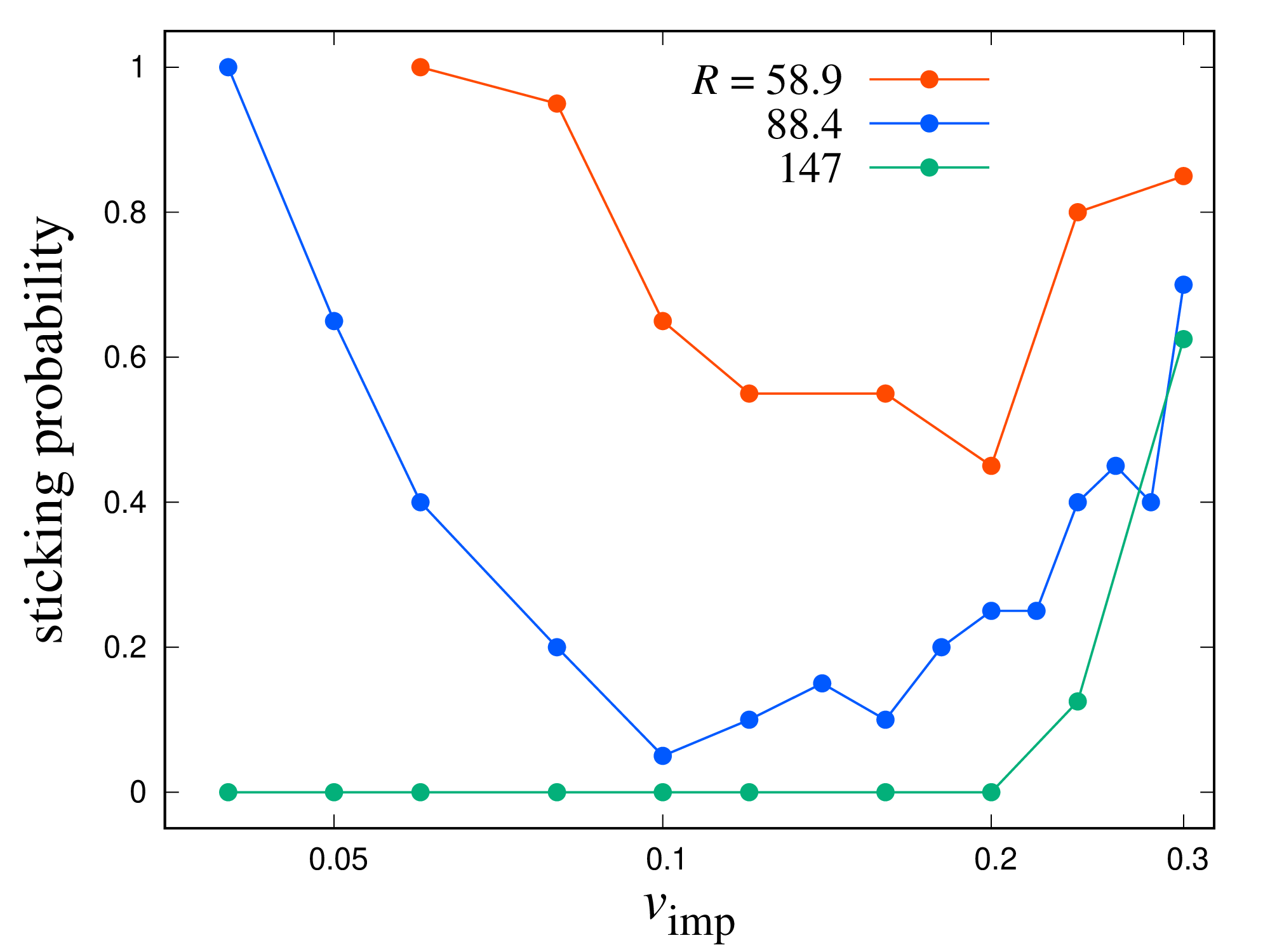}
    \caption{
    Sticking probability as a function of the impact velocity for each particle radius, which is calculated from 20 runs with $R=58.9$ and 88.4, and 8 runs for $R=147$.
    }
    \label{fig:sticking-pro}
\end{figure}

Figure~\ref{fig:sticking-pro} shows the sticking probability for each particle radius as a function of $v_{\rm imp}$.
For low impact velocities, smaller particles easily stick.
The onset impact velocity for bouncing increases with decreasing particle radius.
For $v_{\rm imp}>0.2$, the sticking probability rapidly decreases since coalescence occurs with strong energy dissipation due to plastic deformation.

\subsubsection{Comparison with the Krijt model including the plastic deformation}

In the Krijt model, we set the relaxation time to be $T_{\rm vis}(R) = 0.075 (R/147)$ \cite{2013JPhD...46Q5303K}.
%This value corresponds to the previous study \cite{2013JPhD...46Q5303K}.
The Krijt model agrees well with the MD results for $v_{\rm imp}\lesssim0.10$ with $R=147$ and for $v_{\rm imp}=0.10$ with $R=294$.
%The MD results for $v_{\rm imp}\lesssim0.08$ with $R=88.4$ are reproduced within the error bars and do not match at $R=58.9$.
For $v_{\rm imp}\lesssim0.08$, the Krijt model reproduces the coefficient of restitution $e$ with $R=88.4$ to within the error bars but not that with $R=58.9$.
In other words, the Krijt model reproduces MD results only for low impact velocities and large radii.%, although it cannot agree with the MD results for small radii, even with low impact velocities.
%E1120: 半径が大きくて衝突速度が小さい場合のみ再現すると書けばいいのでは?
%Y: 修正しました。

\begin{figure}[h]
	\includegraphics[width=0.8\columnwidth]{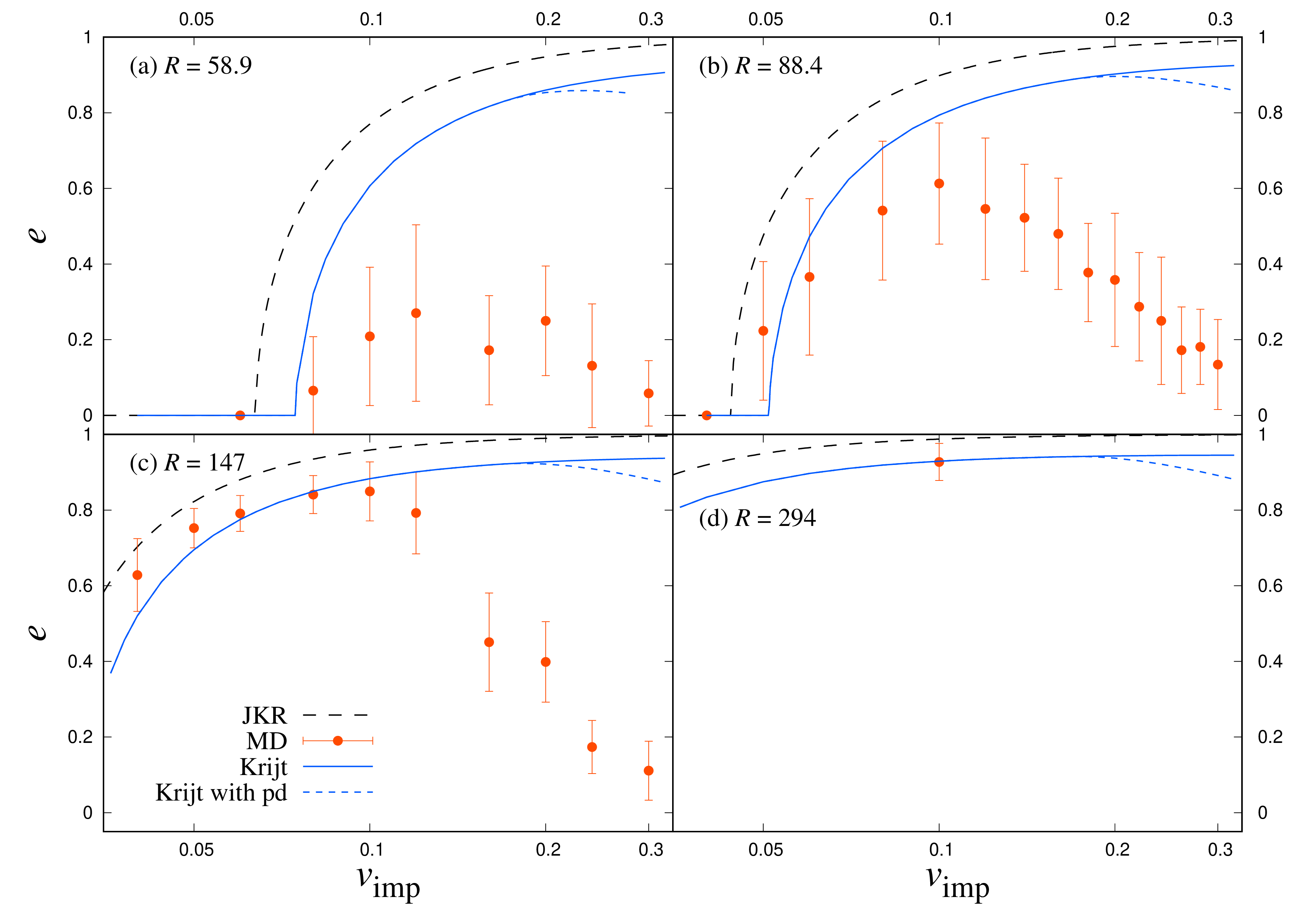}
	\caption{ \label{fig:v-e}
        Variation of the COR as a function of impact velocity.
        The red points with error bars are the RMS and its deviation from 20-run results with $R=58.9$ and 88.4, and 8-run results for $R=147$ and 294.
        We compare the MD results with the JKR theory (dashed black curves) and the Krijt model (solid blue curves).
        The dashed blue curves show the Krijt model with plastic deformation (pd) for $Y=0.12E^*$.
        The relaxation time is $T_{\rm vis}(R)=0.075(R/147)$, which is linear with respect to the particle radius.	
	}
\end{figure}

The Krijt model considers the effect of plastic deformation based on Thornton and Ning \cite{THORNTON1998154}.
The plastic deformation leads to decreased $e$ due to energy loss as shown in eq.~(\ref{eq:TN}).
%E1120: この文と前の文の繋がりが不明。
%Y: 上記のように修正しました。失われるエネルギーが塑性変形の開始圧力に依存し、理論的にはp=1.6Yであるという趣旨です。
%\textcolor{red}{The plastic deformation theoretically occurs when $p_{\rm c}\simeq 1.6Y$, and it is considered that $Y\simeq0.38E^*$.}
%For this yield strength \textcolor{red}{($Y\simeq0.38E^*$)}, the impact velocity at the onset of plastic deformation is $v_{\rm y}\simeq 1.8$, although the COR begins to decrease at $v_{\rm imp}\simeq0.1$, as shown in Fig.~\ref{fig:v-e}.
%For the onset impact velocity to be $v_{\rm y}=0.1$, the yield strength would be $Y\simeq0.12E^*$
%\textcolor{red}{
%Although for $Y \simeq 0.38E^*$ the corresponding impact velocity at the onset of plastic deformation is $v_{\rm y} \simeq 1.8$, the COR of MD simulations decreases for $v_{\rm imp} > 0.1$, as shown in Fig.~\ref{fig:v-e}.
%Hence we set $v_{\rm y}$ to be 0.1 to reproduce the decrease in $e$ with the effect of the plastic deformation.
%The corresponding yield strength $Y=0.12 E^*$, which is similar to the value of the previous works \citep[e.g.,][]{1993PhysRevB.48.6795Q,2014PhRvE..89c3308T}.
%}
%E1120: どの理由?
%Y: defect-free結晶だからという理由でしたので、そう記述しました。
%The dotted blue curve in Fig.~\ref{fig:v-e} shows the results with $Y\simeq0.12E^*$.
%E1120: 図の説明へ移動。 results 具体的に。
%Y: 移動しました。
As discussed in Sec.~\ref{sub:pladef}, we employ $v_Y=0.17$ obtained by Takato for the Krijt model with plastic deformation, and plot it in Figs.~\ref{fig:v-e147}, \ref{fig:v-e}, and \ref{fig:krijt_ex}.
However, as shown in Fig.~\ref{fig:v-e}, the energy dissipation due to the plastic deformation of the model is insufficient to reproduce the MD results although the COR of the Krijt model with plastic deformation decreases with $v_{\rm imp}$ for $v_{\rm imp}\gtrsim0.20$.
The value of $v_Y=0.17$ is estimated from the Hertz theory, and the JKR theory is expected to give $v_Y$ just slightly smaller.
Consequently, the model curves are expected to shift to the lower left.
However, the shift is not enough to be consistent with the MD simulations although the shift increases with decreasing particle radius.
Even when the impact velocity is near or less than the onset velocity of the plastic deformation, more dissipation of the translational kinetic energy occurs in the MD simulations.
In particular, the energy dissipation strongly depends on the particle radius and the Krijt model does not fit with this radius dependence.
The results of this study indicate that collisions between particles with adhesion result in additional dissipation compared to the previous MD studies using particles without adhesion.
%We thus find that conventional models cannot explain the energy dissipation that occurs in the MD simulations.

%図\ref{fig:v-e}は反発係数を衝突速度の関数として表している。
%灰色は結晶方向を変えた各ランの結果である。
%$R=29.4, 58.9$, 88.4で20ラン、$R=147$, 294で8ラン実行し、エラーバー付き点は二乗平均平方根$e_{\rm RMS}=\sqrt{\langle e^2 \rangle}$とその偏差を示す。
%結晶方向は衝突結果に影響を及ぼす。

%図\ref{fig:v-e}よりMDの反発係数はJKR理論値より小さく、MDでよりエネルギー散逸が生じていることを示す。
%そして、MDでは$v_{\rm imp}\sim0.1$で反発係数はピークを持っており、JKR理論と顕著に異なる性質を持つ。
%MD計算で見られたエネルギー散逸は、分子のランダム運動エネルギーの増加や変形による散逸が重要であると考えられ、エネルギーや塑性変形に関する議論はこの後に行う。

%また、図\ref{fig:v-e}ではKrijtモデルとの比較を行っている。
%$R=58.9$では散逸が弱くて、MDと一致しない。
%$R\gtrsim88.4$で、低速度衝突時の反発係数は再現できるようになっている。
%粒子半径$R$が大きくなるほどに、MDと一致する衝突速度領域が増えていく。
%しかし、Krijtモデルでは反発係数のピークは再現できなかった。

%\subsection{Energy distribution}
\subsection{Energetics}
\label{sub:energy}

\begin{figure}[h]
	\includegraphics[width=0.6\columnwidth]{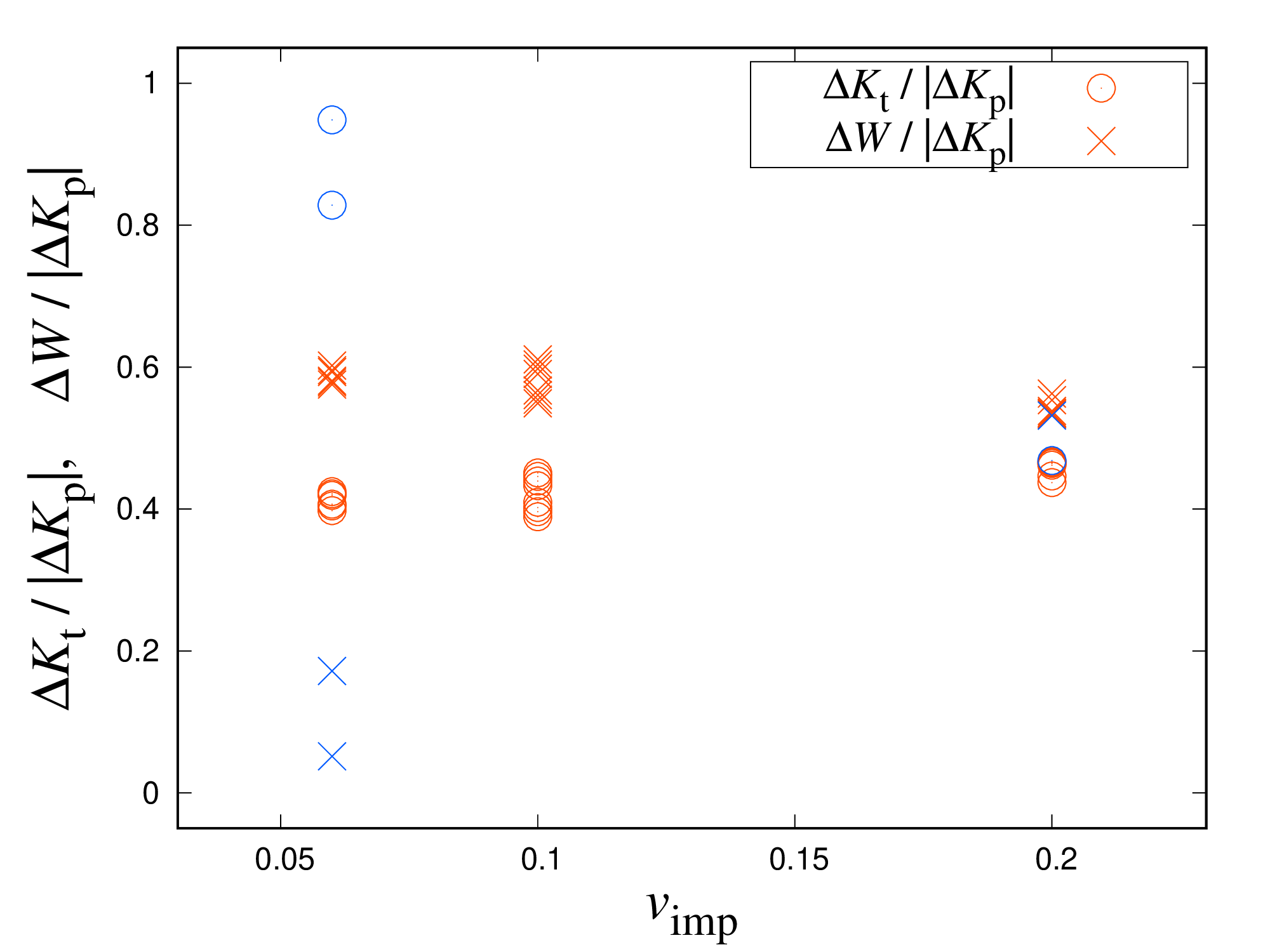}
	\caption{
	The energy conversion $\Delta K_{\rm t}$ and $\Delta W$ as a function of $v_{\rm imp}$.
        The circles and crosses represent $\Delta K_{\rm t}$ and $\Delta W$, respectively.
        %The color bar shows the COR.
        Red represents bouncing results, and blue represents the sticking cases.
%E1122: 図の説明を入れる。分母は K_p ではない? K_t/K_p はΔK_t/K_pにする。合体と反発がどれらわかるようにする。
		\label{fig:ene1}
	}
\end{figure}

In molecular systems, the total energy $K + W$ is conserved, where $K$ and $W$ are the kinetic and potential energies, respectively.
The kinetic energy can be divided into $K=K_{\rm p} + K_{\rm t}$, where $K_{\rm p}$ is the translational kinetic energy of the macroscopic particles and $K_{\rm t}$ is the kinetic energy of random molecular motions; i.e., the thermal energy:
%E1122: 粒子の並進運動のエネルギーと言ったほうがいい?
%Y: translational kinetic energyにしました。
%which can be treated as thermal motion:
\begin{equation}
    K_{\rm t} =  \sum_j^N \frac{1}{2} m_j (v_j - V_{i})^2,
\end{equation}
where $m_j$ and $v_j$ are the mass and velocity of the $j$-th atom, and $V_i$ is the velocity of the center of mass of the macroscopic particle to which the $j$-th atom belongs.
Because the initial temperature of the particle is close to zero, $K_{\rm t}$ at the end of each simulation is almost equal to the increase of the thermal energy $\Delta K_{\rm t}$.
Bulk systems, such as those considered by the JKR theory, mainly discuss the bulk kinetic energy $K_{\rm p}$ and do not consider changes in $K_{\rm t}$ and $W$.
%E1122: bulk systems? JKR は bulk sysmte?
%Y: JKRはbulk systemです。
Here, we investigate the energetics of collisions in the MD simulations.
%energy conversion ratio of $\Delta W$ and $K_{\rm t}$ to the particle kinetic energy change $\Delta K_{p}$ in the molecular system.
%E1122: energy conservation ratio は聞いたことがない。
%E1122: 以下の段落はもう少し丁寧に説明する。合体/反発、高速/低速に分けて説明する。
%Y: 区別した書き方にしてみました。
Figure~\ref{fig:ene1} shows $\Delta K_{\rm t}$ and $\Delta W$ as a function of $v_\mathrm{imp}$.
The respective energy--conversion ratios are $\Delta W/|\Delta K_{\rm p}| \sim 40\%$ and $\Delta K_{\rm t}/|\Delta K_{\rm p}| \sim60\%$ when particles bounce and disconnection occurs.
%E1122: ? raito は使わない。これはいつもではない。ΔK_pで規格化している? K_p ではない?
%Y: 減った並進運動エネルギーがどこへいったのか、について述べるためです。
When particles coalesce, most of $K_\mathrm{p}$ is converted to $K_{\rm t}$ for low impact velocities.
In this case, all of $K_\mathrm{p}$ should be converted, but the change in $W$ is small at low impact velocities because barely any molecular displacement occurs.
Thus, the coalescence results for $v_{\rm imp}=0.06$ are dominated by energy conversion to $K_{\rm t}$.
On the other hand, at high impact velocities, $v_{\rm imp}=0.20$, the energy conversion for the coalescing case is the same as that for a case in which disconnection occurs.
The particles are deformed, and sufficient molecular displacements occur to dissipate the translational kinetic energy into both the potential and thermal energies: $\Delta W/|\Delta K_{\rm p}| \sim 40\%$ and $\Delta K_{\rm t}/|\Delta K_{\rm p}| \sim60\%$.
Assuming that the masses of the two particles after disconnection are the same as those before the collision, the kinetic--energy ratio of the particles is $K_{\rm p, end}/K_{\rm p, ini} = 1 - e^2$.
We have confirmed that $\Delta K_{\rm p}/K_{\rm p, ini} = (\Delta K_{\rm t} + \Delta W)/K_{\rm p, ini} \simeq 1-e^2$ in all results, which means that fewer molecules move between the particles and that the particle masses before and after the collisions are approximately the same.

%エネルギーについて。
%N体系の総エネルギー$K_{\rm m} + E_{\rm T} + W$は保存。
%$K_{\rm m}$はモノマーの運動エネルギー、$E_{\rm T}$は分子のランダム運動のエネルギーで熱エネルギーに相当、$W$はポテンシャルエネルギーでモノマーの形状に関係すると考えられる。
%JKR理論では、モノマーの運動エネルギー$K_{\rm m}$と接触面の表面エネルギー$U_{\rm s}$の和が総エネルギーで、熱や形状変化は考えていない。
%ここではN体系の中で、バルクの運動エネルギーがどこへ変換されたかを見る。

%図\ref{fig:ene1}は、バルクの運動エネルギーの変化量に対する$E_{\rm T}, W$の変化割合。
%$\Delta E_{\rm T}+\Delta W$は$\Delta K_{\rm m}$になっている。
%実際には粒子間の分子の移動がある。
%等質量を仮定すると、エネルギー散逸割合$\Delta K_{\rm m}/K_{\rm m,ini}=1-e^2$になる。
%$(\Delta E_{\rm T}+\Delta W)/K_{\rm m,ini}$と$1-e^2$は同じになることを確認しており、移動する分子の数は全体に対して少なく、あまり重要でない。

%E1122: e のカラーバーは不要では。合体と反発でシンボルを変える? 例えば filled は合体で open は反発とか。
%Y: v_{imp}=0.20時でfilledは見にくくなるので、△にするなどはできますが、凡例が増えて見にくくなるのかなと思います。

\section{New dissipation models} \label{sec:expand}

%In section \ref{sub:restitution}, we check that the JKR theory does not agree with the MD results, and the Krijt model can reproduce a part of the MD results.
%In this section, we introduce new models to reproduce the MD results.
As shown in Sec.~\ref{sub:restitution}, our MD simulations of particle collisions have stronger energy dissipation than those predicted either by the JKR theory or the Krijt model. 
In particular, for high impact velocities with $v_{\rm imp} > 0.1$ and for small particles with $R \lesssim 58.9$, the CORs are much smaller than those predicted by the Krijt model.
Here, we propose two new dissipation models that reproduce the energy dissipation in our MD simulation results.

\subsection{Stress-dependent dissipation model} \label{sub:Krijt_ex}

%正面衝突を考える．その場合，2球の衝突後それらの相対運動運動が停止した時に
%弾性エネルギーや2球間の接触面中心の圧力は最大となる．
%その時の弾性エネルギーは衝突前の相対運動の運動エネルギー π/3 R^3 \rho v_imp^2に等しい．
%この等式より，その時のδ_max，a_max は...で与えられる．
%よって，2球間の接触面中心の最大圧力は(5) 式で与えられる．

The dissipative force of the Krijt model is in qualitative agreement with the ratio of $|F_{\rm loading}-F_{\rm unloading}|/F_{\rm c}$ obtained by the MD simulations, as discussed in Section~\ref{sub:force-Rv}.
However, for the coefficient of restitution $e$, the MD simulation results do not agree with the Krijt model.
Therefore, we discuss what the Krijt model lacks and modify the model to reproduce $e$.

The large energy dissipation that occurs in high-velocity collisions may be due to the high stresses in such collisions.
Using eq.~(\ref{eq:hertz-p}) and (\ref{eq:delta_max}), we find the pressures at maximum compression to be $p_{\rm c, H}\simeq 0.194E^*$ and $0.255E^*$ for $v_{\rm imp}=0.10$ and 0.20, respectively in the case with $R=147$.
The maximum pressures predicted by the JKR theory are $p_{\rm c, J}=0.211 E^*$ and $0.265E^*$ for $v_{\rm imp}=0.10$ and 0.20 for $R=147$.
%It also reaches $p_{\rm c, J}\simeq 0.282E^*$ for $v_{\rm imp}=0.1$ and $R=29.6$ case.
%although we should numerically estimate parameters using the JKR theory; $p_{\rm c}=0.26 E^*$ for $v_{\rm imp}=0.20$ and $R=147$.
%E1128: although 以下の意味がよくわからない。
%Y: 間違いでした。コロン以降を表記させました。
Because Young's modulus $E$ is about twice as large as $E^*$, the induced pressures are $\geq 0.1E$.
Such high pressures can cause plastic deformation and, thus, strong energy dissipation.
This estimate agrees with the strong energy dissipation observed in our MD simulation for high-velocity collisions.
On the other hand, for small particles, the pressure $p_{\rm c}$ is not small even at the stationary contact with $\delta=\delta_0$.
In fact, for $R=58.9$ and $\delta=\delta_0$, the pressure $p_c$ is obtained as $\sim 0.16E^*$ from eq.~(\ref{eq:JKR-p}).
The JKR theory predicts $p_{\rm c, J}\simeq 0.282E^*$ for $v_{\rm imp}=0.1$ and $R=29.6$.
Thus, for small particles with $R\leq 58.9$, the induced pressure can cause plastic deformation even in low-velocity collisions.
This is consistent with the small values of COR shown in Fig.~\ref{fig:v-e}a.

We, therefore, propose a new, stress-dependent dissipation model to reproduce the strong energy dissipation that occurs under high stress in particle collisions.
In this model, we adopt the same formula for the dissipation force as in the Krijt model (eq.~(\ref{eq:krijt-force})), although the relaxation time $T_{\rm vis}$ now depends on the pressure at the center of the contact area, $p_{\rm c}$ as
%We consider that the pressure is related to the dissipation.
%The dissipation force should increase with increasing the pressure $p_{\rm c}$.
%Here, We modify the relaxation time of the Krijt model as
%Krijtモデルを拡張する。
%MD計算から$v_{\rm imp}$大で散逸が大きい。
%$v_{\rm imp}$と圧力$p_{\rm max}$には関連がある。
%なので散逸には圧力依存性があるだろう。
%そこで、$T_{\rm vis}$に$p_{\rm max}$依存性の形式を与えた: $p_{\rm max}=p(0)=E^*(a^2+\delta R)/\pi aR$。
\begin{equation} \label{eq:Tvis}
    T_{\rm vis}(p_{\rm c}) = \left\{ 
    \begin{array}{lc}
	\displaystyle T_{\rm vis,0}(R)\exp \left[ \left(B \frac{p_{\rm c}}{E^*}\right)^{\zeta} \right] & (p_{\rm c}>0), \\ [10pt]
	\displaystyle T_{\rm vis,0}(R) & (p_{\rm c}<0),
    \end{array}
    \right.
\end{equation}
where $T_{\rm vis, 0}(R)$ and $B$ are coefficients and $\zeta$ is a power-law-index.
The pressure $p_{\rm c}$ at the center of the contact area is given by eq.~(\ref{eq:JKR-p}).
The relaxation time is linear in the particle radius:
\begin{equation}
    T_{\rm vis,0}(R) = C R,
\end{equation}
where $C$ is a coefficient.
We adopt the exponential form of $T_{\rm vis}$ on $p_{\rm c}$ to express the strong energy dissipation dependent on the stress.
%E1128: C の説明必要。
%Y: 説明追加しました。
Using eq.~(\ref{eq:Tvis}), we solve eq.~(\ref{eq:adot}) and (\ref{eq:EoS}) to investigate the evolution of the contact radius and the relative motion of two particles.
%E1128: もう少し何を計算するか丁寧に説明する。
%Y: 接触半径と2粒子の相対運動を計算するので、それを追記しました。
%\textcolor{red}{We change $B, C$ and $\zeta$ as $6.0<B<6.4$, $2\times10^{-4}<C<5\times10^{-4}$ and $1<\zeta<5$.}
Then, we search for optimal values of $B, C$, and $\zeta$ in the range of $6.0<B<6.4$, $2\times10^{-4}<C<5\times10^{-4}$ and $1<\zeta<5$ to fit our model to the MD simulation results.

\begin{figure}[h]
    \includegraphics[width=0.8\textwidth]{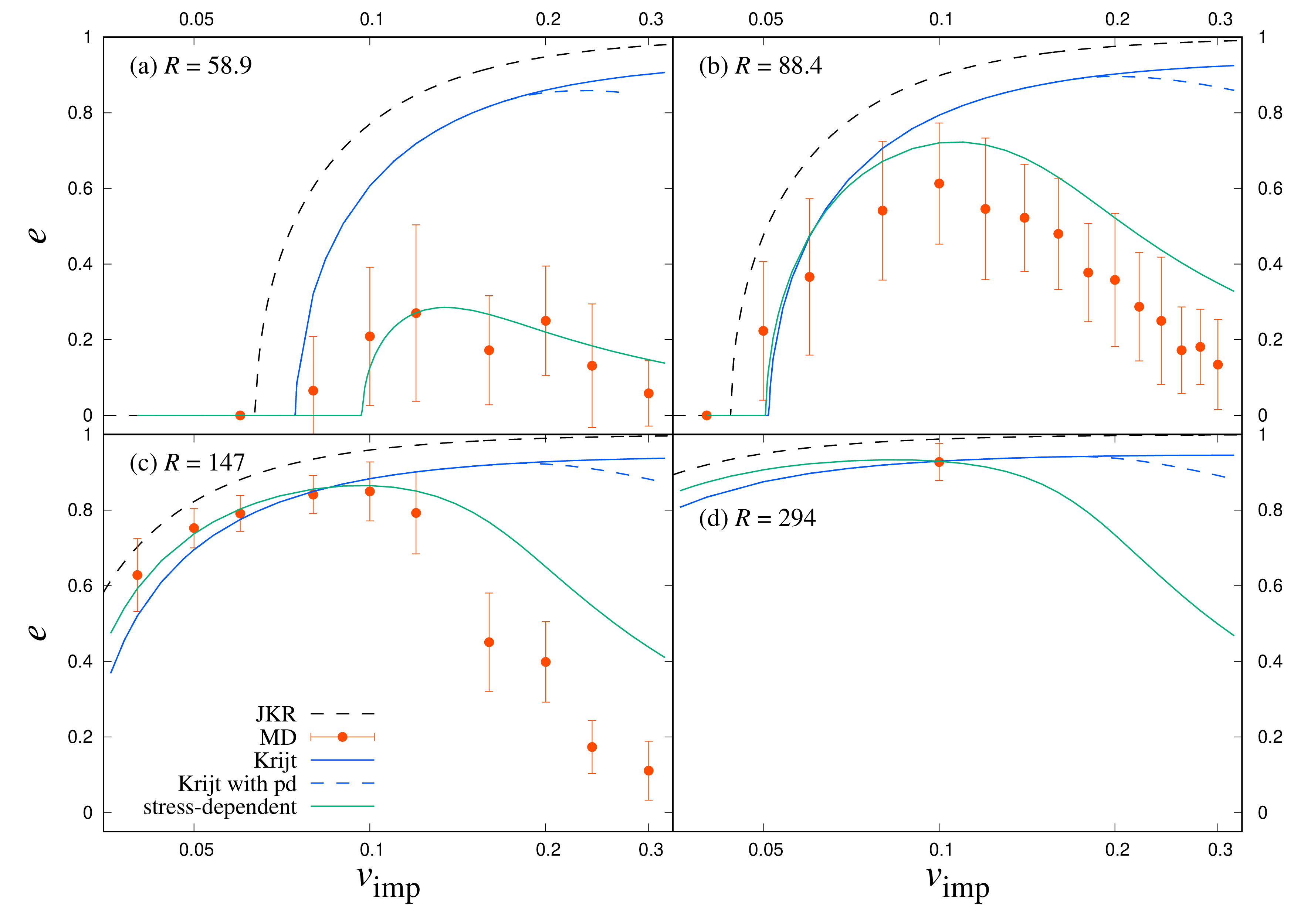}
    \caption{\label{fig:krijt_ex}
Coefficient of restitution for the new model with $T_{\rm vis, 0}(R)\simeq0.031(R/147)$, $B\simeq 6.4$ and $\zeta=4.0$ as a function of $v_{\rm imp}$ compared to the MD simulation results and the Krijt model.
    The red points with error bars represent the MD results, solid blue curves represent the Krijt model, and solid green curves represent the stress-dependent dissipation model.
    The dashed blue curves show the Krijt model with plastic deformation (pd) for $Y=0.12E^*$.
%E1128: 点とエラーバー、線種の説明必要。
%Y: 追加しました。
    }
\end{figure}

Figure~\ref{fig:krijt_ex} shows the fitting results obtained with $B\simeq 6.2$ and $\zeta=4.0$.
Since $T_{\rm vis,0}\simeq0.031$ for $R=147$, we obtain $C\simeq 2.1\times10^{-4}$.
The relaxation time $T_{\rm vis}(p_{\rm c})$ is almost the same as that in the Krijt model.
This new model is consistent with the MD simulation results well for $R=58.9$ and 294.
The new model also reproduces the MD simulation results in the range $v_{\rm imp}\lesssim0.2$ for $R=88.4$ and $v_{\rm imp}\lesssim 0.12$ for $R=147$.
Compared with the Krijt model, the new model better reproduces the MD simulation results for $R=58.9$ and the COR peak for $R=88.4$ and $147$ for low impact velocities.
These fitted values are expected to depend on the potential.
For example, in the case of silica, the pressure at which deformation occurs is expected to be greater than for LJ atoms, and then $B$ is expected to be smaller than that for LJ atoms.
At high impact velocities, there appear to be some discrepancies between the model and the MD simulations.
To achieve a more accurate reproduction at high-velocity collisions, it may be necessary to handle plastic deformation in a more precise way.

\begin{figure}[h]
    \includegraphics[width=0.5\columnwidth]{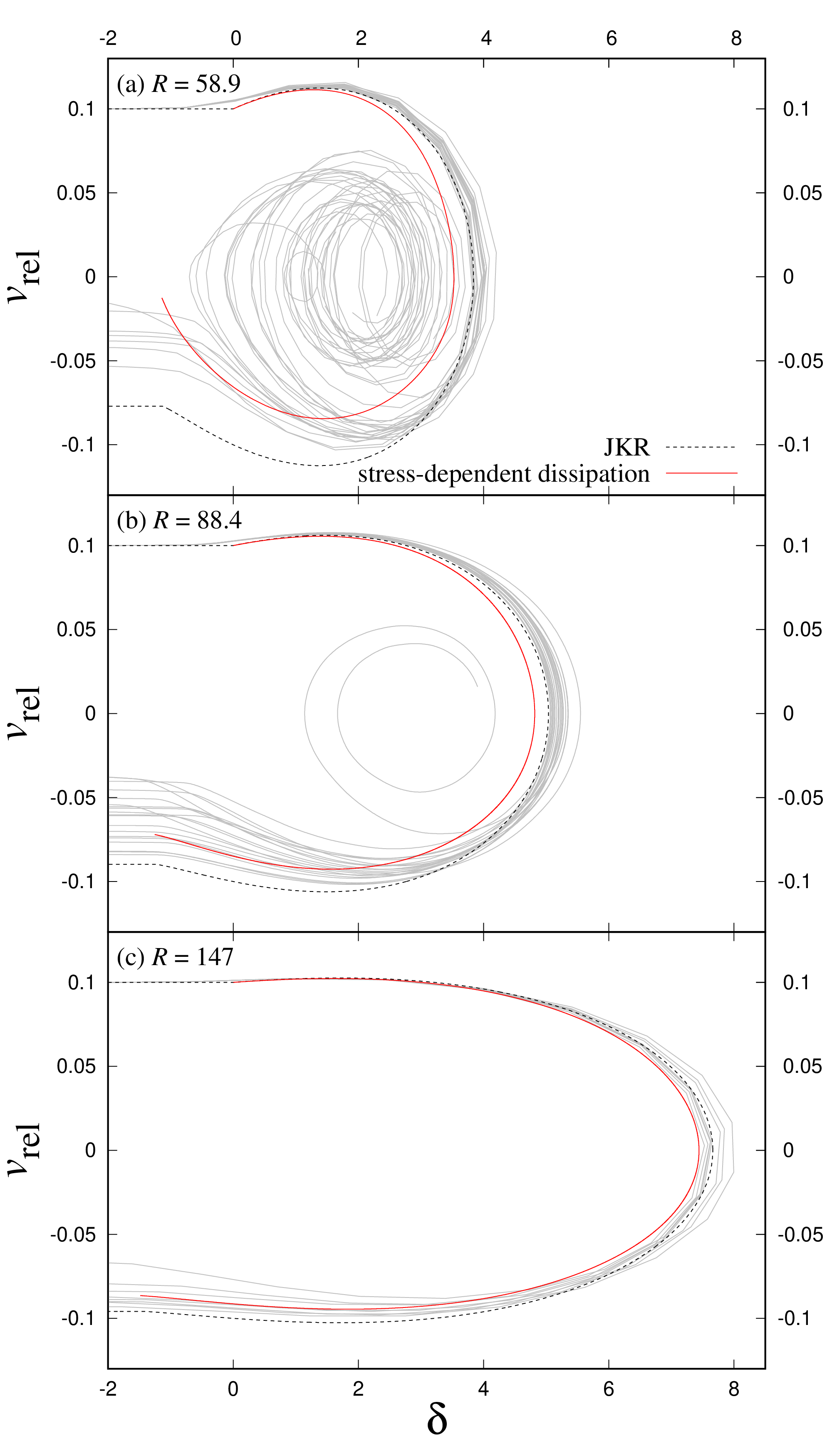}
    \caption{\label{fig:del-vrel}
    Relation between $\delta$ and $v_{\rm rel}$ for $v_{\rm imp}=0.1$ and $R=58.9, 88.4$ and 147.
    The black dotted curves represent the JKR theory, the gray curves represent each MD simulation result, and the red curves represent the new model.
    The mutual approach in the new model is smaller than in the MD simulation results.
    }
\end{figure}

We also examine the trajectories of particles in the $\delta$-$v_\mathrm{rel}$ phase plane to check whether the new model can reproduce the contact model by comparing the MD results.
%E1128: 何のためにこれを調べるのか述べる。
%Y: 軌道の観点からもMDと比較してモデルを検証するため、と述べます。
Figure~\ref{fig:del-vrel} shows the $\delta$-$v_{\rm rel}$ relation with $v_{\rm imp} = 0.10$.
%The red lines show the new model, and 
We find that $\delta$ in the new model is smaller than those in either the JKR theory or the MD simulation results.
%Because the difference between the new model and the MD simulation results decreases with increasing $R$, 
Although the new model reproduces the motion of large particles well, the displacement of $\delta$ cannot be reproduced perfectly.
%From Fig.~\ref{fig:interaction}, the interparticle force agrees with the JKR theory during the loading phase ($v_{\rm rel}>0$) but deviates from it during the unloading phase ($v_{\rm rel}<0$).
%Y0202: 以下の内容に変更
%As discussed in Sec.~\ref{subsub:hysteresis}, the plastic deformation may decrease the resilience of the particles after the maximum compression, creating hysteresis.
%The Krijt and our models include the effects of delayed contact radius and dissipation forces but do not fully represent the effect of plastic deformation.
At $R=58.9$ in Fig.~\ref{fig:del-vrel}, the difference of $\delta$ between the MD simulations and the new model at maximum compression is about 0.4, which may indicate plastic deformation.
%This suggests that the dissipation force is active during the unloading phase but hardly during the loading phase.
%In contrast, the dissipation force of the new model is active for both phases.
%The new model cannot perfectly reproduce the trajectories of $\delta$ due to the difference in the force from the MD simulation results.
%Y0131: 上記のように力の図を用いて説明
%Since smaller particles lead to stronger pressures in the contact area, as shown in Sec.~\ref{sub:pladef}, the dissipation force becomes strong even if $\delta$ is small for small $R$.
%E1128: このR依存性の理由を述べる。
%Y: Rが小さいと圧力が大きくなるので、散逸力が増大する、という旨を記載しました。
%The Krijt and the new models do not consider the effect of deformation directly because both treat the effect of deformation as a dissipation force.
%Therefore, the displacement of $\delta$ cannot be reproduced perfectly.

%$T_{\rm vis,0}$と$B$は係数、$\zeta$は指数でこれらはフリーパラメータ。
%ただし、$T_{\rm vis,0}$はKrijt et alが半径に比例することを示しており、それに従い本研究では$T_{\rm vis, 0}\simeq 0.03$を採用した。
%この$T_{\rm vis}$を用いて、$B$と$\zeta$を変化させ、接触半径進化式(\ref{eq:adot})と運動方程式(\ref{eq:EoS})を解きMD計算結果と比較した。

%図\ref{fig:modified_model}は最もfitするモデル($B\simeq 6.4$, $\zeta=4.0$)の結果を示している。
%$R=58.9$における再現はKrijtより優れている。
%そして他の半径についても$v_{\rm imp}\lesssim 0.12$と幅広い衝突速度範囲ででよく合う。
%しかし、それ以上の衝突速度では散逸が不十分である。
%これは大規模な塑性変形が原因であると考えられる。

%図\ref{fig:del-vrel}は$\delta-v_{\rm rel}$関係を示す。
%図よりモデルの変位はMDと比べて少し小さい。

\subsection{Simple dissipation model}

%E1128: なぜ簡単なモデルを考えるのかの理由を述べる。
%Y: Krijtモデルなどは複雑なので、という説明を入れました。
The Krijt and the new models described above are too complex to apply in powder simulations.
We, therefore, propose a simple power-law dissipation model in which the dissipation force is given by
%次のシンプルモデルを導入。
\begin{equation} \label{eq:expand}
	F_{\rm dis} = {\rm sgn}(v_{\rm rel})D E^* |v_{\rm rel}|^{\alpha} a^{\beta},
\end{equation}
where $D$ is a coefficient and $\alpha$ and $\beta$ are power-law-indices.
We include the factor ${\rm sgn} (v_{\rm rel})$ because the dissipation force and relative velocity point in opposite directions.
We searched for optimal values of $D$, $\alpha$, and $\beta$ to fit the MD simulation results by solving the equation of motion:
%$D$は係数、$\alpha$と$\beta$は指数。
%${\rm sgn}(\dot{\delta})$は散逸力と相対速度は反対方向のためセットする。
%$D$と$\alpha$、$\beta$を変化させ、運動方程式
\begin{equation} \label{eq:motion}
	m^*\frac{{\rm d}^2 \delta}{{\rm d} t^2} = -(F_{\rm JKR, n} + F_{\rm dis}).
\end{equation}
In this model, we do not consider the time evolution of the contact radius but instead use the contact radius based on eq~(\ref{eq:delta-a}).
%を解き、MD計算結果にフィットするモデルを探した。
%ここで換算質量$m^*$は$1/m^*=1/m_1+1/m_2$、$m_{1,2}$は粒子質量。

\begin{figure}[ht]
	\includegraphics[width=0.8\textwidth]{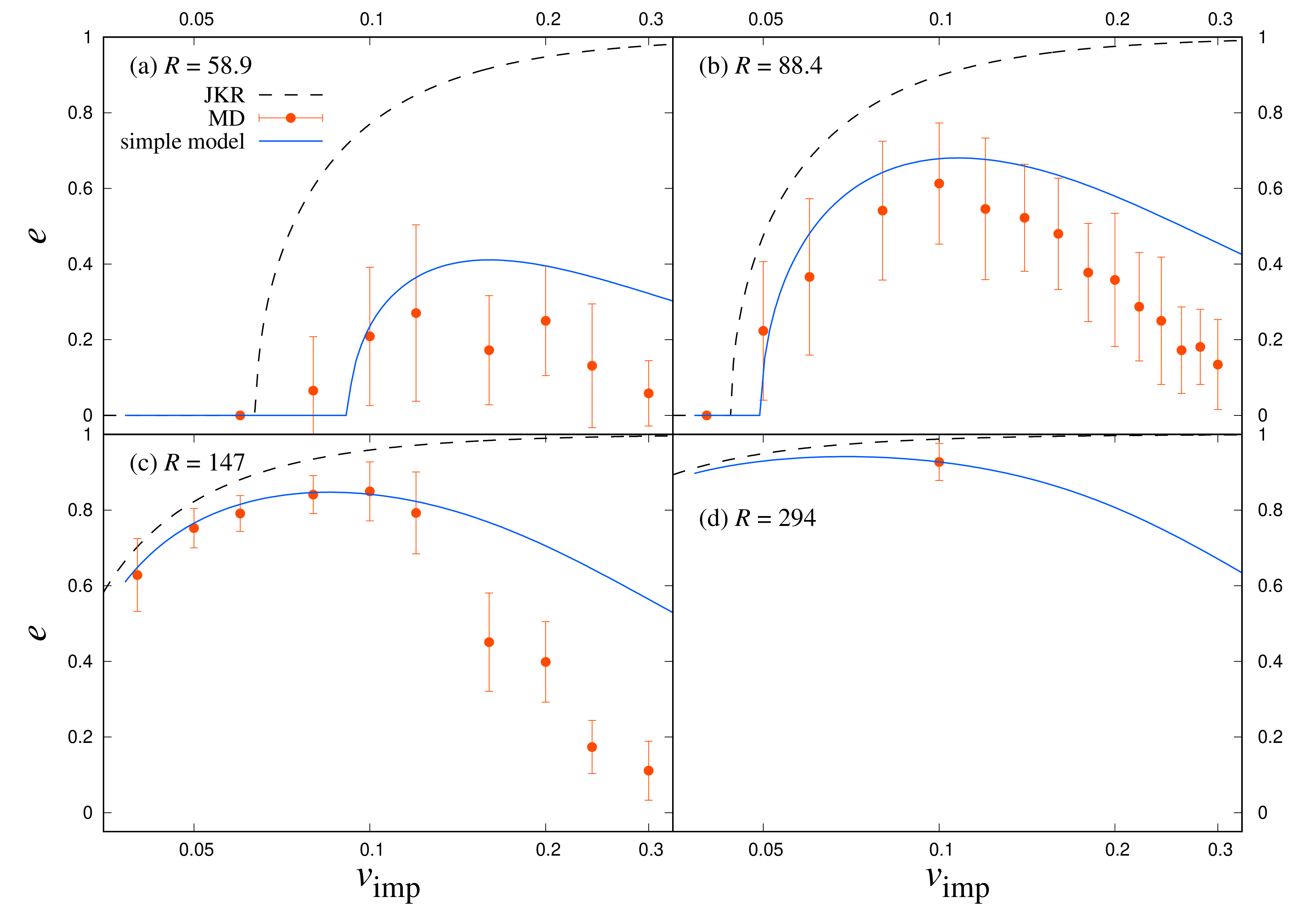}
\caption{
    Same as Fig.\ref{fig:krijt_ex} but for the simple model with ($F_{\rm dis} \propto D |v_{\rm rel}|^{3} a^{3/2}$).
    \label{fig:expansion_v}
    }
\end{figure}

Figure~\ref{fig:expansion_v} shows the COR for this simple model with $D=118$, $\alpha=3.0$ and $\beta=1.5$.
This model reproduces the MD simulation results for $v_{\rm imp}\lesssim0.12$.
Small $\alpha$ and large $\beta$ result in weak dissipation.
We find that the appropriate values are $2.0<\alpha<3.0$ and $1.0<\beta<1.5$.
Because we simply add the dissipative force to the JKR force, it is easy to use this model for powder simulations with low-velocity collisions.
%図\ref{fig:expansion_v}は、よく合うモデル$C=118, \alpha=3.0$、$\beta=1.5$の結果とMDとの比較を示している。
%このモデルでも$v_{\rm imp}\lesssim 0.12$でよく合う。
%$\alpha$は大きいと散逸が弱くなり、小さいと強くなる。
%粒子半径が小さい時に、$\beta$が小さいと散逸があまり効かなくなり、逆に$\beta$が大きいと散逸が強すぎる結果になる。
%E1128: このモデルの意義や適用制限について述べる。
%Y: この節の最後に記載しました。

%\section{Discussion} \label{sec:discussion}

%\subsection{Plastic deformation}

%The theoretical model of plastic sphere collisions gives the onset of plastic deformation.
%The impact velocity of deformation onset is given as 
%塑性変形が生じる衝突速度$v_y$は次で与えられる:
%\begin{equation}
%	v_y = 10.2 \frac{R^*(1.6Y)^5}{m^*E^{*4}},
%\end{equation}
%where $Y$ is the yield strength, which is $Y/E\geq0.2$ theoretically.
%When using $\nu=0.37$ and $Y/E=0.07$, we can explain $v_y\simeq0.1$, where the peak of $e$ exists.
%However, the model cannot explain the decline of the COR.
%The model of plastic deformation considers only the contact area, but does not consider the deformation over the particle seen in Fig.~\ref{fig:deform}.
%$Y$は yield strengthで、$Y/E\geq0.2$。
%衝突速度ー反発係数グラフのピークは塑性変形のしきい値で説明できる。
%$\nu=0.347$として、$Y/E\sim0.07$の時に$v_y\simeq0.1$となり、反発係数の減少は説明できる。
%しかし、減少の傾きは説明できない事に注意する必要がある。
%塑性変形モデルはあくまで接触面での塑性変形しか考慮していないが、スナップショットで見られるように粒子全体に渡る変形は含まれていないためである。

%Some results suggest evidence of plastic deformation.
%We can check easily the plastic deformation in Figure~\ref{fig:interaction_high}.
%The significant shift of the curve represents the plastic deformation.
%This hysteresis can be seen in Figure~\ref{fig:interaction}, so we can consider that plastic deformation occurs even for low-velocity impact.

%塑性変形は相互作用の図のシフトから読み取れる。
%また、スナップショットの高速度衝突を見ればわかりやすい。
%エネルギーに関して、熱エネルギー増加分だけポテンシャルエネルギーも増加するはず。
%つまり$\Delta W - \Delta K_{\rm T}$が変形によるエネルギー変化分。

%E1128ここまで
\section{Summary} \label{sec:Summary}

In the present work, we have studied head-on collisions between two equal-mass particles by using Molecular Dynamics (MD) simulations with the Lennard-Jones (LJ) potential as the intermolecular potential.
We have investigated the normal interparticle force between the macroscopic particles and the coefficient of restitution (COR) $e$, and we have constructed a new contact model that includes energy dissipation to reproduce the simulation results.
Our main findings are summarized as follows:

\begin{enumerate}

    \item 
    In the unloading phase of collisions between two particles, the interparticle force deviates from that of the JKR theory, whereas in the loading phase, it agrees with the JKR theory (Figs.~\ref{fig:interaction147} and \ref{fig:interaction}).
    The difference in the force between the loading and unloading phases represents the hysteresis in the particle interaction, which dissipates the kinetic energy of the motion of each particle's center of mass.
%E1128: 論理が逆では。エネルギー散逸があるからヒステリシスが見られる。以下の文も改訂必要。
%Y: indicateに変更しました。
    The contact radius also has hysteresis; $a$ of the MD simulations is smaller than that of the JKR theory in the loading phase and larger in the unloading phase.
    %The delay in $\dot{a}$ does not contribute to the hysteresis well.
    The hysteresis in the force is greater for smaller particle sizes or higher collision velocities.
    In particular, for high-velocity collisions with $v_{\rm imp}\gtrsim0.2$, plastic deformation of the particle is observed (Figs.~\ref{fig:interaction_high} and \ref{fig:deform}).
    The plastic deformation for $v_{\rm imp}\gtrsim0.2$ can be explained by the yield strength estimated by the previous studies.
    %The force at the loading phase is different from the unloading phase, although it agrees with the JKR theory at the loading phase.
    %This difference shows the hysteresis in the collision process.
    %The significant hysteresis can be seen in high-velocity collisions, where the compression length $\delta$ remains large at the unloading phase.
    %This displacement of $\delta$ suggests particle deformation, and the deformation causes the hysteresis.
    %The displacement is large for high impact velocity, so the high impact velocity results in the large hysteresis.
    %The degree of hysteresis is evaluated by the ratio of the difference of the force to the force at the loading.
    %The force at the loading phase is small for a small radius, so the small particle results in a large hysteresis.
    %The hysteresis leads to the difference in relative velocity between before and after the collisions, which causes inelasticity of collisions.
%   相互作用はloading時はJKRと一致し、loadingとunloadingにヒステリシスが見られる。
%	unloadingではloadingより重心間距離が小さくなっており、変位が小さくなっていることが分かった。
%	これは接触面付近での塑性変形によるものと考えられる。
%	また衝突速度が大きい場合には、変形はモノマー全体に渡る。

    \item 
    Energy dissipation in a collision reduces $e$.
    The CORs obtained in our MD simulations are smaller than those of the JKR theory, which is natural because energy dissipation occurs at any time during contact in the MD simulations.
    %the JKR theory does not have the hysteresis of the interaction force during the contact of the particles.
%E1128: ここもヒステリシスを出すのは少し変。    大事なのはエネルギー散逸。以下も改訂必要。
%Y: 接触中はどこでもエネルギー散逸が起きているためとしました。
    The Krijt model includes such hysteresis in the force to accurately describe the energy dissipation.
    For high-velocity collisions ($v_{\rm imp}>0.1$), our MD simulations show much higher energy dissipation than that predicted by the Krijt model, although the Krijt model reproduces our simulations well for low-velocity collisions ($v_{\rm imp}<0.1$) in Fig.~\ref{fig:v-e}.
    %The hysteresis causes energy dissipation and makes the COR small.
    %The COR of the MD simulations is smaller than that of the JKR theory because the JKR theory does not consider the hysteresis (Fig.~\ref{fig:v-e}).
    %The COR significantly decreases compared to the JKR theory for high impact velocity and small radius because the hysteresis is large for these conditions.
    %Compared to the Ktijt model, which includes the hysteresis, we find that the Krijt model agrees with the MD results for small impact, although it does not agree with the MD results for small radii.
%	どの衝突速度やモノマー半径でもMD計算で得られた反発係数はJKR理論より低いことが示された。
%	MD計算結果は衝突速度依存性ではv=0.1を超えると反発係数は減少し始め、これは塑性変形で説明できるかもしれない。
%	しかし、高速度衝突時の塑性変形によるエネルギー散逸量は多大であり、接触面のみを考えた塑性変形モデルでは散逸量は不十分である。
%	Krijtモデルは、半径が大きいほど、低速度衝突側から合うようになる。
    In contrast, the strong energy dissipation in the MD simulations for $v_{\rm imp}>0.1$ cannot be explained even by the Krijt model with plastic deformation using the yield strength obtained by the previous MD simulations of collisions between non-adhesive particles.
    Collisions of adhesive particles produce additional dissipation rather than non-adhesive particles.
    %%%%%%%%%%%%%%%%%%%%%%%%%%%%%%%%%%%%%%%%%%%%%%%%%%%v>0.1でのMDの強い散逸は、plastic defromationでは説明できない (Sec IV.B.3).

    \item 
    To reproduce the strong energy dissipation observed in our MD simulations for high-velocity collisions, we have proposed a new, stress-dependent dissipation model in which the relaxation time, $T_{\rm vis}$, increases rapidly with the pressure at the center of the contact surface, $p_{\rm c}$, for $p_{\rm c}>0.1E^*$ (eq.~(\ref{eq:Tvis})).
    We found that the stress-dependent dissipation model successfully reproduces the CORs in our MD simulations well for $v_{\rm imp}<0.2$.
    We have also proposed another simple dissipation model (eq.~(\ref{eq:expand})) that can reproduce the CORs of our MD simulations for $v_{\rm imp}<0.2$ and which is expected to be useful for powder simulations.
    %We consider that the gap between the Krijt model and the MD results comes from the large pressure at the contact area for high impact velocity.
    %Therefore, we propose a new model as eq.~(\ref{eq:Tvis}) so that $T_{\rm vis}(p_{\rm max})$ increases with increasing pressure at the center of contact $p_{\rm max}$.  
    %We find that the new model with $\zeta=4$ can reproduce the MD results for $v_{\rm imp} \lesssim 0.12$.
    %The model of rapidly increasing $T_{\rm vis}$ with respect to $p_{\rm max}$ is required to reproduce the MD results well.
    %We also propose another new model in which the simple dissipation force of eq.~(\ref{eq:expand}) is added to the JKR theory.
    %We find that this simple new model can also reproduce the MD results for $v_{\rm imp} \lesssim 0.12$.
%	Krijtモデルの$T_{\rm vis}$に接触面の中心圧力に依存する形式を与えた。
%	結果、$\exp[(Bp_{\rm max}/E^*)^4]$モデルが$v_{\rm imp}\lesssim 0.12$でMD計算をよく再現できた。
%	また、簡素な散逸力モデルを検証し、相対速度の3乗、接触半径の3/2乗に比例する散逸力モデルが$v_{\rm imp}\lesssim 0.12$でのMD計算の再現に成功した。
%	しかし、いずれも高速度衝突時の大きなエネルギー散逸を再現するには至らず、接触面以外の変形が要因である。

\end{enumerate}

The proposed new models successfully reproduce the MD results for $v_{\rm imp}<0.2$, although the energy dissipation is insufficient for higher impact velocities.
This is due to particle deformation at high impact velocities, which leads to significantly small values of the COR.
Neither the Krijt model nor the proposed models include this effect.
Thus, we cannot reproduce the COR of our MD simulations at high impact velocities.
Both the new stress-dependent model and the Krijt model solve for the evolution of the contact radius independently, although the dissipation due to delay of $a$ is negligibly small in these models.
%E1128: これここに必要? これと以下の議論は結果のところでしては。
%Y: 結果のところにいれるとわかりにくくなる内容なので、future workと同列にしています。disuccion節を作るには内容がなさすぎるというのも理由です。
We, therefore, used the same relaxation time $T_{\rm vis}$ in the dissipation force and contact radius evolution.
If $T_{\rm vis}$ in eq.~(\ref{eq:adot}) is much larger than that of dissipation force, the delay of $a$ is more significant and can be effective in decreasing COR.
This is one of the possibilities to explain our MD simulations.
We thus need to understand better the property of $T_{\rm vis}$.

We also need to consider the effects of different particle properties.
In this study, we used particles with an ideal FCC structure.
However, we should also investigate the effects of structures such as body-centered cubic (BCC) or amorphous structures and their dependence on the filling factors of the macroscopic particles.
The particle temperature also affects the collision results, and we should explore this as well in a future paper.
We should also simulate particle collisions with more realistic molecules, such as water, to understand dust growth in planet-forming regions.
These subjects are all planned for future work.

%Future workは、本研究は欠陥のないFCC結晶のモノマーを用いたが、欠陥がある場合や、別の結晶構造やアモルファス構造の場合に結果がどのようなるのかを調べる必要性が考えられる。
%また、実際の粒子衝突への応用を考えると、LJ分子ではなく、水分子を用いて調べる必要がある。
%そして、本研究は正面衝突を取り扱ったが、斜め衝突による回転などの他の相互作用についてもJKR理論とのずれを明確にする必要がある。
%We demonstrated the expanded model can reproduce the MD results and it is appropriate for the particle interaction.
%In future work, we plan to investigate other interactions such as rolling, sliding and twisting.

\begin{acknowledgments}

The authors thank Sota Arakawa for his valuable comments on our results.
Y.Y. was supported by JSPS KAKENHI Grant Number 22KJ0859, and E.K. and H.T. were supported by JSPS KAKENHI Grant Number 18H05438.
H.T was supported by JSPS KAKENHI Grant No. 23K22549.
%E1119: 渦状腕論文と同じ番号でお願いします。
%Y: 承知しました。
Our MD simulations were performed on Cray XC50 and the GPU cluster at the Center for Computational Astrophysics, National Astronomical Observatory of Japan.
The authors also thank Enago (www.enago.jp) for the English language review.

\end{acknowledgments}

\appendix

\section{Determination of $E^*$ and $R$} \label{App:Hertz}

\begin{figure*}[ht]
    \includegraphics[width=0.8\textwidth]{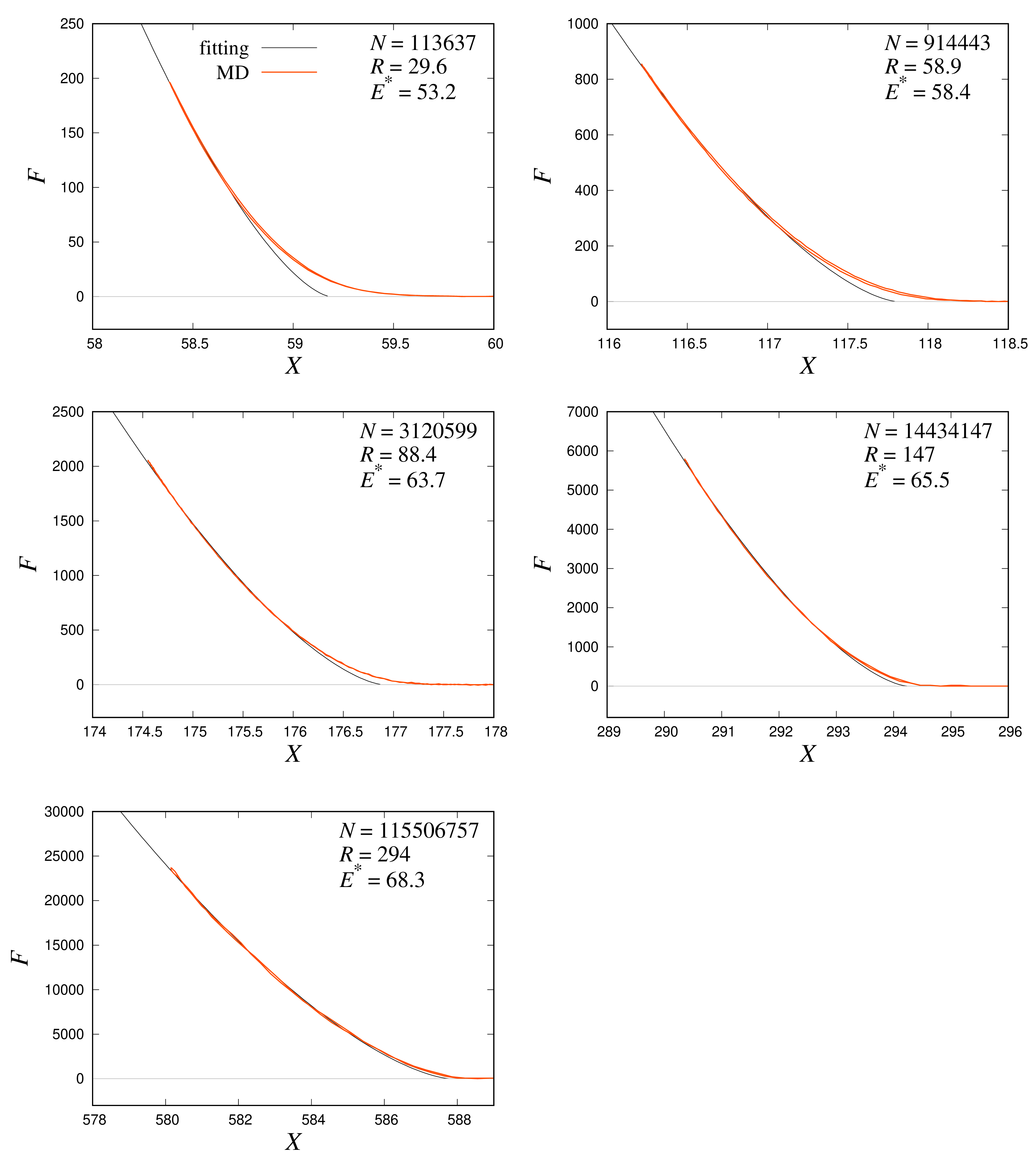}
    \caption{
    The force acts on the particles.
    The red lines are the MD results, and the black lines are the fitting results.
    \label{fig:hertz}
    }
\end{figure*}

We used MD simulations to determine the particle radius $R$ and the reduced Young's modulus $E^*$.
We prepared five particles by hollowing out atoms so that $R=$29.4, 58.8, 88.2, 147, and 294.
We also remove the attractive term $-(r/\sigma)^{-6}$ from the molecular potential between atoms belonging to different particles.
The modified LJ potential then becomes
\begin{equation}
    u_{\rm LJ, mod}(r_{ij})=4\epsilon \left[ \left(\frac{r_{ij}}{\sigma}\right)^{-12} - \delta_{k\ell}\left(\frac{r_{ij}}{\sigma}\right)^{-6} \right],
\end{equation}
where $\delta_{k\ell}$ is the Kronecker delta, and $k$ and $\ell$ represent the macroscopic particles to which the $i$- and $j$-th atoms belong, respectively.
In this case, the force between particles corresponds to a Hertzian contact.
%This method has been used to investigate the onset of plastic deformation \cite{2014PhRvE..89c3308T,2015PhRvE..92c2403T}.
We calculate the force between the particles and fit the parameters $R$ and $E^*$ using eq.~(\ref{eq:hertz}).
When we fit the parameters, we ignore the data for small $F$ since a small force works due to the intermolecular force existing even if the two particles are not in contact.
Figure.~\ref{fig:hertz} shows the forces and fitting lines as a function of $X$, and Table.~\ref{tab:model} summarizes the fitting results.
The fitting results are slightly larger than the initially prepared radii, with a difference of less than 1.
Here, we simulated only one run for each radius with the same orientation as that shown in Fig.~\ref{fig:interaction}.
Note that the fitting results may change slightly depending on the orientations.

\section{Model of crack propagation} \label{App:a}

Here, we summarize the crack--propagation model used in the Krijt model.
The apparent surface energy is expressed as \cite{2013JPhD...46Q5303K}
\begin{equation}
    G_{\rm eff} = \frac{E^*}{4\pi a R^{*2}} (a^2-\delta R^*)^2.
\end{equation}
If $\delta$ and $a$ have the relationship shown in eq.~(\ref{eq:delta-a}), which is satisfied by the JKR theory, then $G_{\rm eff} = \gamma$.
Because of viscoelasticity, however, the evolution of the contact area differs slightly from that of the JKR theory \cite[e.g.,][]{2007PhRvE..76e1302B, 2013JPhD...46Q5303K}.
The crack speed $\dot{a}$ can be written as \cite{2006JCIS..296..284G, 2004JPhD...37.2557G}
\begin{widetext}
\begin{equation} \label{eq:adot}
	\frac{\sigma_0^2 T_{\rm vis}}{2E^* \gamma} \dot{a} = \left\{ 
	\begin{array}{lc}
		\displaystyle 0.15\left[ \beta \ln \left( \frac{0.98}{1-\beta}\right) \right]^{-1}, & (\beta\leq 0.29), \\ [12pt]
		(0.1035x+0.3421)x^{1.1160}, & (0.29<\beta<1,\ {\rm where}\ x=1/\beta-1), \\ [7pt]
		-(0.2112x+0.3939)x^{1.1403}, & (1<\beta<3.7,\ {\rm where}\ x = \beta-1), \\ [4pt]
		\displaystyle -0.24 \beta \left[ \ln \left( \frac{0.98}{1-1/\beta} \right) \right]^{-1}, & (3.7\leq \beta),
	\end{array}
	\right.
\end{equation}
\end{widetext}
where $\beta=G_{\rm eff}/\gamma$ and $\sigma_0 \simeq \gamma/$(spacing between atoms).
We also assume the initial condition given in Appendix.~B of Krijt et al. \cite{2013JPhD...46Q5303K}.

\section{Surface energy} \label{App:surface_energy}

We derive the surface energy in our MD simulations using the potential energy of a spherical particle with an FCC structure of the atoms.
The surface energy of a spherical particle is given by
\begin{equation}
    \gamma = -\frac{N e_1 - U}{4\pi R^2},
\end{equation}
where $N$ is the number of atoms in the system, and $e_1$ is the potential energy per atom in the FCC solid.
The total potential energy of the particle, $U$, is obtained from our MD simulation.
The energy $e_1$ is the value when the atoms extend to infinity, meaning there is no surface or boundary.

The energy $e_1$ is obtained as follows.
If the LJ potential has no cutoff, the potential energy per atom in the FCC solid, $e_1$, is given by \cite{1976itss.book.....K}
\begin{eqnarray}
    e_1 & = & \frac{1}{2} \times 4\epsilon \left[ \sum_{j} \left( p_{j}\frac{D}{\sigma} \right)^{-12} - \sum_j \left( p_{j}\frac{D}{\sigma} \right)^{-6} \right] \nonumber \\
    & =&  2\epsilon \left[a\left(D/\sigma\right)^{-12} - b \left(D/\sigma\right)^{-6}\right],
\end{eqnarray}
where $D$ is the nearest neighbor distance and $p_{j} D$ is the distance between a reference atom and any other atom $j$.
For the FCC structure, the coefficients $a$ and $b$ are given by
\begin{equation}
    a=\sum_j p_j^{-12} = 12.13,\ b=\sum_j p_j^{-6} = 14.45.
\end{equation}
However, in this study, we set the cutoff $r_{\rm cut}=5.0\sigma$, so we need to include the effects of the cutoff in determining $e_1$.
We can neglect the cutoff effect on the coefficient $a$ due to the rapid decreases of $p_j^{-12}$.
There are two cutoff effects on the coefficient $b$.
One is to set the potential to be zero at $r(=p_j D) = r_{\rm cut}$ by replacing $p_j^{-6}$ with $p_{j}^{-6} - (r_{\rm cut}/D)^{-6}$.
The resulting correction $\Delta b_1$ is given by
\begin{equation}
    \Delta b_{1} = -\left(\frac{r_{\rm cut}}{D}\right)^{-6} \frac{4\pi}{3}\frac{r_{\rm cut}^3}{V_1} = - \frac{4\sqrt{2}\pi}{3} \left(\frac{r_{\rm cut}}{D}\right)^{-3},
\end{equation}
where $V_1=D^3/\sqrt{2}$ is the volume per atom in the FCC structure.
The other cutoff effect on $b$ is neglecting terms with $r>r_{\rm cut}$.
We find that the correction due to the neglect, $\Delta b_2$, equals $\Delta b_1$.
Thus, including these corrections, we obtain $e_1$ as
\begin{eqnarray} \label{eq:e1_corr}
    e_1 & = 2 & \epsilon \left[ 12.13 \left( \frac{D}{\sigma} \right)^{-12} \right. \nonumber \\
    & & \left. - \left\{ 14.45 - \frac{8\sqrt{2}\pi}{3} \left(\frac{r_{\rm cut}}{D}\right)^{-3} \right\} \left( \frac{D}{\sigma} \right)^{-6} \right].
\end{eqnarray}
The nearest neighbor distance $D$ is determined by the equilibrium condition ${\rm d}e_1/{\rm d}D=0$.
From the equilibrium condition with eq.~(\ref{eq:e1_corr}), we obtain $D\simeq1.091\sigma$ and $e_1\simeq-8.464\epsilon$.
The obtained $D$ agrees with that of the spherical particles in our MD simulations.

The potential $U$ is obtained from the MD simulation.
We obtain $U=-9.96\times10^5, -7.88\times10^6, -2.61\times 10^7, -1.22\times10^8$ for $R=29.6, 58.9, 88.4$ and 147 respectively in the LJ units.
Finally, we get $\gamma=3.18, 3.18, 3.15, 3.06$ respectively.
Thus, We take $\gamma=3.17$ as the nominal value in this paper.
This value agrees with that of the previous work $\gamma=3.18$ \cite{2012PThPS.195..101T}.

\bibliography{apssamp}

\providecommand{\noopsort}[1]{}\providecommand{\singleletter}[1]{#1}%
\begin{thebibliography}{54}%
\makeatletter
\providecommand \@ifxundefined [1]{%
 \@ifx{#1\undefined}
}%
\providecommand \@ifnum [1]{%
 \ifnum #1\expandafter \@firstoftwo
 \else \expandafter \@secondoftwo
 \fi
}%
\providecommand \@ifx [1]{%
 \ifx #1\expandafter \@firstoftwo
 \else \expandafter \@secondoftwo
 \fi
}%
\providecommand \natexlab [1]{#1}%
\providecommand \enquote  [1]{``#1''}%
\providecommand \bibnamefont  [1]{#1}%
\providecommand \bibfnamefont [1]{#1}%
\providecommand \citenamefont [1]{#1}%
\providecommand \href@noop [0]{\@secondoftwo}%
\providecommand \href [0]{\begingroup \@sanitize@url \@href}%
\providecommand \@href[1]{\@@startlink{#1}\@@href}%
\providecommand \@@href[1]{\endgroup#1\@@endlink}%
\providecommand \@sanitize@url [0]{\catcode `\\12\catcode `\$12\catcode `\&12\catcode `\#12\catcode `\^12\catcode `\_12\catcode `\%12\relax}%
\providecommand \@@startlink[1]{}%
\providecommand \@@endlink[0]{}%
\providecommand \url  [0]{\begingroup\@sanitize@url \@url }%
\providecommand \@url [1]{\endgroup\@href {#1}{\urlprefix }}%
\providecommand \urlprefix  [0]{URL }%
\providecommand \Eprint [0]{\href }%
\providecommand \doibase [0]{https://doi.org/}%
\providecommand \selectlanguage [0]{\@gobble}%
\providecommand \bibinfo  [0]{\@secondoftwo}%
\providecommand \bibfield  [0]{\@secondoftwo}%
\providecommand \translation [1]{[#1]}%
\providecommand \BibitemOpen [0]{}%
\providecommand \bibitemStop [0]{}%
\providecommand \bibitemNoStop [0]{.\EOS\space}%
\providecommand \EOS [0]{\spacefactor3000\relax}%
\providecommand \BibitemShut  [1]{\csname bibitem#1\endcsname}%
\let\auto@bib@innerbib\@empty
%</preamble>
\bibitem [{\citenamefont {Reynolds}\ \emph {et~al.}(2005)\citenamefont {Reynolds}, \citenamefont {Fu}, \citenamefont {Cheong}, \citenamefont {Hounslow},\ and\ \citenamefont {Salman}}]{REYNOLDS20053969}%
  \BibitemOpen
  \bibfield  {author} {\bibinfo {author} {\bibfnamefont {G.}~\bibnamefont {Reynolds}}, \bibinfo {author} {\bibfnamefont {J.}~\bibnamefont {Fu}}, \bibinfo {author} {\bibfnamefont {Y.}~\bibnamefont {Cheong}}, \bibinfo {author} {\bibfnamefont {M.}~\bibnamefont {Hounslow}},\ and\ \bibinfo {author} {\bibfnamefont {A.}~\bibnamefont {Salman}},\ }\href {https://doi.org/https://doi.org/10.1016/j.ces.2005.02.029} {\bibfield  {journal} {\bibinfo  {journal} {Chemical Engineering Science}\ }\textbf {\bibinfo {volume} {60}},\ \bibinfo {pages} {3969} (\bibinfo {year} {2005})},\ \bibinfo {note} {granulation across the length scales - 2nd International Workshop on Granulation}\BibitemShut {NoStop}%
\bibitem [{\citenamefont {Jutzi}\ \emph {et~al.}(2008)\citenamefont {Jutzi}, \citenamefont {Benz},\ and\ \citenamefont {Michel}}]{JUTZI2008242}%
  \BibitemOpen
  \bibfield  {author} {\bibinfo {author} {\bibfnamefont {M.}~\bibnamefont {Jutzi}}, \bibinfo {author} {\bibfnamefont {W.}~\bibnamefont {Benz}},\ and\ \bibinfo {author} {\bibfnamefont {P.}~\bibnamefont {Michel}},\ }\href {https://doi.org/https://doi.org/10.1016/j.icarus.2008.06.013} {\bibfield  {journal} {\bibinfo  {journal} {Icarus}\ }\textbf {\bibinfo {volume} {198}},\ \bibinfo {pages} {242} (\bibinfo {year} {2008})}\BibitemShut {NoStop}%
\bibitem [{\citenamefont {Mishra}\ and\ \citenamefont {Thornton}(2001)}]{MISHRA2001225}%
  \BibitemOpen
  \bibfield  {author} {\bibinfo {author} {\bibfnamefont {B.}~\bibnamefont {Mishra}}\ and\ \bibinfo {author} {\bibfnamefont {C.}~\bibnamefont {Thornton}},\ }\href {https://doi.org/https://doi.org/10.1016/S0301-7516(00)00065-X} {\bibfield  {journal} {\bibinfo  {journal} {International Journal of Mineral Processing}\ }\textbf {\bibinfo {volume} {61}},\ \bibinfo {pages} {225} (\bibinfo {year} {2001})}\BibitemShut {NoStop}%
\bibitem [{\citenamefont {{Hestroffer}}\ \emph {et~al.}(2019)\citenamefont {{Hestroffer}}, \citenamefont {{S{\'a}nchez}}, \citenamefont {{Staron}}, \citenamefont {{Bagatin}}, \citenamefont {{Eggl}}, \citenamefont {{Losert}}, \citenamefont {{Murdoch}}, \citenamefont {{Opsomer}}, \citenamefont {{Radjai}}, \citenamefont {{Richardson}}, \citenamefont {{Salazar}}, \citenamefont {{Scheeres}}, \citenamefont {{Schwartz}}, \citenamefont {{Taberlet}},\ and\ \citenamefont {{Yano}}}]{2019A&ARv..27....6H}%
  \BibitemOpen
  \bibfield  {author} {\bibinfo {author} {\bibfnamefont {D.}~\bibnamefont {{Hestroffer}}}, \bibinfo {author} {\bibfnamefont {P.}~\bibnamefont {{S{\'a}nchez}}}, \bibinfo {author} {\bibfnamefont {L.}~\bibnamefont {{Staron}}}, \bibinfo {author} {\bibfnamefont {A.~C.}\ \bibnamefont {{Bagatin}}}, \bibinfo {author} {\bibfnamefont {S.}~\bibnamefont {{Eggl}}}, \bibinfo {author} {\bibfnamefont {W.}~\bibnamefont {{Losert}}}, \bibinfo {author} {\bibfnamefont {N.}~\bibnamefont {{Murdoch}}}, \bibinfo {author} {\bibfnamefont {E.}~\bibnamefont {{Opsomer}}}, \bibinfo {author} {\bibfnamefont {F.}~\bibnamefont {{Radjai}}}, \bibinfo {author} {\bibfnamefont {D.~C.}\ \bibnamefont {{Richardson}}}, \bibinfo {author} {\bibfnamefont {M.}~\bibnamefont {{Salazar}}}, \bibinfo {author} {\bibfnamefont {D.~J.}\ \bibnamefont {{Scheeres}}}, \bibinfo {author} {\bibfnamefont {S.}~\bibnamefont {{Schwartz}}}, \bibinfo {author} {\bibfnamefont {N.}~\bibnamefont {{Taberlet}}},\ and\ \bibinfo {author} {\bibfnamefont {H.}~\bibnamefont {{Yano}}},\
  }\href {https://doi.org/10.1007/s00159-019-0117-5} {\bibfield  {journal} {\bibinfo  {journal} {The Astronomy and Astrophysics Review}\ }\textbf {\bibinfo {volume} {27}},\ \bibinfo {eid} {6} (\bibinfo {year} {2019})},\ \Eprint {https://arxiv.org/abs/1907.02615} {arXiv:1907.02615 [astro-ph.EP]} \BibitemShut {NoStop}%
\bibitem [{\citenamefont {{Suyama}}\ \emph {et~al.}(2008)\citenamefont {{Suyama}}, \citenamefont {{Wada}},\ and\ \citenamefont {{Tanaka}}}]{2008ApJ...684.1310S}%
  \BibitemOpen
  \bibfield  {author} {\bibinfo {author} {\bibfnamefont {T.}~\bibnamefont {{Suyama}}}, \bibinfo {author} {\bibfnamefont {K.}~\bibnamefont {{Wada}}},\ and\ \bibinfo {author} {\bibfnamefont {H.}~\bibnamefont {{Tanaka}}},\ }\href {https://doi.org/10.1086/590143} {\bibfield  {journal} {\bibinfo  {journal} {Astrophysical Journal}\ }\textbf {\bibinfo {volume} {684}},\ \bibinfo {pages} {1310} (\bibinfo {year} {2008})}\BibitemShut {NoStop}%
\bibitem [{\citenamefont {{Wada}}\ \emph {et~al.}(2013)\citenamefont {{Wada}}, \citenamefont {{Tanaka}}, \citenamefont {{Okuzumi}}, \citenamefont {{Kobayashi}}, \citenamefont {{Suyama}}, \citenamefont {{Kimura}},\ and\ \citenamefont {{Yamamoto}}}]{2013A&A...559A..62W}%
  \BibitemOpen
  \bibfield  {author} {\bibinfo {author} {\bibfnamefont {K.}~\bibnamefont {{Wada}}}, \bibinfo {author} {\bibfnamefont {H.}~\bibnamefont {{Tanaka}}}, \bibinfo {author} {\bibfnamefont {S.}~\bibnamefont {{Okuzumi}}}, \bibinfo {author} {\bibfnamefont {H.}~\bibnamefont {{Kobayashi}}}, \bibinfo {author} {\bibfnamefont {T.}~\bibnamefont {{Suyama}}}, \bibinfo {author} {\bibfnamefont {H.}~\bibnamefont {{Kimura}}},\ and\ \bibinfo {author} {\bibfnamefont {T.}~\bibnamefont {{Yamamoto}}},\ }\href {https://doi.org/10.1051/0004-6361/201322259} {\bibfield  {journal} {\bibinfo  {journal} {Astronomy \& Astrophysics}\ }\textbf {\bibinfo {volume} {559}},\ \bibinfo {eid} {A62} (\bibinfo {year} {2013})}\BibitemShut {NoStop}%
\bibitem [{\citenamefont {{Hasegawa}}\ \emph {et~al.}(2023)\citenamefont {{Hasegawa}}, \citenamefont {{Suzuki}}, \citenamefont {{Tanaka}}, \citenamefont {{Kobayashi}},\ and\ \citenamefont {{Wada}}}]{2023ApJ...944...38H}%
  \BibitemOpen
  \bibfield  {author} {\bibinfo {author} {\bibfnamefont {Y.}~\bibnamefont {{Hasegawa}}}, \bibinfo {author} {\bibfnamefont {T.~K.}\ \bibnamefont {{Suzuki}}}, \bibinfo {author} {\bibfnamefont {H.}~\bibnamefont {{Tanaka}}}, \bibinfo {author} {\bibfnamefont {H.}~\bibnamefont {{Kobayashi}}},\ and\ \bibinfo {author} {\bibfnamefont {K.}~\bibnamefont {{Wada}}},\ }\href {https://doi.org/10.3847/1538-4357/acadda} {\bibfield  {journal} {\bibinfo  {journal} {Astrophysical Journal}\ }\textbf {\bibinfo {volume} {944}},\ \bibinfo {eid} {38} (\bibinfo {year} {2023})}\BibitemShut {NoStop}%
\bibitem [{\citenamefont {{Hasegawa}}\ \emph {et~al.}(2021)\citenamefont {{Hasegawa}}, \citenamefont {{Suzuki}}, \citenamefont {{Tanaka}}, \citenamefont {{Kobayashi}},\ and\ \citenamefont {{Wada}}}]{2021ApJ...915...22H}%
  \BibitemOpen
  \bibfield  {author} {\bibinfo {author} {\bibfnamefont {Y.}~\bibnamefont {{Hasegawa}}}, \bibinfo {author} {\bibfnamefont {T.~K.}\ \bibnamefont {{Suzuki}}}, \bibinfo {author} {\bibfnamefont {H.}~\bibnamefont {{Tanaka}}}, \bibinfo {author} {\bibfnamefont {H.}~\bibnamefont {{Kobayashi}}},\ and\ \bibinfo {author} {\bibfnamefont {K.}~\bibnamefont {{Wada}}},\ }\href {https://doi.org/10.3847/1538-4357/abf6cf} {\bibfield  {journal} {\bibinfo  {journal} {Astrophysical Journal}\ }\textbf {\bibinfo {volume} {915}},\ \bibinfo {eid} {22} (\bibinfo {year} {2021})}\BibitemShut {NoStop}%
\bibitem [{\citenamefont {{Tatsuuma}}\ \emph {et~al.}(2019)\citenamefont {{Tatsuuma}}, \citenamefont {{Kataoka}},\ and\ \citenamefont {{Tanaka}}}]{2019ApJ...874..159T}%
  \BibitemOpen
  \bibfield  {author} {\bibinfo {author} {\bibfnamefont {M.}~\bibnamefont {{Tatsuuma}}}, \bibinfo {author} {\bibfnamefont {A.}~\bibnamefont {{Kataoka}}},\ and\ \bibinfo {author} {\bibfnamefont {H.}~\bibnamefont {{Tanaka}}},\ }\href {https://doi.org/10.3847/1538-4357/ab09f7} {\bibfield  {journal} {\bibinfo  {journal} {Astrophysical Journal}\ }\textbf {\bibinfo {volume} {874}},\ \bibinfo {eid} {159} (\bibinfo {year} {2019})}\BibitemShut {NoStop}%
\bibitem [{\citenamefont {{Wada}}\ \emph {et~al.}(2008)\citenamefont {{Wada}}, \citenamefont {{Tanaka}}, \citenamefont {{Suyama}}, \citenamefont {{Kimura}},\ and\ \citenamefont {{Yamamoto}}}]{2008ApJ...677.1296W}%
  \BibitemOpen
  \bibfield  {author} {\bibinfo {author} {\bibfnamefont {K.}~\bibnamefont {{Wada}}}, \bibinfo {author} {\bibfnamefont {H.}~\bibnamefont {{Tanaka}}}, \bibinfo {author} {\bibfnamefont {T.}~\bibnamefont {{Suyama}}}, \bibinfo {author} {\bibfnamefont {H.}~\bibnamefont {{Kimura}}},\ and\ \bibinfo {author} {\bibfnamefont {T.}~\bibnamefont {{Yamamoto}}},\ }\href {https://doi.org/10.1086/529511} {\bibfield  {journal} {\bibinfo  {journal} {\apj}\ }\textbf {\bibinfo {volume} {677}},\ \bibinfo {pages} {1296} (\bibinfo {year} {2008})}\BibitemShut {NoStop}%
\bibitem [{\citenamefont {{Johnson}}\ \emph {et~al.}(1971)\citenamefont {{Johnson}}, \citenamefont {{Kendall}},\ and\ \citenamefont {{Roberts}}}]{1971RSPSA.324..301J}%
  \BibitemOpen
  \bibfield  {author} {\bibinfo {author} {\bibfnamefont {K.~L.}\ \bibnamefont {{Johnson}}}, \bibinfo {author} {\bibfnamefont {K.}~\bibnamefont {{Kendall}}},\ and\ \bibinfo {author} {\bibfnamefont {A.~D.}\ \bibnamefont {{Roberts}}},\ }\href {https://doi.org/10.1098/rspa.1971.0141} {\bibfield  {journal} {\bibinfo  {journal} {Proceedings of the Royal Society of London Series A}\ }\textbf {\bibinfo {volume} {324}},\ \bibinfo {pages} {301} (\bibinfo {year} {1971})}\BibitemShut {NoStop}%
\bibitem [{\citenamefont {{Dominik}}\ and\ \citenamefont {{Tielens}}(1995)}]{1995PMagA..72..783D}%
  \BibitemOpen
  \bibfield  {author} {\bibinfo {author} {\bibfnamefont {C.}~\bibnamefont {{Dominik}}}\ and\ \bibinfo {author} {\bibfnamefont {A.~G.~G.~M.}\ \bibnamefont {{Tielens}}},\ }\href {https://doi.org/10.1080/01418619508243800} {\bibfield  {journal} {\bibinfo  {journal} {Philosophical Magazine, Part A}\ }\textbf {\bibinfo {volume} {72}},\ \bibinfo {pages} {783} (\bibinfo {year} {1995})}\BibitemShut {NoStop}%
\bibitem [{\citenamefont {{Dominik}}\ and\ \citenamefont {{Tielens}}(1996)}]{1996PMagA..73.1279D}%
  \BibitemOpen
  \bibfield  {author} {\bibinfo {author} {\bibfnamefont {C.}~\bibnamefont {{Dominik}}}\ and\ \bibinfo {author} {\bibfnamefont {A.~G.~G.~M.}\ \bibnamefont {{Tielens}}},\ }\href {https://doi.org/10.1080/01418619608245132} {\bibfield  {journal} {\bibinfo  {journal} {Philosophical Magazine, Part A}\ }\textbf {\bibinfo {volume} {73}},\ \bibinfo {pages} {1279} (\bibinfo {year} {1996})}\BibitemShut {NoStop}%
\bibitem [{\citenamefont {{Dominik}}\ and\ \citenamefont {{Tielens}}(1997)}]{1997ApJ...480..647D}%
  \BibitemOpen
  \bibfield  {author} {\bibinfo {author} {\bibfnamefont {C.}~\bibnamefont {{Dominik}}}\ and\ \bibinfo {author} {\bibfnamefont {A.~G.~G.~M.}\ \bibnamefont {{Tielens}}},\ }\href {https://doi.org/10.1086/303996} {\bibfield  {journal} {\bibinfo  {journal} {Astrophysical Journal}\ }\textbf {\bibinfo {volume} {480}},\ \bibinfo {pages} {647} (\bibinfo {year} {1997})}\BibitemShut {NoStop}%
\bibitem [{\citenamefont {{Johnson}}(1987)}]{1987come.book.....J}%
  \BibitemOpen
  \bibfield  {author} {\bibinfo {author} {\bibfnamefont {K.~L.}\ \bibnamefont {{Johnson}}},\ }\href@noop {} {\emph {\bibinfo {title} {{Contact Mechanics}}}}\ (\bibinfo  {publisher} {Cambridge University Press},\ \bibinfo {year} {1987})\BibitemShut {NoStop}%
\bibitem [{\citenamefont {{Derjaguin}}\ \emph {et~al.}(1975)\citenamefont {{Derjaguin}}, \citenamefont {{Muller}},\ and\ \citenamefont {{Toporov}}}]{1975JCIS...53..314D}%
  \BibitemOpen
  \bibfield  {author} {\bibinfo {author} {\bibfnamefont {B.~V.}\ \bibnamefont {{Derjaguin}}}, \bibinfo {author} {\bibfnamefont {V.~M.}\ \bibnamefont {{Muller}}},\ and\ \bibinfo {author} {\bibfnamefont {Y.~P.}\ \bibnamefont {{Toporov}}},\ }\href {https://doi.org/10.1016/0021-9797(75)90018-1} {\bibfield  {journal} {\bibinfo  {journal} {Journal of Colloid and Interface Science}\ }\textbf {\bibinfo {volume} {53}},\ \bibinfo {pages} {314} (\bibinfo {year} {1975})}\BibitemShut {NoStop}%
\bibitem [{\citenamefont {{Dugdale}}(1960)}]{1960JMPSo...8..100D}%
  \BibitemOpen
  \bibfield  {author} {\bibinfo {author} {\bibfnamefont {D.~S.}\ \bibnamefont {{Dugdale}}},\ }\href {https://doi.org/10.1016/0022-5096(60)90013-2} {\bibfield  {journal} {\bibinfo  {journal} {Journal of Mechanics Physics of Solids}\ }\textbf {\bibinfo {volume} {8}},\ \bibinfo {pages} {100} (\bibinfo {year} {1960})}\BibitemShut {NoStop}%
\bibitem [{\citenamefont {{Maugis}}(1992)}]{1992JCIS..150..243M}%
  \BibitemOpen
  \bibfield  {author} {\bibinfo {author} {\bibfnamefont {D.}~\bibnamefont {{Maugis}}},\ }\href {https://doi.org/10.1016/0021-9797(92)90285-T} {\bibfield  {journal} {\bibinfo  {journal} {Journal of Colloid and Interface Science}\ }\textbf {\bibinfo {volume} {150}},\ \bibinfo {pages} {243} (\bibinfo {year} {1992})}\BibitemShut {NoStop}%
\bibitem [{\citenamefont {{Tanaka}}\ \emph {et~al.}(2012)\citenamefont {{Tanaka}}, \citenamefont {{Wada}}, \citenamefont {{Suyama}},\ and\ \citenamefont {{Okuzumi}}}]{2012PThPS.195..101T}%
  \BibitemOpen
  \bibfield  {author} {\bibinfo {author} {\bibfnamefont {H.}~\bibnamefont {{Tanaka}}}, \bibinfo {author} {\bibfnamefont {K.}~\bibnamefont {{Wada}}}, \bibinfo {author} {\bibfnamefont {T.}~\bibnamefont {{Suyama}}},\ and\ \bibinfo {author} {\bibfnamefont {S.}~\bibnamefont {{Okuzumi}}},\ }\href {https://doi.org/10.1143/PTPS.195.101} {\bibfield  {journal} {\bibinfo  {journal} {Progress of Theoretical Physics Supplement}\ }\textbf {\bibinfo {volume} {195}},\ \bibinfo {pages} {101} (\bibinfo {year} {2012})}\BibitemShut {NoStop}%
\bibitem [{\citenamefont {{Takato}}\ \emph {et~al.}(2014)\citenamefont {{Takato}}, \citenamefont {{Sen}},\ and\ \citenamefont {{Lechman}}}]{2014PhRvE..89c3308T}%
  \BibitemOpen
  \bibfield  {author} {\bibinfo {author} {\bibfnamefont {Y.}~\bibnamefont {{Takato}}}, \bibinfo {author} {\bibfnamefont {S.}~\bibnamefont {{Sen}}},\ and\ \bibinfo {author} {\bibfnamefont {J.~B.}\ \bibnamefont {{Lechman}}},\ }\href {https://doi.org/10.1103/PhysRevE.89.033308} {\bibfield  {journal} {\bibinfo  {journal} {\pre}\ }\textbf {\bibinfo {volume} {89}},\ \bibinfo {eid} {033308} (\bibinfo {year} {2014})}\BibitemShut {NoStop}%
\bibitem [{\citenamefont {Takato}\ \emph {et~al.}(2015)\citenamefont {Takato}, \citenamefont {Benson},\ and\ \citenamefont {Sen}}]{PhysRevE.92.0324032015}%
  \BibitemOpen
  \bibfield  {author} {\bibinfo {author} {\bibfnamefont {Y.}~\bibnamefont {Takato}}, \bibinfo {author} {\bibfnamefont {M.~E.}\ \bibnamefont {Benson}},\ and\ \bibinfo {author} {\bibfnamefont {S.}~\bibnamefont {Sen}},\ }\href {https://doi.org/10.1103/PhysRevE.92.032403} {\bibfield  {journal} {\bibinfo  {journal} {Phys. Rev. E}\ }\textbf {\bibinfo {volume} {92}},\ \bibinfo {pages} {032403} (\bibinfo {year} {2015})}\BibitemShut {NoStop}%
\bibitem [{\citenamefont {{Krijt}}\ \emph {et~al.}(2013)\citenamefont {{Krijt}}, \citenamefont {{G{\"u}ttler}}, \citenamefont {{Hei{\ss}elmann}}, \citenamefont {{Dominik}},\ and\ \citenamefont {{Tielens}}}]{2013JPhD...46Q5303K}%
  \BibitemOpen
  \bibfield  {author} {\bibinfo {author} {\bibfnamefont {S.}~\bibnamefont {{Krijt}}}, \bibinfo {author} {\bibfnamefont {C.}~\bibnamefont {{G{\"u}ttler}}}, \bibinfo {author} {\bibfnamefont {D.}~\bibnamefont {{Hei{\ss}elmann}}}, \bibinfo {author} {\bibfnamefont {C.}~\bibnamefont {{Dominik}}},\ and\ \bibinfo {author} {\bibfnamefont {A.~G.~G.~M.}\ \bibnamefont {{Tielens}}},\ }\href {https://doi.org/10.1088/0022-3727/46/43/435303} {\bibfield  {journal} {\bibinfo  {journal} {Journal of Physics D Applied Physics}\ }\textbf {\bibinfo {volume} {46}},\ \bibinfo {eid} {435303} (\bibinfo {year} {2013})}\BibitemShut {NoStop}%
\bibitem [{\citenamefont {{Greenwood}}(2004)}]{2004JPhD...37.2557G}%
  \BibitemOpen
  \bibfield  {author} {\bibinfo {author} {\bibfnamefont {J.~A.}\ \bibnamefont {{Greenwood}}},\ }\href {https://doi.org/10.1088/0022-3727/37/18/011} {\bibfield  {journal} {\bibinfo  {journal} {Journal of Physics D Applied Physics}\ }\textbf {\bibinfo {volume} {37}},\ \bibinfo {pages} {2557} (\bibinfo {year} {2004})}\BibitemShut {NoStop}%
\bibitem [{\citenamefont {{Brilliantov}}\ \emph {et~al.}(2007)\citenamefont {{Brilliantov}}, \citenamefont {{Albers}}, \citenamefont {{Spahn}},\ and\ \citenamefont {{P{\"o}schel}}}]{2007PhRvE..76e1302B}%
  \BibitemOpen
  \bibfield  {author} {\bibinfo {author} {\bibfnamefont {N.~V.}\ \bibnamefont {{Brilliantov}}}, \bibinfo {author} {\bibfnamefont {N.}~\bibnamefont {{Albers}}}, \bibinfo {author} {\bibfnamefont {F.}~\bibnamefont {{Spahn}}},\ and\ \bibinfo {author} {\bibfnamefont {T.}~\bibnamefont {{P{\"o}schel}}},\ }\href {https://doi.org/10.1103/PhysRevE.76.051302} {\bibfield  {journal} {\bibinfo  {journal} {\pre}\ }\textbf {\bibinfo {volume} {76}},\ \bibinfo {eid} {051302} (\bibinfo {year} {2007})}\BibitemShut {NoStop}%
\bibitem [{\citenamefont {{Kim}}\ and\ \citenamefont {{Dunn}}(2008)}]{2008JAerS..39..373K}%
  \BibitemOpen
  \bibfield  {author} {\bibinfo {author} {\bibfnamefont {O.~V.}\ \bibnamefont {{Kim}}}\ and\ \bibinfo {author} {\bibfnamefont {P.~F.}\ \bibnamefont {{Dunn}}},\ }\href {https://doi.org/10.1016/j.jaerosci.2007.12.007} {\bibfield  {journal} {\bibinfo  {journal} {Journal of Aerosol Science}\ }\textbf {\bibinfo {volume} {39}},\ \bibinfo {pages} {373} (\bibinfo {year} {2008})}\BibitemShut {NoStop}%
\bibitem [{\citenamefont {{Wall}}\ \emph {et~al.}(1990)\citenamefont {{Wall}}, \citenamefont {{John}}, \citenamefont {{Wang}},\ and\ \citenamefont {{Goren}}}]{1990AerST..12..926W}%
  \BibitemOpen
  \bibfield  {author} {\bibinfo {author} {\bibfnamefont {S.}~\bibnamefont {{Wall}}}, \bibinfo {author} {\bibfnamefont {W.}~\bibnamefont {{John}}}, \bibinfo {author} {\bibfnamefont {H.-C.}\ \bibnamefont {{Wang}}},\ and\ \bibinfo {author} {\bibfnamefont {S.~L.}\ \bibnamefont {{Goren}}},\ }\href {https://doi.org/10.1080/02786829008959404} {\bibfield  {journal} {\bibinfo  {journal} {Aerosol Science Technology}\ }\textbf {\bibinfo {volume} {12}},\ \bibinfo {pages} {926} (\bibinfo {year} {1990})}\BibitemShut {NoStop}%
\bibitem [{\citenamefont {{Anders}}\ and\ \citenamefont {{Urbassek}}(2021)}]{2021A&A...647L..13A}%
  \BibitemOpen
  \bibfield  {author} {\bibinfo {author} {\bibfnamefont {C.}~\bibnamefont {{Anders}}}\ and\ \bibinfo {author} {\bibfnamefont {H.~M.}\ \bibnamefont {{Urbassek}}},\ }\href {https://doi.org/10.1051/0004-6361/202140295} {\bibfield  {journal} {\bibinfo  {journal} {Astronomy \& Astrophysics}\ }\textbf {\bibinfo {volume} {647}},\ \bibinfo {eid} {L13} (\bibinfo {year} {2021})}\BibitemShut {NoStop}%
\bibitem [{\citenamefont {{Nietiadi}}\ \emph {et~al.}(2020{\natexlab{a}})\citenamefont {{Nietiadi}}, \citenamefont {{Valencia}}, \citenamefont {{Gonzalez}}, \citenamefont {{Bringa}},\ and\ \citenamefont {{Urbassek}}}]{2020A&A...641A.159N}%
  \BibitemOpen
  \bibfield  {author} {\bibinfo {author} {\bibfnamefont {M.~L.}\ \bibnamefont {{Nietiadi}}}, \bibinfo {author} {\bibfnamefont {F.}~\bibnamefont {{Valencia}}}, \bibinfo {author} {\bibfnamefont {R.~I.}\ \bibnamefont {{Gonzalez}}}, \bibinfo {author} {\bibfnamefont {E.~M.}\ \bibnamefont {{Bringa}}},\ and\ \bibinfo {author} {\bibfnamefont {H.~M.}\ \bibnamefont {{Urbassek}}},\ }\href {https://doi.org/10.1051/0004-6361/202038183} {\bibfield  {journal} {\bibinfo  {journal} {Astronomy \& Astrophysics}\ }\textbf {\bibinfo {volume} {641}},\ \bibinfo {eid} {A159} (\bibinfo {year} {2020}{\natexlab{a}})}\BibitemShut {NoStop}%
\bibitem [{\citenamefont {{Nietiadi}}\ \emph {et~al.}(2020{\natexlab{b}})\citenamefont {{Nietiadi}}, \citenamefont {{Rosandi}},\ and\ \citenamefont {{Urbassek}}}]{2020NRL....15...67N}%
  \BibitemOpen
  \bibfield  {author} {\bibinfo {author} {\bibfnamefont {M.~L.}\ \bibnamefont {{Nietiadi}}}, \bibinfo {author} {\bibfnamefont {Y.}~\bibnamefont {{Rosandi}}},\ and\ \bibinfo {author} {\bibfnamefont {H.~M.}\ \bibnamefont {{Urbassek}}},\ }\href {https://doi.org/10.1186/s11671-020-03296-y} {\bibfield  {journal} {\bibinfo  {journal} {Nanoscale Research Letters}\ }\textbf {\bibinfo {volume} {15}},\ \bibinfo {eid} {67} (\bibinfo {year} {2020}{\natexlab{b}})}\BibitemShut {NoStop}%
\bibitem [{\citenamefont {{Takato}}\ \emph {et~al.}(2018)\citenamefont {{Takato}}, \citenamefont {{Benson}},\ and\ \citenamefont {{Sen}}}]{2018RSPSA.47470723T}%
  \BibitemOpen
  \bibfield  {author} {\bibinfo {author} {\bibfnamefont {Y.}~\bibnamefont {{Takato}}}, \bibinfo {author} {\bibfnamefont {M.~E.}\ \bibnamefont {{Benson}}},\ and\ \bibinfo {author} {\bibfnamefont {S.}~\bibnamefont {{Sen}}},\ }\href {https://doi.org/10.1098/rspa.2017.0723} {\bibfield  {journal} {\bibinfo  {journal} {Proceedings of the Royal Society of London Series A}\ }\textbf {\bibinfo {volume} {474}},\ \bibinfo {eid} {20170723} (\bibinfo {year} {2018})}\BibitemShut {NoStop}%
\bibitem [{\citenamefont {{Mill{\'a}n}}\ \emph {et~al.}(2016)\citenamefont {{Mill{\'a}n}}, \citenamefont {{Tramontina}}, \citenamefont {{Urbassek}},\ and\ \citenamefont {{Bringa}}}]{2016PhRvE..93f3004M}%
  \BibitemOpen
  \bibfield  {author} {\bibinfo {author} {\bibfnamefont {E.~N.}\ \bibnamefont {{Mill{\'a}n}}}, \bibinfo {author} {\bibfnamefont {D.~R.}\ \bibnamefont {{Tramontina}}}, \bibinfo {author} {\bibfnamefont {H.~M.}\ \bibnamefont {{Urbassek}}},\ and\ \bibinfo {author} {\bibfnamefont {E.~M.}\ \bibnamefont {{Bringa}}},\ }\href {https://doi.org/10.1103/PhysRevE.93.063004} {\bibfield  {journal} {\bibinfo  {journal} {\pre}\ }\textbf {\bibinfo {volume} {93}},\ \bibinfo {eid} {063004} (\bibinfo {year} {2016})}\BibitemShut {NoStop}%
\bibitem [{\citenamefont {{Nietiadi}}\ \emph {et~al.}(2017{\natexlab{a}})\citenamefont {{Nietiadi}}, \citenamefont {{Umst{\"a}tter}}, \citenamefont {{Alabd Alhafez}}, \citenamefont {{Rosandi}}, \citenamefont {{Bringa}},\ and\ \citenamefont {{Urbassek}}}]{2017GeoRL..4410822N}%
  \BibitemOpen
  \bibfield  {author} {\bibinfo {author} {\bibfnamefont {M.~L.}\ \bibnamefont {{Nietiadi}}}, \bibinfo {author} {\bibfnamefont {P.}~\bibnamefont {{Umst{\"a}tter}}}, \bibinfo {author} {\bibfnamefont {I.}~\bibnamefont {{Alabd Alhafez}}}, \bibinfo {author} {\bibfnamefont {Y.}~\bibnamefont {{Rosandi}}}, \bibinfo {author} {\bibfnamefont {E.~M.}\ \bibnamefont {{Bringa}}},\ and\ \bibinfo {author} {\bibfnamefont {H.~M.}\ \bibnamefont {{Urbassek}}},\ }\href {https://doi.org/10.1002/2017GL075395} {\bibfield  {journal} {\bibinfo  {journal} {Geophysical Research Letters}\ }\textbf {\bibinfo {volume} {44}},\ \bibinfo {pages} {10,822} (\bibinfo {year} {2017}{\natexlab{a}})}\BibitemShut {NoStop}%
\bibitem [{\citenamefont {{Nietiadi}}\ \emph {et~al.}(2020{\natexlab{c}})\citenamefont {{Nietiadi}}, \citenamefont {{Rosandi}},\ and\ \citenamefont {{Urbassek}}}]{2020Icar..35213996N}%
  \BibitemOpen
  \bibfield  {author} {\bibinfo {author} {\bibfnamefont {M.~L.}\ \bibnamefont {{Nietiadi}}}, \bibinfo {author} {\bibfnamefont {Y.}~\bibnamefont {{Rosandi}}},\ and\ \bibinfo {author} {\bibfnamefont {H.~M.}\ \bibnamefont {{Urbassek}}},\ }\href {https://doi.org/10.1016/j.icarus.2020.113996} {\bibfield  {journal} {\bibinfo  {journal} {Icarus}\ }\textbf {\bibinfo {volume} {352}},\ \bibinfo {eid} {113996} (\bibinfo {year} {2020}{\natexlab{c}})}\BibitemShut {NoStop}%
\bibitem [{\citenamefont {{Nietiadi}}\ \emph {et~al.}(2017{\natexlab{b}})\citenamefont {{Nietiadi}}, \citenamefont {{Umst{\"a}tter}}, \citenamefont {{Tjong}}, \citenamefont {{Rosandi}}, \citenamefont {{Mill{\'a}n}}, \citenamefont {{Bringa}},\ and\ \citenamefont {{Urbassek}}}]{2017PCCP...1916555N}%
  \BibitemOpen
  \bibfield  {author} {\bibinfo {author} {\bibfnamefont {M.~L.}\ \bibnamefont {{Nietiadi}}}, \bibinfo {author} {\bibfnamefont {P.}~\bibnamefont {{Umst{\"a}tter}}}, \bibinfo {author} {\bibfnamefont {T.}~\bibnamefont {{Tjong}}}, \bibinfo {author} {\bibfnamefont {Y.}~\bibnamefont {{Rosandi}}}, \bibinfo {author} {\bibfnamefont {E.~N.}\ \bibnamefont {{Mill{\'a}n}}}, \bibinfo {author} {\bibfnamefont {E.~M.}\ \bibnamefont {{Bringa}}},\ and\ \bibinfo {author} {\bibfnamefont {H.~M.}\ \bibnamefont {{Urbassek}}},\ }\href {https://doi.org/10.1039/C7CP02106B} {\bibfield  {journal} {\bibinfo  {journal} {Physical Chemistry Chemical Physics (Incorporating Faraday Transactions)}\ }\textbf {\bibinfo {volume} {19}},\ \bibinfo {pages} {16555} (\bibinfo {year} {2017}{\natexlab{b}})}\BibitemShut {NoStop}%
\bibitem [{\citenamefont {{Nietiadi}}\ \emph {et~al.}(2019)\citenamefont {{Nietiadi}}, \citenamefont {{Mill{\'a}n}}, \citenamefont {{Bringa}},\ and\ \citenamefont {{Urbassek}}}]{2019PhRvE..99c2904N}%
  \BibitemOpen
  \bibfield  {author} {\bibinfo {author} {\bibfnamefont {M.~L.}\ \bibnamefont {{Nietiadi}}}, \bibinfo {author} {\bibfnamefont {E.~N.}\ \bibnamefont {{Mill{\'a}n}}}, \bibinfo {author} {\bibfnamefont {E.~M.}\ \bibnamefont {{Bringa}}},\ and\ \bibinfo {author} {\bibfnamefont {H.~M.}\ \bibnamefont {{Urbassek}}},\ }\href {https://doi.org/10.1103/PhysRevE.99.032904} {\bibfield  {journal} {\bibinfo  {journal} {\pre}\ }\textbf {\bibinfo {volume} {99}},\ \bibinfo {eid} {032904} (\bibinfo {year} {2019})}\BibitemShut {NoStop}%
\bibitem [{\citenamefont {{Umst{\"a}tter}}\ and\ \citenamefont {{Urbassek}}(2021)}]{2021NatSR..1114591U}%
  \BibitemOpen
  \bibfield  {author} {\bibinfo {author} {\bibfnamefont {P.}~\bibnamefont {{Umst{\"a}tter}}}\ and\ \bibinfo {author} {\bibfnamefont {H.~M.}\ \bibnamefont {{Urbassek}}},\ }\href {https://doi.org/10.1038/s41598-021-93984-1} {\bibfield  {journal} {\bibinfo  {journal} {Scientific Reports}\ }\textbf {\bibinfo {volume} {11}},\ \bibinfo {eid} {14591} (\bibinfo {year} {2021})}\BibitemShut {NoStop}%
\bibitem [{\citenamefont {{Awasthi}}\ \emph {et~al.}(2007)\citenamefont {{Awasthi}}, \citenamefont {{Hendy}}, \citenamefont {{Zoontjens}}, \citenamefont {{Brown}},\ and\ \citenamefont {{Natali}}}]{2007PhRvB..76k5437A}%
  \BibitemOpen
  \bibfield  {author} {\bibinfo {author} {\bibfnamefont {A.}~\bibnamefont {{Awasthi}}}, \bibinfo {author} {\bibfnamefont {S.~C.}\ \bibnamefont {{Hendy}}}, \bibinfo {author} {\bibfnamefont {P.}~\bibnamefont {{Zoontjens}}}, \bibinfo {author} {\bibfnamefont {S.~A.}\ \bibnamefont {{Brown}}},\ and\ \bibinfo {author} {\bibfnamefont {F.}~\bibnamefont {{Natali}}},\ }\href {https://doi.org/10.1103/PhysRevB.76.115437} {\bibfield  {journal} {\bibinfo  {journal} {\prb}\ }\textbf {\bibinfo {volume} {76}},\ \bibinfo {eid} {115437} (\bibinfo {year} {2007})},\ \Eprint {https://arxiv.org/abs/0709.0994} {arXiv:0709.0994 [cond-mat.other]} \BibitemShut {NoStop}%
\bibitem [{Note1()}]{Note1}%
  \BibitemOpen
  \bibinfo {note} {Note that $\gamma $ in this paper is equal to $\gamma _L$ and twice $\gamma $ of Krijt et al. \cite {2013JPhD...46Q5303K}.}\BibitemShut {Stop}%
\bibitem [{\citenamefont {{Wada}}\ \emph {et~al.}(2007)\citenamefont {{Wada}}, \citenamefont {{Tanaka}}, \citenamefont {{Suyama}}, \citenamefont {{Kimura}},\ and\ \citenamefont {{Yamamoto}}}]{2007ApJ...661..320W}%
  \BibitemOpen
  \bibfield  {author} {\bibinfo {author} {\bibfnamefont {K.}~\bibnamefont {{Wada}}}, \bibinfo {author} {\bibfnamefont {H.}~\bibnamefont {{Tanaka}}}, \bibinfo {author} {\bibfnamefont {T.}~\bibnamefont {{Suyama}}}, \bibinfo {author} {\bibfnamefont {H.}~\bibnamefont {{Kimura}}},\ and\ \bibinfo {author} {\bibfnamefont {T.}~\bibnamefont {{Yamamoto}}},\ }\href {https://doi.org/10.1086/514332} {\bibfield  {journal} {\bibinfo  {journal} {Astrophysical Journal}\ }\textbf {\bibinfo {volume} {661}},\ \bibinfo {pages} {320} (\bibinfo {year} {2007})}\BibitemShut {NoStop}%
\bibitem [{\citenamefont {Brilliantov}\ \emph {et~al.}(1996)\citenamefont {Brilliantov}, \citenamefont {Spahn}, \citenamefont {Hertzsch},\ and\ \citenamefont {P\"oschel}}]{1996PhysRevE.53.5382B}%
  \BibitemOpen
  \bibfield  {author} {\bibinfo {author} {\bibfnamefont {N.~V.}\ \bibnamefont {Brilliantov}}, \bibinfo {author} {\bibfnamefont {F.}~\bibnamefont {Spahn}}, \bibinfo {author} {\bibfnamefont {J.-M.}\ \bibnamefont {Hertzsch}},\ and\ \bibinfo {author} {\bibfnamefont {T.}~\bibnamefont {P\"oschel}},\ }\href {https://doi.org/10.1103/PhysRevE.53.5382} {\bibfield  {journal} {\bibinfo  {journal} {Phys. Rev. E}\ }\textbf {\bibinfo {volume} {53}},\ \bibinfo {pages} {5382} (\bibinfo {year} {1996})}\BibitemShut {NoStop}%
\bibitem [{\citenamefont {{Muller}}\ \emph {et~al.}(1980)\citenamefont {{Muller}}, \citenamefont {{Yushchenko}},\ and\ \citenamefont {{Derjaguin}}}]{1980JCIS...77...91M}%
  \BibitemOpen
  \bibfield  {author} {\bibinfo {author} {\bibfnamefont {V.~M.}\ \bibnamefont {{Muller}}}, \bibinfo {author} {\bibfnamefont {V.~S.}\ \bibnamefont {{Yushchenko}}},\ and\ \bibinfo {author} {\bibfnamefont {B.~V.}\ \bibnamefont {{Derjaguin}}},\ }\href {https://doi.org/10.1016/0021-9797(80)90419-1} {\bibfield  {journal} {\bibinfo  {journal} {Journal of Colloid and Interface Science}\ }\textbf {\bibinfo {volume} {77}},\ \bibinfo {pages} {91} (\bibinfo {year} {1980})}\BibitemShut {NoStop}%
\bibitem [{\citenamefont {{Arakawa}}\ and\ \citenamefont {{Krijt}}(2021)}]{2021ApJ...910..130A}%
  \BibitemOpen
  \bibfield  {author} {\bibinfo {author} {\bibfnamefont {S.}~\bibnamefont {{Arakawa}}}\ and\ \bibinfo {author} {\bibfnamefont {S.}~\bibnamefont {{Krijt}}},\ }\href {https://doi.org/10.3847/1538-4357/abe61d} {\bibfield  {journal} {\bibinfo  {journal} {Astrophysical Journal}\ }\textbf {\bibinfo {volume} {910}},\ \bibinfo {eid} {130} (\bibinfo {year} {2021})}\BibitemShut {NoStop}%
\bibitem [{\citenamefont {{Griffith}}(1921)}]{1921RSPTA.221..163G}%
  \BibitemOpen
  \bibfield  {author} {\bibinfo {author} {\bibfnamefont {A.~A.}\ \bibnamefont {{Griffith}}},\ }\href {https://doi.org/10.1098/rsta.1921.0006} {\bibfield  {journal} {\bibinfo  {journal} {Philosophical Transactions of the Royal Society of London Series A}\ }\textbf {\bibinfo {volume} {221}},\ \bibinfo {pages} {163} (\bibinfo {year} {1921})}\BibitemShut {NoStop}%
\bibitem [{\citenamefont {Wahl}\ \emph {et~al.}(2006)\citenamefont {Wahl}, \citenamefont {Asif}, \citenamefont {Greenwood},\ and\ \citenamefont {Johnson}}]{WAHL2006178}%
  \BibitemOpen
  \bibfield  {author} {\bibinfo {author} {\bibfnamefont {K.}~\bibnamefont {Wahl}}, \bibinfo {author} {\bibfnamefont {S.}~\bibnamefont {Asif}}, \bibinfo {author} {\bibfnamefont {J.}~\bibnamefont {Greenwood}},\ and\ \bibinfo {author} {\bibfnamefont {K.}~\bibnamefont {Johnson}},\ }\href {https://doi.org/https://doi.org/10.1016/j.jcis.2005.08.028} {\bibfield  {journal} {\bibinfo  {journal} {Journal of Colloid and Interface Science}\ }\textbf {\bibinfo {volume} {296}},\ \bibinfo {pages} {178} (\bibinfo {year} {2006})}\BibitemShut {NoStop}%
\bibitem [{\citenamefont {Thornton}\ and\ \citenamefont {Ning}(1998)}]{THORNTON1998154}%
  \BibitemOpen
  \bibfield  {author} {\bibinfo {author} {\bibfnamefont {C.}~\bibnamefont {Thornton}}\ and\ \bibinfo {author} {\bibfnamefont {Z.}~\bibnamefont {Ning}},\ }\href {https://doi.org/https://doi.org/10.1016/S0032-5910(98)00099-0} {\bibfield  {journal} {\bibinfo  {journal} {Powder Technology}\ }\textbf {\bibinfo {volume} {99}},\ \bibinfo {pages} {154} (\bibinfo {year} {1998})}\BibitemShut {NoStop}%
\bibitem [{\citenamefont {Thompson}\ \emph {et~al.}(2022)\citenamefont {Thompson}, \citenamefont {Aktulga}, \citenamefont {Berger}, \citenamefont {Bolintineanu}, \citenamefont {Brown}, \citenamefont {Crozier}, \citenamefont {in~'t Veld}, \citenamefont {Kohlmeyer}, \citenamefont {Moore}, \citenamefont {Nguyen}, \citenamefont {Shan}, \citenamefont {Stevens}, \citenamefont {Tranchida}, \citenamefont {Trott},\ and\ \citenamefont {Plimpton}}]{LAMMPS}%
  \BibitemOpen
  \bibfield  {author} {\bibinfo {author} {\bibfnamefont {A.~P.}\ \bibnamefont {Thompson}}, \bibinfo {author} {\bibfnamefont {H.~M.}\ \bibnamefont {Aktulga}}, \bibinfo {author} {\bibfnamefont {R.}~\bibnamefont {Berger}}, \bibinfo {author} {\bibfnamefont {D.~S.}\ \bibnamefont {Bolintineanu}}, \bibinfo {author} {\bibfnamefont {W.~M.}\ \bibnamefont {Brown}}, \bibinfo {author} {\bibfnamefont {P.~S.}\ \bibnamefont {Crozier}}, \bibinfo {author} {\bibfnamefont {P.~J.}\ \bibnamefont {in~'t Veld}}, \bibinfo {author} {\bibfnamefont {A.}~\bibnamefont {Kohlmeyer}}, \bibinfo {author} {\bibfnamefont {S.~G.}\ \bibnamefont {Moore}}, \bibinfo {author} {\bibfnamefont {T.~D.}\ \bibnamefont {Nguyen}}, \bibinfo {author} {\bibfnamefont {R.}~\bibnamefont {Shan}}, \bibinfo {author} {\bibfnamefont {M.~J.}\ \bibnamefont {Stevens}}, \bibinfo {author} {\bibfnamefont {J.}~\bibnamefont {Tranchida}}, \bibinfo {author} {\bibfnamefont {C.}~\bibnamefont {Trott}},\ and\ \bibinfo {author} {\bibfnamefont {S.~J.}\ \bibnamefont {Plimpton}},\ }\href
  {https://doi.org/10.1016/j.cpc.2021.108171} {\bibfield  {journal} {\bibinfo  {journal} {Comp. Phys. Comm.}\ }\textbf {\bibinfo {volume} {271}},\ \bibinfo {pages} {108171} (\bibinfo {year} {2022})}\BibitemShut {NoStop}%
\bibitem [{\citenamefont {Michels}\ \emph {et~al.}(1949)\citenamefont {Michels}, \citenamefont {Wijker},\ and\ \citenamefont {Wijker}}]{MICHELS1949627}%
  \BibitemOpen
  \bibfield  {author} {\bibinfo {author} {\bibfnamefont {A.}~\bibnamefont {Michels}}, \bibinfo {author} {\bibfnamefont {H.}~\bibnamefont {Wijker}},\ and\ \bibinfo {author} {\bibfnamefont {H.}~\bibnamefont {Wijker}},\ }\href {https://doi.org/https://doi.org/10.1016/0031-8914(49)90119-6} {\bibfield  {journal} {\bibinfo  {journal} {Physica}\ }\textbf {\bibinfo {volume} {15}},\ \bibinfo {pages} {627} (\bibinfo {year} {1949})}\BibitemShut {NoStop}%
\bibitem [{\citenamefont {Hirschfelder}\ \emph {et~al.}(1967)\citenamefont {Hirschfelder}, \citenamefont {Curtiss},\ and\ \citenamefont {Bird}}]{1967Mtog}%
  \BibitemOpen
  \bibfield  {author} {\bibinfo {author} {\bibfnamefont {J.~O.}\ \bibnamefont {Hirschfelder}}, \bibinfo {author} {\bibfnamefont {C.~F.}\ \bibnamefont {Curtiss}},\ and\ \bibinfo {author} {\bibfnamefont {R.~B.}\ \bibnamefont {Bird}},\ }\href@noop {} {\emph {\bibinfo {title} {Molecular theory of gases and liquids}}}\ (\bibinfo  {publisher} {Wiley, New York},\ \bibinfo {year} {1967})\BibitemShut {NoStop}%
\bibitem [{\citenamefont {Lu}\ and\ \citenamefont {Pound}(1975)}]{1975PhysSS..30..619H}%
  \BibitemOpen
  \bibfield  {author} {\bibinfo {author} {\bibfnamefont {T.~H.}\ \bibnamefont {Lu}}\ and\ \bibinfo {author} {\bibfnamefont {G.~M.}\ \bibnamefont {Pound}},\ }\href {https://doi.org/10.1002/PSSA.2210300223} {\bibfield  {journal} {\bibinfo  {journal} {Physica Status Solidi (a)}\ }\textbf {\bibinfo {volume} {30}},\ \bibinfo {pages} {619} (\bibinfo {year} {1975})}\BibitemShut {NoStop}%
\bibitem [{\citenamefont {Quesnel}\ \emph {et~al.}(1993)\citenamefont {Quesnel}, \citenamefont {Rimai},\ and\ \citenamefont {DeMejo}}]{PhysRevB.48.6795}%
  \BibitemOpen
  \bibfield  {author} {\bibinfo {author} {\bibfnamefont {D.~J.}\ \bibnamefont {Quesnel}}, \bibinfo {author} {\bibfnamefont {D.~S.}\ \bibnamefont {Rimai}},\ and\ \bibinfo {author} {\bibfnamefont {L.~P.}\ \bibnamefont {DeMejo}},\ }\href {https://doi.org/10.1103/PhysRevB.48.6795} {\bibfield  {journal} {\bibinfo  {journal} {Phys. Rev. B}\ }\textbf {\bibinfo {volume} {48}},\ \bibinfo {pages} {6795} (\bibinfo {year} {1993})}\BibitemShut {NoStop}%
\bibitem [{\citenamefont {{Stukowski}}(2010)}]{ovito}%
  \BibitemOpen
  \bibfield  {author} {\bibinfo {author} {\bibfnamefont {A.}~\bibnamefont {{Stukowski}}},\ }\href {https://doi.org/10.1088/0965-0393/18/1/015012} {\bibfield  {journal} {\bibinfo  {journal} {Modelling Simul. Mater. Sci. Eng.}\ }\textbf {\bibinfo {volume} {18}},\ \bibinfo {eid} {015012} (\bibinfo {year} {2010})}\BibitemShut {NoStop}%
\bibitem [{\citenamefont {Petch}(1953)}]{1953Iron.174..25P}%
  \BibitemOpen
  \bibfield  {author} {\bibinfo {author} {\bibfnamefont {N.~J.}\ \bibnamefont {Petch}},\ }\href@noop {} {\bibfield  {journal} {\bibinfo  {journal} {J. Iron Steel Inst.}\ }\textbf {\bibinfo {volume} {174}},\ \bibinfo {pages} {25} (\bibinfo {year} {1953})}\BibitemShut {NoStop}%
\bibitem [{\citenamefont {{Greenwood}}\ and\ \citenamefont {{Johnson}}(2006)}]{2006JCIS..296..284G}%
  \BibitemOpen
  \bibfield  {author} {\bibinfo {author} {\bibfnamefont {J.~A.}\ \bibnamefont {{Greenwood}}}\ and\ \bibinfo {author} {\bibfnamefont {K.~L.}\ \bibnamefont {{Johnson}}},\ }\href {https://doi.org/10.1016/j.jcis.2005.08.069} {\bibfield  {journal} {\bibinfo  {journal} {Journal of Colloid and Interface Science}\ }\textbf {\bibinfo {volume} {296}},\ \bibinfo {pages} {284} (\bibinfo {year} {2006})}\BibitemShut {NoStop}%
\bibitem [{\citenamefont {{Kittel}}(1976)}]{1976itss.book.....K}%
  \BibitemOpen
  \bibfield  {author} {\bibinfo {author} {\bibfnamefont {C.}~\bibnamefont {{Kittel}}},\ }\href@noop {} {\emph {\bibinfo {title} {{Introduction to solid state physics}}}},\ \bibinfo {edition} {5th}\ ed.\ (\bibinfo  {publisher} {Wiley, New York},\ \bibinfo {year} {1976})\BibitemShut {NoStop}%
\end{thebibliography}%

\end{document}